\documentclass[preprint2, times, tighten, twocolappendix]{aastex631}

\usepackage{CJK}
\usepackage[caption=false]{subfig}

\newcommand{\Msun}{\ensuremath{{M_{\odot}}}}

\newcommand{\logM}{\ensuremath{{M_{\star}/M_{\odot}}}}

\providecommand{\e}[1]{\ensuremath{\times 10^{#1}}}

\begin{document}

\title{To high redshift and low mass: exploring the emergence of quenched galaxies and their environments at $3<z<6$ in the ultra-deep JADES MIRI F770W parallel}

\correspondingauthor{Stacey Alberts}
\email{salberts@arizona.edu}

\author[0000-0002-0786-7307]{Stacey Alberts}
\affiliation{Steward Observatory, University of Arizona, 933 North Cherry Avenue, Tucson, AZ 85721, USA}

\author[0000-0003-2919-7495]{Christina C.\ Williams}
\affiliation{NSF’s National Optical-Infrared Astronomy Research Laboratory, 950 North Cherry Avenue, Tucson, AZ 85719, USA}
\affiliation{Steward Observatory, University of Arizona, 933 North Cherry Avenue, Tucson, AZ 85721, USA}

\author[0000-0003-4337-6211]{Jakob M. Helton}
\affiliation{Steward Observatory, University of Arizona, 933 North Cherry Avenue, Tucson, AZ 85721, USA}

\author[0000-0002-1714-1905]{Katherine A. Suess}\altaffiliation{NHFP Hubble Fellow}\affiliation{Kavli Institute for Particle Astrophysics and Cosmology and Department of Physics, Stanford University, Stanford, CA 94305, USA}

\author[0000-0001-7673-2257]{Zhiyuan Ji}
\affiliation{Steward Observatory, University of Arizona, 933 North Cherry Avenue, Tucson, AZ 85721, USA}

\author[0000-0003-4702-7561]{Irene Shivaei} \affiliation{Centro de Astrobiolog\'ia (CAB), CSIC-INTA, Ctra. de Ajalvir km 4, Torrej\'on de Ardoz, E-28850, Madrid, Spain}

\author[0000-0002-6221-1829]{Jianwei Lyu} 
%(\begin{CJK}{UTF8}{gbsn}吕建伟\end{CJK})}
\affiliation{Steward Observatory, University of Arizona,
933 North Cherry Avenue, Tucson, AZ 85721, USA}

\author[0000-0003-2303-6519]{George Rieke} \affiliation{Steward Observatory and Dept of Planetary Sciences, University of Arizona 933 North Cherry Avenue Tucson AZ 85721 USA}

\author[0000-0003-0215-1104]{William M.\ Baker} \affiliation{Kavli Institute for Cosmology, University of Cambridge, Madingley Road, Cambridge CB3 0HA, UK} \affiliation{Cavendish Laboratory, University of Cambridge, 19 JJ Thomson Avenue, Cambridge CB3 0HE, UK}

\author[0000-0001-8470-7094]{Nina Bonaventura}
\affiliation{Steward Observatory, University of Arizona, 933 North Cherry Avenue, Tucson, AZ 85721, USA}

\author[0000-0002-8651-9879]{Andrew J. Bunker}
\affiliation{Department of Physics, University of Oxford, Denys Wilkinson Building, Keble Road, Oxford OX13RH, U.K.}

\author[0000-0002-6719-380X]{Stefano Carniani}
\affiliation{Scuola Normale Superiore, Piazza dei Cavalieri 7, I-56126 Pisa, Italy}

\author[0000-0003-3458-2275]{Stephane Charlot}
\affiliation{Sorbonne Universit\'e, CNRS, UMR 7095, Institut d'Astrophysique de Paris, 98 bis bd Arago, 75014 Paris, France}

\author[0000-0002-9551-0534]{Emma Curtis-Lake}
\affiliation{Centre for Astrophysics Research, Department of Physics, Astronomy and Mathematics, University of Hertfordshire, Hatfield AL10 9AB, UK}

\author[0000-0003-2388-8172]{Francesco D'Eugenio}
\affiliation{Kavli Institute for Cosmology, University of Cambridge, Madingley Road, Cambridge, CB3 0HA, UK}
\affiliation{Cavendish Laboratory, University of Cambridge, 19 JJ Thomson Avenue, Cambridge, CB3 0HE, UK}

\author[0000-0002-2929-3121]{Daniel J. Eisenstein}
\affiliation{Center for Astrophysics $\vert$ Harvard \& Smithsonian, 60 Garden Street, Cambridge, MA 02138, USA}

\author[0000-0002-2380-9801]{Anna de Graaff}
\affiliation{Max-Planck-Institut f\"ur Astronomie, K\"onigstuhl 17, D-69117, Heidelberg, Germany}

\author[0000-0001-9262-9997]{Kevin N. Hainline}
\affiliation{Steward Observatory, University of Arizona, 933 North Cherry Avenue, Tucson, AZ 85721, USA}

\author[0000-0002-8543-761X]{Ryan Hausen} \affiliation{Department of Physics and Astronomy, The Johns Hopkins University, 3400 N. Charles St., Baltimore, MD 21218}

\author[0000-0002-9280-7594]{Benjamin D. Johnson}
\affiliation{Center for Astrophysics $\vert$ Harvard \& Smithsonian, 60 Garden Street, Cambridge, MA 02138, USA}

\author[0000-0002-4985-3819]{Roberto Maiolino}
\affiliation{Kavli Institute for Cosmology, University of Cambridge, Madingley Road, Cambridge, CB3 0HA, UK; Cavendish Laboratory, University of Cambridge, 19 JJ Thomson Avenue, Cambridge, CB3 0HE, UK}

\author[0000-0002-7392-7814]{Eleonora Parlanti}
\affiliation{Scuola Normale Superiore, Piazza dei Cavalieri 7, I-56126 Pisa, Italy}

\author[0000-0002-7893-6170]{Marcia J. Rieke}\affiliation{Steward Observatory, University of Arizona, 933 North Cherry Avenue, Tucson, AZ 85721, USA}

\author[0000-0002-4271-0364]{Brant E. Robertson}
\affiliation{Department of Astronomy and Astrophysics, University of California, Santa Cruz, 1156 High Street, Santa Cruz, CA 95064, USA}

\author{Yang Sun}
\affiliation{Steward Observatory, University of Arizona, 933 North Cherry Avenue, Tucson, AZ 85721, USA}

\author[0000-0002-8224-4505]{Sandro Tacchella}
\affiliation{Kavli Institute for Cosmology, University of Cambridge, Madingley Road, Cambridge, CB3 0HA, UK}
\affiliation{Cavendish Laboratory, University of Cambridge, 19 JJ Thomson Avenue, Cambridge, CB3 0HE, UK}

\author[0000-0001-9262-9997]{Christopher N. A. Willmer}
\affiliation{Steward Observatory, University of Arizona, 933 North Cherry Avenue, Tucson, AZ 85721, USA}

\author[0000-0002-4201-7367]{Chris J. Willott}
\affil{NRC Herzberg, 5071 West Saanich Rd, Victoria, BC V9E 2E7, Canada}

%\affiliation{Affil2}

%\collaboration{20}{(AAS Journals Data Editors)}

%\author{Author3}
%\affiliation{Affil3}



\begin{abstract}
We present the robust selection of quiescent (QG) and post-starburst (PSB) galaxies using ultra-deep NIRCam and MIRI imaging from the JWST Advanced Deep Extragalactic Survey (JADES). Key to this is MIRI 7.7$\mu$m imaging which breaks the degeneracy between old stellar populations and dust attenuation at $3<z<6$ by providing rest-frame $J$-band. Using this, we identify 23 passively evolving galaxies in UVJ color space in a mass-limited (log $\logM\geq8.5$) sample over 8.8 arcmin$^2$. Evaluation of this selection with and without 7.7$\,\mu$m shows that dense wavelength coverage with NIRCam ($8-11$ bands including $1-4$ medium-bands) can compensate for lacking the $J-$band anchor, meaning that robust selection of high-redshift QGs is possible with NIRCam alone. Our sample is characterized by rapid quenching timescales ($\sim100-600$ Myr) with formation redshifts $z_{\rm f}\lesssim8.5$ and includes a potential record-holding massive QG at $z_{\rm phot}=5.33_{-0.17}^{+0.16}$ and two QGs with evidence for significant residual dust content ($A_{\rm V}\sim1-2$).  In addition, we present a large sample of 12 log $\logM=8.5-9.5$ PSBs, demonstrating that UVJ selection can be extended to low mass.  Analysis of the environment of our sample reveals that the group known as the Cosmic Rose contains a massive QG and a dust-obscured star-forming galaxy (a so-called Jekyll and Hyde pair) plus three additional QGs within $\sim20$ kpc.  Moreover, the Cosmic Rose is part of a larger overdensity at $z\sim3.7$ which contains 7/12 of our low-mass PSBs.  Another 4 low-mass PSBs are members of an overdensity at $z\sim3.4$; this result strongly indicates low-mass PSBs are preferentially associated with overdense environments at $z>3$.
\end{abstract}



\section{Introduction}\label{sec:intro}
 
A persistent challenge to a complete picture of galaxy evolution is explaining the cessation of star formation in galaxies. It is one of the most transformational events in the life cycles of galaxies, giving rise to galaxy bimodality \citep[e.g.][]{kauffmann2003, baldry2004}, underpinning the Hubble Sequence \citep{hubble1926}, and creating the distinctly passive populations that inhabit local galaxy clusters 
%and the evolution of galaxy cluster members
\citep{butcher1978}. 
Yet we lack a comprehensive picture of the astrophysics that halts galaxy growth \citep[e.g.][]{man2018}. Large-scale extragalactic surveys have demonstrated that a number of physical processes are likely at play, impacting preferentially both the most massive galaxies and galaxies in the densest environments  \citep[e.g.][]{peng2010}. However, the relative importance of these quenching processes over cosmic time remains mostly unconstrained, in part due to the difficulty in performing uniform identification of quiescent galaxies\footnote{In this work, we refer to passively evolving galaxies by the terms quiescent galaxy and post-starburst galaxy in different contexts for convenience (see Section~\ref{sec:selection}), where the latter is typically defined spectroscopically as a young passively evolving galaxy with a spectrum still dominated by A-type stars. At the redshifts relevant to this study ($z>3$), most massive passively evolving galaxies are likely post-starbursts \citep{deugenio2020}.} (QGs). Additionally, environment plays an ever larger role as the growth of cosmic structure proceeds, which creates new challenges as secular quenching and environmental quenching processes may operate simultaneously. 

The remarkable discovery and spectroscopic confirmation of massive quiescent galaxies beyond $z>3$ ($<2$ Gyr after the Big Bang) has now brought us closer to the epoch when quiescent galaxies first emerged. JWST spectroscopy has enabled new powerful constraints on the timescales over which $z>3$ quiescent galaxies form and ``quench'' (stop forming stars), indicating that some massive galaxies may have formed extremely rapidly (formation era $z>9$), and quenched their star formation quickly \citep[growth lifetime $<$200-700 Myr;][]{carnall2023a, glazebrook2023, nanayakkara2022}. These timescales and their stellar masses are extreme enough to cause tension with the expectation of typical baryonic growth efficiencies \citep{labbe2023a, xiao2023}. Regardless of this tension, given the short cosmic timescales (within the first billion years after the Big Bang), these early quiescent sources represent key opportunities to place constraints on quenching mechanisms in a more straightforward way than at later times. Unfortunately, the majority of spectroscopic studies to date have targeted candidates selected from wide-area surveys with relatively limited and shallow photometric coverage. This has limited detailed characterization of QGs to only the most massive and brightest systems at $z<4$, with gravitational lensing paving the way in enabling spectroscopic analysis at $10<\mathrm{log}\,\logM<11$ for small lensed and eventually unlensed samples at Cosmic Noon \citep[$1.5 < z < 3$; e.g.][]{newman2018, deugenio2020, akhshik2023, marchesini2023, park2023}. As such, little is known about the evolution and population statistics of old, massive quiescent galaxies at cosmic noon, nor younger, more recently quenched post-starbursts$^1$ \citep[PSBs; e.g.][]{glazebrook2017, schreiber2018a, strait2023, looser2023, carnall2023, deugenio2023, antwi-danso2023a} at redshifts $\gtrsim3-5$. And the lower mass (log $\logM<10$) quiescent galaxy population remains largely unstudied beyond the low redshift Universe.

The abundance of quiescent galaxies across cosmic time and their typical timescales for quenching are key constraints on the prevalence of specific quenching mechanisms.  In massive galaxies, Active Galactic Nuclei (AGN) are often invoked in simulations to reproduce the bimodality of galaxies \citep[e.g][and references therein]{somerville2015a} and black hole mass has been shown to be a strong predictor of quiescence \citep[e.g.][]{bluck2022, piotrowska2022, bluck2023a}.  How this proceeds remains unclear, however.  Rapid quenching may be induced by strong AGN feedback and outflows \citep[e.g.][]{peng2015, trussler2020} or more gradual quenching may result from moderate AGN feedback that prevents gas inflows and results in starvation \citep[e.g.][]{piotrowska2022, bluck2022, baker2023}.  Similarly, environmental mechanisms capable of quenching galaxies can proceed rapidly $(\sim$ few hundred Myr) $-$ i.e. ram pressure stripping (RPS) of cold, dense gas \citep[][]{boselli2022, cortese2021} $-$ or more slowly ($>1$ Gyr), as in the case of starvation and/or RPS of hot halo gas \citep{larson1980, balogh2000}.  Environment may also trigger or enhance internal quenching mechanisms through efficiently cutting off gas inflows or through gravitational interactions and/or galaxy mergers \citep[see][for a review]{alberts2022a}.

Strong observational constraints on quenching timescales have been hard-won.  Observations of the left-over gas reserves in quiescent galaxies  \citep{bezanson2019, williams2021, whitaker2021b, belli2021, suzuki2022} point to both extremely rapid and effective destruction of star-forming fuel \emph{and} to significant lingering gas reservoirs \citep{suess2017, spilker2022, french2015, rowlands2015,alatalo2016}. The reconstruction of star formation histories (SFHs) from detailed rest-frame optical spectroscopy has also proven to provide powerful constraints \citep{kriek2016, glazebrook2017, schreiber2018a, belli2019, forrest2020a, tacchella2022a, suess2022, setton2023, kriek2023}.  These approaches, however, are exceedingly costly and yield small samples, making selection of promising targets and supplemental statistical studies with photometric datasets extremely important. With JWST, we are now moving into the high redshift regime where the limited age of the Universe may ease interpretation of SFHs.  An additional intriguing new opportunity is the possibility of isolating environmental mechanisms through the study of dwarf (log $\logM<9-9.5$) galaxies. In the low redshift Universe, secular quenching in dwarf galaxies is expected to occur over long timescales, with $<1\%$ of log $\logM\sim8$ galaxies expected to quench without environmental influence \citep{geha2012}.  We are now in a position to test if this is also the case at higher redshifts with JWST.

In this work, we take advantage of ultra-deep MIRI imaging in F770W (reaching 28 mag, $5\sigma$) to supply rest-frame $J$-band at $3<z<6$, breaking the degeneracy between old stellar populations and reddening from dust \citep[e.g][]{labbe2005, williams2009}.  We identify signs of quenching in a mass-limited sample down to log $\logM=8.5$, assess the robustness of our HST+NIRCam+MIRI selection and the need for the MIRI anchor, and present the properties of quiescent and post-starburst galaxies to high redshift and low mass.  This work will provide guidance for the wider extragalactic surveys focused on NIRCam, with no or relatively shallow MIRI coverage, such as CEERS \citep[reaching 25.3-26.5 in F770W over $\sim14$ sq arcmin;][]{yang2023}, COSMOS-Web \citep[24-25 AB in F770W over 0.19 sq degree;][]{casey2022a}, and PRIMER (25.6 AB in F770W over 0.066 sq degree; PI: J. Dunlop, GO 1837).  In Section~\ref{sec:data}, we present the data used in this study and in Section~\ref{sec:sample}, the selection of our mass-limited parent sample and measurement of its properties.  Section~\ref{sec:qgs} describes the selection of our quiescent and post-starburst samples and how this selection would change given color derived with NIRCam only or with 3-band (observed) color selections proposed in the literature.  In the discussion (Section~\ref{sec:disc}), we examine the completeness and contamination in our selection (Section~\ref{sec:completeness}-\ref{sec:contamination}), the nature of our sample (Section~\ref{sec:qgdisc}), the relation between quenching in low-mass galaxies and environment (Section~\ref{sec:rose}), and the abundance of QGs at high redshift (Section~\ref{sec:abundance}).  Section~\ref{sec:conclusions} presents our conclusions.  All magnitudes are quoted in the AB system \citep{oke1983}.  We adopt concordance cosmology ($\Omega_{\rm M}=0.3$, $\Omega_{\lambda}$=0.7, $H_0=70$ km s$^{-1}$ Mpc$^{-1}$), and a \citet{kroupa2001} initial mass function.

\section{Data}\label{sec:data}

The primary dataset for this work is NIRCam and MIRI imaging from the JWST Advanced Deep Extragalactic Survey \citep[JADES;][]{eisenstein2023} in the region where deep MIRI imaging in a single band, F770W, was obtained in parallel with deep NIRCam imaging in Oct 2022 (PID 1180; PI D. Eisenstein).  The MIRI imaging includes four pointings just south of the Hubble Ultra Deep Field \citep[HUDF;][]{beckwith2006} in GOODS-S and totals 61-94 ks of exposure time per pointing. A majority of the MIRI parallel area ($\sim8.8$ sq arcmin) is covered by JADES medium-depth NIRcam imaging in eight filters (F090W, F115W, F150W, F200W, F277W, F356W, F410M, F444W), with additional partial coverage ($\sim4$ sq arcmin) in F335M (PID 1210; N. Luetzgendorf).  A small region ($\sim1.1$ sq arcmin) is additionally covered by F182M, F210M, and F444W imaging from the public First Reionization Epoch Spectroscopic COmpete Survey \citep[FRESCO;][]{oesch2023}.  We further incorporate HST imaging over the full area at 0.4-0.85$\mu$m (F435W, F606W, F775W, F814W, F850LP) from the Advanced Camera for Survey (ACS) from the deep composite images compiled by the Hubble Legacy Field \citep[HLF;][]{illingworth2016, whitaker2019}.  The NIRCam data reduction and construction of the JADES HST+NIRCam photometric catalog follow the description in \citet{rieke2023}.  
%\SA{Due to deblending issues around the Cosmic Rose in later catalogs, we use v0.8.1. This raises the question of what to do about source IDs, since they do not always match up with the public releases.}

A full description of the reduction of the MIRI parallel will be presented in Alberts et al., (in prep).  For a summary, the reader is referred to \citet{lyu2023} and \citet{williams2023a}.  As this work relies heavily on the measurement of accurate colors, we adopt HST and NIRCam photometry extracted from images convolved to the F444W PSF (FWHM $0.145\arcsec$).  Rather than convolve HST and NIRCam further to the resolution at F770W ($0.26\arcsec$), we rebin and convolve only the F444W image to the MIRI pixel size ($0.06\arcsec$) and PSF\footnote{Due to the presence of the cross-artifact in the F560W and F770W bands \citep{gaspar2021}, we adopt an empirical F770W PSF constructed from high dynamic range imaging of stars taken during JWST commissioning (A. Gaspar, private communication).} using {\sc astropy} routines {\tt convolve\_fft} and {\tt reproject} \citep{astropycollaboration2022}. We then measure the F444W - F770W color of sources in the JADES NIRCam detection image \citep{rieke2023} in an aperture with diameter $d=0.7\arcsec$ (covering $\sim65\%$ encircled energy at F770W) with no aperture corrections applied. For $z>3$ galaxies, we assume that a $d=0.7\arcsec$ aperture is more than sufficient to encompass the entire galaxy given their typical sizes \citep[$R_e\lesssim0.25\arcsec$ at $z\sim3-5$;][]{shibuya2015, ormerod2023a}.  For the remainder of this work, unless otherwise specified, we use aperture-corrected HST + NIRCam + MIRI photometry extracted using a $d=0.5\arcsec$ aperture, deriving the F770W in this smaller aperture based on the ($d=0.5\arcsec$ aperture corrected) total F444W flux and the convolved F444W - F770W color.  By doing this, we preserve the advantage of the higher NIRCam resolution by adopting an aperture size appropriate to compact galaxies at high redshift, though still large enough to measure an integrated color robust against potential color gradients.

In total, we have 17 bands of deep photometry covering $0.4-7.7\,\mu$m for the majority of our area.  The 5 HST/ACS bands reach $5\sigma$ point source sensitivities of $\sim28-29$ mag, the 9 JADES NIRCam bands reach  $29-30$ mag \citep[in a $0.2\arcsec$ aperture][]{hainline2023}, MIRI F770W reaches 27.6-27.9 mag ($0.8\arcsec$ aperture, Alberts et al., in prep.).  Over a smaller area, we additionally have FRESCO F182M and F210M, reaching depths of 28.2 mag \citep[$0.3\arcsec$ aperture;][]{oesch2023}.
%\SA{Add the depths for FRESCO and HST.}

\begin{figure}[h!]
    \centering
    \includegraphics[width=\columnwidth]{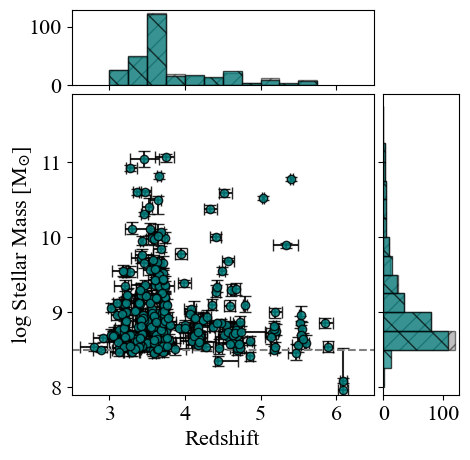}
    \caption{Photometric redshift and stellar mass distribution of the final, mass-selected sample, cut at log $M_{\star}/\Msun\geq8.5$.  The main panel shows the properties (green circles) derived from fitting including the F770W datapoint. The histograms show the distributions with (green solid) and without (gray hatched) the F770W flux density. }

    \label{fig:sample}
\end{figure}

\section{Sample and Properties}\label{sec:sample}

For our parent sample, we build a mass-limited (log $\logM\geq8.5$)  catalog of $3<z<6$ galaxies within the JADES MIRI footprint.  To do this, we start with an initial sample of 1,350 galaxies with a JADES photometric redshift \citep[see][for details]{hainline2023} measured with EAZY \citep{brammer2008} between $z_{\rm phot}=3$ and $z_{\rm phot}=6$  and with a $F_{\rm F444W}$ flux density greater than 28.9 mag; this low limit is chosen to ensure completeness down to our mass limit.  We perform a first pass of spectral energy distribution (SED) fitting (described in the next section) using JADES HST+NIRCam photometry (Section~\ref{sec:data}) to measure and make a cut on stellar mass.  

\subsection{SED Fitting}\label{sec:sedfitting}

As our goal is to identify quenched galaxies, we adopt the Bayesian SED fitting code {\sc BAGPIPES}\footnote{{\sc Bagpipes} uses the {\sc Multinest} nest sampling algorithm \citep{feroz2019} via {\sc PYMULTINEST} \citep{buchner2014}.} \citep{carnall2018}, which has been used extensively in modeling quiescent galaxies \citep{carnall2019, carnall2020, carnall2023a, antwi-danso2023, kaushal2023, hamadouche2022, hamadouche2023, leung2023}. Our fits use the default {\sc Bagpipes} stellar population models, namely the 2016 update of \citet{bruzual2003} from \citet{chevallard2016}. For the assumed SFH, we adopt a parametric double power-law \citep{carnall2018, carnall2019a}.  By separately treating the rising and falling slopes, a double power-law SFH allows for rapid and recent quenching \citep{merlin2018}, which is the expected dominant mode of quenching at high redshift, observable in a post-starburst or young quiescent galaxy phase \citep[e.g.][]{whitaker2012, wild2016, rowlands2018, belli2019, park2023}. %\SA{(Move this to the discussion?) We note, however, that this SFH parameterization cannot reproduce multiple individual bursts of star formation.  This is most relevant for this work in the case of recent rejuvenation of quenched systems, which would require a combined declining SFH plus recent burst.}

Dust attenuation is modeled using \citet{noll2009, salim2018}, which is parameterized as a power-law deviation from the \citet{calzetti2000} attenuation law.  As our sample will contain galaxies ranging from quiescent to dusty and star forming, we allow a large variation in the V-band attenuation ($A_{\rm V}=0-10$).  
%Dust emission is modeled using \citet{draine2007}; however, we fix the parameters to standard values ($Q_{\rm PAH}=2$, $U_{\rm min}=1$, $\Gamma=0.01$) given that we have no constraints past rest-frame $5\mu$m.  
Nebular and continuum emission are included based on the {\sc CLOUDY} photoionization code \citep{ferland2013, byler2017} with a fixed ionization parameter ($U=10^{-3}$) and a stellar birth cloud lifetime of 10 Myr.  Stellar and gas-phase metallicity are assumed to be identical and the metallicity parameter is allowed to vary between 0.2 and 2.5 times solar metallicity. The JADES EAZY photometric redshifts are used as priors and we impose a $5\%$ error floor on all photometric bands.  

With this setup, we fit our initial, flux-limited sample of 1,350 galaxies and use the median of the stellar mass posterior distributions to define our $3<z<6$ parent sample as mass-limited at log $M_{\star}/\Msun\geq8.5$\footnote{The lowest mass galaxies in our parent sample reach a minimum flux of $35$ nJy, 3.5x our initial flux cut and 7x the F444W point source sensitivity, with signal-to-noise (SNR) $\gtrsim10$.  Assuming stellar mass scales with the rest-frame 1$\mu$m, it is likely our parent sample is fully mass-limited.  However, full completeness testing is beyond the scope of this work.}.  We identify 
%106 
304 galaxies above this mass cut in our 8.8 sq. arcmin area; 
%104 
288 (293) have a 
%$>10\sigma$ 
SNR$>5\sigma$ ($>3\sigma$) detection in F770W. 

SED fitting is performed twice on the mass-limited parent sample: once with HST+NIRCam photometry only and then again adding in the F770W.  Final fits using the HST+NIRCam+MIRI photometry with $\chi^2_{\nu}$ greater than $1\sigma$ of their expected $\chi^2$ distribution given their degrees of freedom (number of bands minus the number of free parameters) are visually inspected and 7 are rejected (one star in GAIA DR2 \citep{gaiacollaboration2018}, one probable star with a saturated core, five improperly deblended sub-structures within extended, low redshift galaxies).  
We also note double peaked posteriors for metallicity in 9 high-mass (log $\logM>9.7$) galaxies.  As metallicity is not robustly constrained by photometry \citep{tacchella2022a, nersesian2023}, % reference checked
we refit all galaxies above this mass with metallicity fixed to 1/3 $Z_{\odot}$, appropriate for log $\logM\sim10$ galaxies at $z\gtrsim3$ \citep{cullen2019, sanders2021}, and fixed to $Z_{\odot}$ \citep{maiolino2019}.  We adopt the fit that resolves the double-peaked posteriors with the lowest $\chi^2_{\nu}$\footnote{The revised fits are all subsolar except in the case of 179465 and 172799, which have log $\logM\sim11$.  
%All sources with revised fits are marked in Table~\ref{tbl:classification}.
}.
The resulting redshift and stellar mass distributions are shown in Figure~\ref{fig:sample}.  The histograms show the difference in distributions between the HST+NIRCam and the HST+NIRCam+MIRI fitted redshifts and masses; they are nearly indistinguishable as the redshifts and stellar masses with and without the F770W are in good agreement (see the next section for further discussion). Of our final 297 sources in our parent sample, $96\%$ are fit with $\geq13$ bands of photometry. The percentages with medium band photometry in F182M, F210M, F335M, and F410M are $17, 16, 44$, and $100\%$, respectively.

The {\sc Bagpipes} fits provide measurements of basic properties (redshift, stellar mass, colors, specific-SFR [SSFR], mass-weighted age) as well as higher-order properties (formation redshift, quenching timescales, see discussion in Section~\ref{sec:timescales}).  As was shown in \citet{suess2022b} using mock recovery tests, basic property measurements of post-starburst galaxies are robust when using both parametric \citep[delayed-$\tau$ models, double power-law models; e.g.][]{carnall2019a} % ref checked
and non-parametric \citep[continuity prior; e.g.][]{leja2019b} % ref checked
SFHs.  Specifically, double power-law SFHs in {\sc Bagpipes} were recently shown in \citet{kaushal2023} to recover consistent late-time SFHs as the non-parametric continuity prior used in the Prospector modeling code \citep{johnson2021} in massive galaxies.
Higher order properties are known to be sensitive to the assumed priors, particularly in the case of complex intrinsic SFHs \citep{suess2022b, kaushal2023}; % ref checked
for example, a double power-law SFH cannot capture multiple bursts of star formation, such as expected from a rejuvenation event \citep{akhshik2021, woodrum2022}.  The accurate measurement of formation and quenching timescales for quiescent galaxies using a double power-law with {\sc Bagpipes} was tested in \citet[][see also \citealt{carnall2019a}]{carnall2018} against simulated galaxies with a range of SFHs, finding median systematic offsets of 100-200 Myr but significant scatter.

\subsection{Measuring stellar masses with MIRI}\label{sec:masses}

\begin{figure}[h!]
    \centering
    \includegraphics[width=\columnwidth]{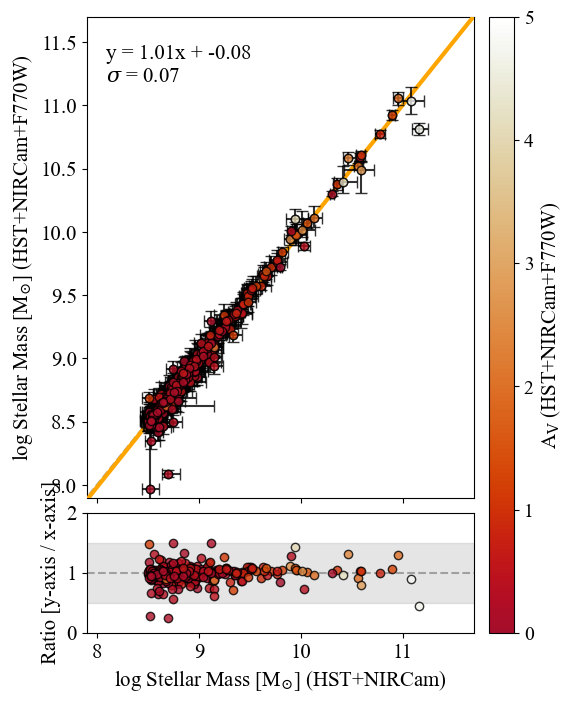}
    \caption{Comparison of the stellar mass measurements down to log $\logM=8.5$ with and without the F770W datapoint. The colorbar indicates $A_{\rm V}$.  The vast majority, save for a few at $A_{\rm V}\gtrsim2$, are well fit by a linear relation (orange line) with a scatter of 0.07 dex. }

    \label{fig:mass}
\end{figure}

With our parent sample, we now investigate whether the addition of deep MIRI F770W photometry significantly changes the inferred stellar masses by providing rest-frame near-infrared constraints and mitigating uncertainties from e.g. dust attenuation or recent star formation.
%Similar to the selection of quenching galaxies, the degeneracy in optical colors caused by age, metallicity, and dust attenuation can drive uncertainties in stellar mass. 
Even with JWST, the ideal coverage past the peak in stellar emission at $\gtrsim1\mu$m (rest-frame) redshifts out of NIRCam at $z\gtrsim3$.  Recent work presented by the CEERS team \citep{papovich2023} found that MIRI coverage at F560W and F770W significantly reduced the stellar masses of high redshift galaxies with sparse $<1\mu$m (rest-frame) coverage. Potential drivers of this difference are young stellar populations, which have been shown to easily outshine older populations in spatially resolved and integrated studies \citep[e.g.][] {gimenez-arteaga2023, baker2023a}, and galaxies where emission lines boost emission in broadband filters \citep[][but see \citealt{desprez2023}]{endsley2023, perez-gonzalez2023, tacchella2023, arrabalharo2023}.  

\begin{figure*}[t!]
    \centering
    \includegraphics[width=2\columnwidth]{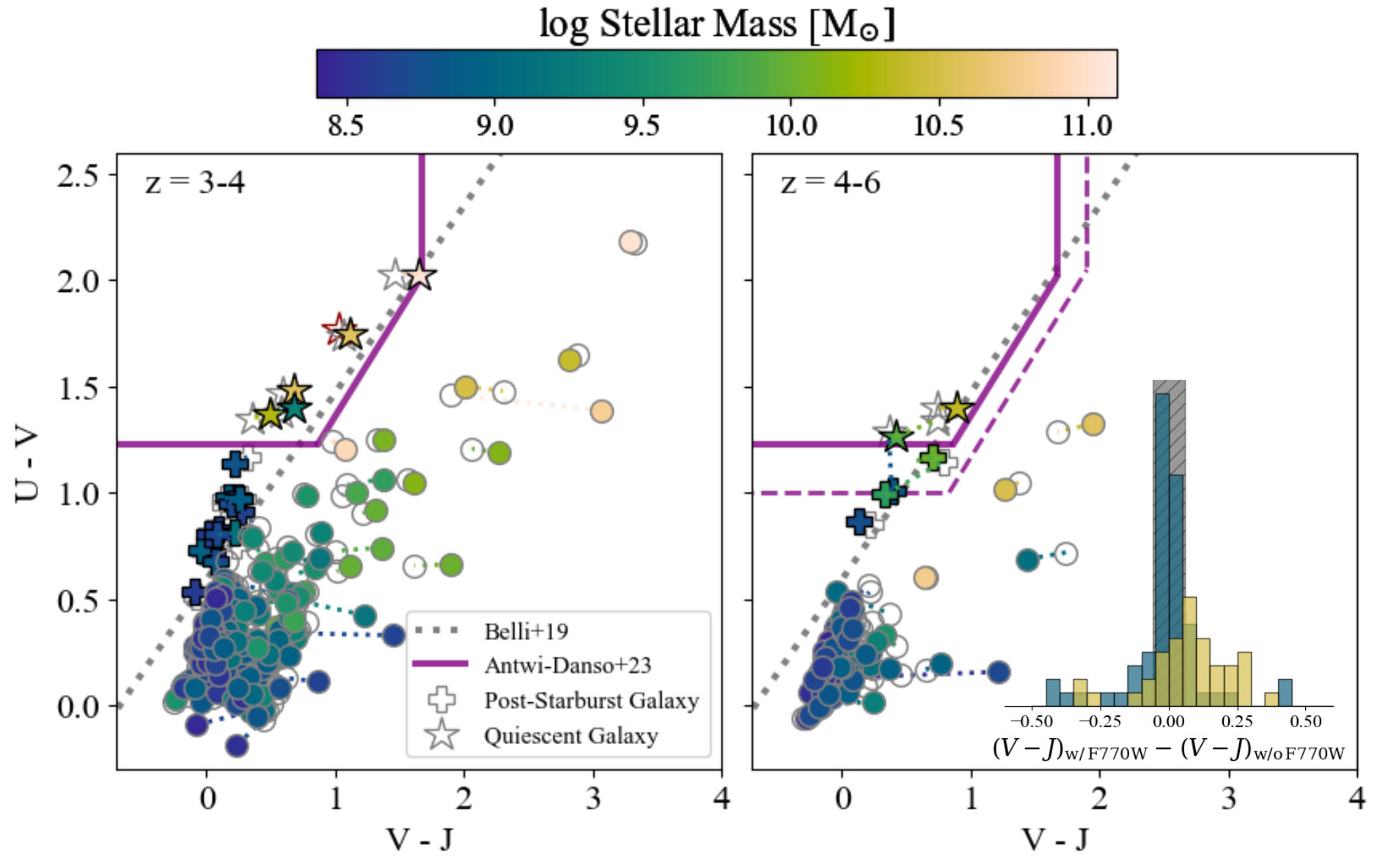}
    \caption{The rest-frame UVJ colors of log $\logM\geq8.5$ galaxies in the JADES MIRI parallel footprint at $z=3-4$ (left) and $z=4-6$ (right).  Closed symbols colored by stellar mass are colors derived from SED modeling including the F770W datapoint.  Open gray symbols are colors derived from fits excluding F770W. The connecting lines show where sources move in UVJ space when MIRI is added. The red open star is the close companion source to JADES 172799 (172799b; Section~\ref{sec:rose}). The purple solid (dashed) shows the main (expanded) UVJ selection region for QGs from \citet{antwi-danso2023}. The gray dotted line is the selection from B19, which extends past the standard $U-V$ boundary. The inset histogram shows that the color shifts are consistent within the measurement uncertainties (gray hatched region) for log $\logM=8.5-9.5$ (blue), but show a small systematic shift redward at higher masses (orange).}

    \label{fig:uvj}
\end{figure*}

In Figure~\ref{fig:mass}, we show the comparison of stellar masses measured with and without the F770W band (tracing rest-frame $1.9-1.1\,\mu$m at $z=3-6$).  We find a remarkably tight correlation across our mass range, with a scatter of $\sim0.07$ dex.  This tight, linear correlation is also seen in the redshifts derived with and without MIRI, which has a $1\sigma$ scatter of $0.14$.  We note that we also find good agreement between the {\sc Bagpipes}-derived redshifts and our EAZY redshift priors, with a scatter of $0.19$. This agreement is likely due to the $8-11$ bands of deep NIRCam coverage; the dense coverage including $1-4$ medium bands with high SNR can accurately establish the shape of the optical continuum without strong susceptibility to emission line boosting 1-2 filters \citep{desprez2023, arrabalharo2023}.  This holds even for our most dust-obscured sources (up to $A_{\rm V}\sim4$), in contrast with the 
%The strongest (though still weak) discrepancies in stellar mass when adding in the F770W are in galaxies with evidence for higher dust attenuation ($A_{\rm V}\gtrsim2$).  
stronger differences ($\sim0.6$ dex) in stellar mass found when selecting specifically for the so-called HST-dark galaxies at $z>3$ \citep{williams2023a}.  Similar to this work, weak to no difference is found for most $z\sim8$ galaxies in the JADES MIRI parallel (Helton et al., in prep) which are expected to generally be blue star-forming galaxies with little dust \citep[e.g.][]{stanway2005, wilkins2011}.

\section{Quiescent Galaxy Candidates with Deep MIRI}\label{sec:qgs}

\subsection{Rest-frame UVJ colors and specific-SFR thresholds}\label{sec:selection}

The commonly used UVJ color selection for massive quiescent galaxies \citep[e.g.][]{williams2009, whitaker2011, muzzin2013a, straatman2014, straatman2016} hinges on having a long wavelength anchor, typically rest-frame $J$-band, to break the degeneracy between stellar age and dust reddening.  Prior to JWST, access to this anchor quickly redshifted out of sensitive ground-based and HST photometry and was either supplied by the lower resolution, lower sensitivity Spitzer/IRAC bands or extrapolation from SED fitting.  Unfortunately, such sparse coverage and/or extrapolation is known to greatly increase contamination from reddened star-forming galaxies at $3<z<6$ \citep[e.g.][]{antwi-danso2023}.

JWST improves selection in multiple ways: denser wavelength coverage of the rest-frame near-infrared, higher sensitivity which can greatly decrease uncertainties when measuring rest-frame colors \citep{merlin2018}, and, with MIRI imaging, interpolation (rather than extrapolation) to rest-frame $J$-band.
%eliminate the need for extrapolation to $J$-band.  
In Figure~\ref{fig:uvj}, we show the inferred rest-frame $U-V$ and $V-J$ colors for our mass-limited parent sample, measured from our SED modeling, in two redshift bins. The median uncertainties in the colors are $\sigma(U-V)=0.05\,(0.04)$ and $\sigma(V-J)=0.05\,(0.06)$ at $z=3-4$ ($z=4-6$). Closed symbols are derived from fits including F770W, which are tied by dotted lines to open symbols derived from fits excluding F770W.  In most cases, colors derived with and without F770W are in good agreement; the median color shifts in $U-V$ are negligible, 
while the color shifts in $V-J$ are largely comparable to the measurement uncertainties for $8.5\leq\mathrm{log}\,\logM<9.5$ and are small (median $\Delta(V-J)=0.06$) but systematically redder for log $\logM>9.5$ (Figure~\ref{fig:uvj}, right, inset).

\begin{table*}
\centering
\begin{scriptsize}
\caption{UVJ-Selected Quiescent Galaxy Candidates\label{tbl:properties}}\begin{tabular}{lccccccccccc}
\hline\hline
ID & RA & Dec & $z$ & log $M_{\star}/\Msun$ & $U-V$ & $V-J$ &  age$_{\rm MW}$ & $z_{\rm f}$ & $z_{\rm q}$ & $\Delta t_{\rm q}^a$ & $\tau_{\rm q}^a$  \\
&  &  &  &  &  &  & Gyr & & & Gyr &  \\
 \hline
\multicolumn{12}{c}{\underline{Robust}}\\
176606 & 53.076502 & -27.864167 & $3.36_{-0.05}^{+0.06}$ & $10.6_{-0.04}^{+0.04}$ & $1.75_{-0.03}^{+0.03}$ & $1.11_{-0.05}^{+0.05}$ & $0.64_{-0.27}^{+0.37}$ & $4.9_{-1.47}^{+1.47}$ & $3.8_{-0.24}^{+0.24}$ & 0.4 & 0.3 \\
175039$^b$ & 53.082581 & -27.866812 & $3.46_{-0.04}^{+0.05}$ & $10.3_{-0.02}^{+0.02}$ & $1.37_{-0.01}^{+0.02}$ & $0.5_{-0.04}^{+0.04}$ & $0.51_{-0.04}^{+0.04}$ & $4.7_{-0.16}^{+0.16}$ & $4.5_{-0.19}^{+0.19}$ & 0.1 & 0.1  \\
16170$^b$ & 53.083613 & -27.887586 & $3.47_{-0.08}^{+0.08}$ & $10.6_{-0.03}^{+0.03}$ & $1.48_{-0.02}^{+0.03}$ & $0.68_{-0.06}^{+0.05}$ & $0.51_{-0.09}^{+0.41}$ & $4.8_{-1.58}^{+1.58}$ & $4.1_{-0.24}^{+0.24}$ & 0.2 & 0.2 \\
35453 & 53.057030 & -27.874375 & $5.33_{-0.17}^{+0.16}$ & $9.9_{-0.04}^{+0.03}$ & $1.26_{-0.03}^{+0.04}$ & $0.41_{-0.07}^{+0.06}$ & $0.47_{-0.11}^{+0.15}$ & $8.6_{-2.09}^{+2.09}$ & $6.6_{-0.95}^{+0.95}$ & 0.2 & 0.3  \\
\\
\hline
\multicolumn{12}{c}{\underline{Tentative}}\\
172809$^{cd}$ & 53.047848 & -27.870220 & $3.66_{-0.23}^{+0.20}$ & $9.4_{-0.04}^{+0.03}$ & $1.4_{-0.06}^{+0.08}$ & $0.68_{-0.09}^{+0.08}$ & $0.84_{-0.38}^{+0.17}$ & $6.4_{-0.99}^{+0.99}$ & $4.2_{-0.48}^{+0.48}$ & 0.6 & 0.4  \\
172799$^{cd}$ & 53.047509 & -27.870503 & $3.75_{-0.15}^{+0.11}$ & $11.1_{-0.06}^{+0.05}$ & $2.03_{-0.05}^{+0.05}$ & $1.65_{-0.07}^{+0.07}$ & $0.63_{-0.22}^{+0.26}$ & $5.7_{-1.33}^{+1.33}$ & $4.7_{-0.70}^{+0.70}$ & 0.3 & 0.2 \\
172799b$^{cd}$ & 53.047627 & -27.870576 & $3.93_{-0.11}^{+0.11}$ & $9.9_{-0.04}^{+0.05}$ & $1.77_{-0.06}^{+0.06}$ & $1.03_{-0.10}^{+0.12}$ & $0.57_{-0.12}^{+0.25}$ & $5.9_{-1.23}^{+1.23}$ & $5.2_{-0.59}^{+0.59}$ & 0.1 & 0.1  \\
170932 & 53.062269 & -27.875047 & $4.33_{-0.09}^{+0.09}$ & $10.4_{-0.04}^{+0.05}$ & $1.4_{-0.04}^{+0.04}$ & $0.89_{-0.06}^{+0.07}$ & $0.27_{-0.08}^{+0.41}$ & $5.3_{-1.96}^{+1.96}$ & $4.8_{-0.28}^{+0.28}$ & 0.2 & 0.1  \\
172811$^e$ & 53.048109 & -27.870184 & $4.53_{-0.20}^{+0.22}$ & $8.8_{-0.04}^{+0.05}$ & $1.29_{-0.09}^{+0.09}$ & $0.36_{-0.09}^{+0.13}$ & $0.54_{-0.16}^{+0.21}$ & $7.1_{-1.77}^{+1.77}$ & $6.0_{-0.93}^{+0.93}$ & 0.2 & 0.2  \\
\\
\hline
\multicolumn{12}{p{1.8\columnwidth}}{Notes: $^a$ The quenching timescale, $\Delta t_{\rm q}$, is the difference between the time at which a galaxy quenched ($t_{\rm q}$) and its formation time ($t_{\rm f}$). $\tau_{\rm q}\equiv \Delta t_{\rm q}/t_{\rm q}$ is the normalized quenching timescale.  See Section~\ref{sec:timescales}. $^b$Member of the $z\sim3.4$ overdensity.
$^c$Members of the Cosmic Rose. $^d$Member of the $z\sim3.7$ overdensity (see Sections~\ref{sec:rose}, Appendix~\ref{app:overdensity} for further details on the overdensities). $^e$Measured parameters from fit not including F770W.}
\end{tabular}
\end{scriptsize}
\end{table*}

\begin{table*}
\centering
\begin{scriptsize}
\caption{B19-selected Post-Starburst Candidates\label{tbl:properties2}}\begin{tabular}{lcccccccccccc}
\hline\hline
ID & RA & Dec & $z$ & log $M_{\star}/\Msun$ & $U-V$ & $V-J$ &  age$_{\rm MW}$ & $z_{\rm f}$ & $z_{\rm q}$ & $\Delta t_{\rm q}$$^a$ & $\tau_{\rm q}$$^a$ & $\Delta$MS \\
 & & &  &  &  &  & Gyr & & & Gyr & &  \\
 \hline
\multicolumn{13}{c}{\underline{Robust}}\\
%176608$^b$ & 53.076909 & -27.864007 & $3.43_{-0.07}^{+0.11}$ & $8.7_{-0.02}^{+0.03}$ & $0.95_{-0.03}^{+0.04}$ & $0.2_{-0.08}^{+0.07}$ & $0.25_{-0.04}^{+0.05}$ & $4.0_{-0.16}^{+0.16}$ & $3.8_{-0.15}^{+0.15}$ & 0.1 & 0.1 & $<-2$ \\
177522$^b$ & 53.082093 & -27.862845 & $3.46_{-0.14}^{+0.19}$ & $8.9_{-0.04}^{+0.04}$ & $0.97_{-0.04}^{+0.05}$ & $0.26_{-0.05}^{+0.06}$ & $0.25_{-0.05}^{+0.12}$ & $4.1_{-0.24}^{+0.24}$ & $3.8_{-0.25}^{+0.25}$ & 0.2 & 0.1 & $<-2$ \\
171534$^d$ & 53.079921 & -27.870211 & $3.67_{-0.14}^{+0.11}$ & $8.9_{-0.04}^{+0.04}$ & $0.98_{-0.03}^{+0.04}$ & $0.17_{-0.07}^{+0.07}$ & $0.28_{-0.04}^{+0.07}$ & $4.4_{-0.23}^{+0.23}$ & $4.1_{-0.21}^{+0.21}$ & 0.1 & 0.1 & $<-2$ \\
170254$^d$ & 53.060745 & -27.876153 & $3.68_{-0.18}^{+0.11}$ & $8.6_{-0.05}^{+0.04}$ & $0.83_{-0.09}^{+0.06}$ & $0.09_{-0.07}^{+0.11}$ & $0.2_{-0.04}^{+0.05}$ & $4.2_{-0.17}^{+0.17}$ & $3.9_{-0.25}^{+0.25}$ & 0.1 & 0.1 & $<-2$ \\
173604$^{cd}$ & 53.046729 & -27.869632 & $3.92_{-0.17}^{+0.10}$ & $8.8_{-0.03}^{+0.04}$ & $1.14_{-0.06}^{+0.07}$ & $0.22_{-0.07}^{+0.09}$ & $0.46_{-0.10}^{+0.21}$ & $5.4_{-0.61}^{+0.61}$ & $4.8_{-0.46}^{+0.46}$ & 0.2 & 0.1 & $<-2$ \\
79086 & 53.044180 & -27.842933 & $4.74_{-0.24}^{+0.32}$ & $8.7_{-0.06}^{+0.08}$ & $0.87_{-0.09}^{+0.10}$ & $0.13_{-0.11}^{+0.14}$ & $0.22_{-0.06}^{+0.14}$ & $5.8_{-0.71}^{+0.71}$ & $5.1_{-0.54}^{+0.54}$ & 0.2 & 0.1 & $<-2$ \\
\\
\hline
\multicolumn{13}{c}{\underline{Tentative}}\\
181568$^b$ & 53.085809 & -27.857762 & $3.28_{-0.05}^{+0.05}$ & $8.7_{-0.04}^{+0.04}$ & $0.8_{-0.02}^{+0.03}$ & $0.08_{-0.05}^{+0.05}$ & $0.19_{-0.04}^{+0.08}$ & $3.7_{-0.15}^{+0.15}$ & $-$ & $-$ & $-$ & -1.1\\
172306$^b$ & 53.049332 & -27.872567 & $3.34_{-0.15}^{+0.21}$ & $8.6_{-0.05}^{+0.04}$ & $0.91_{-0.07}^{+0.07}$ & $0.27_{-0.10}^{+0.09}$ & $0.69_{-0.36}^{+0.30}$ & $5.0_{-1.05}^{+1.05}$ & $-$ & $-$ & $-$ & -1.2\\
174413$^b$ & 53.082623 & -27.868384 & $3.55_{-0.13}^{+0.08}$ & $9.2_{-0.03}^{+0.03}$ & $0.81_{-0.02}^{+0.03}$ & $0.21_{-0.05}^{+0.04}$ & $0.19_{-0.03}^{+0.05}$ & $4.0_{-0.12}^{+0.12}$ & $-$ & $-$ & $-$ & -1.0\\
174444$^d$ & 53.078690 & -27.868339 & $3.56_{-0.12}^{+0.10}$ & $8.6_{-0.04}^{+0.04}$ & $0.81_{-0.02}^{+0.03}$ & $0.02_{-0.04}^{+0.05}$ & $0.22_{-0.03}^{+0.08}$ & $4.1_{-0.16}^{+0.16}$ & $-$ & $-$ & $-$ & -1.2\\
174098$^d$ & 53.075869 & -27.868852 & $3.62_{-0.07}^{+0.05}$ & $8.9_{-0.02}^{+0.03}$ & $0.73_{-0.01}^{+0.02}$ & $-0.04_{-0.02}^{+0.03}$ & $0.17_{-0.02}^{+0.02}$ & $4.1_{-0.06}^{+0.06}$ & $-$ & $-$ & $-$ & -1.1\\
172569$^d$ & 53.051312 & -27.872026 & $3.62_{-0.12}^{+0.11}$ & $9.0_{-0.05}^{+0.04}$ & $0.98_{-0.04}^{+0.03}$ & $0.24_{-0.06}^{+0.05}$ & $0.3_{-0.07}^{+0.21}$ & $4.4_{-0.43}^{+0.43}$ & $-$ & $-$ & $-$ & -1.4\\
40882$^d$ & 53.065272 & -27.869663 & $3.81_{-0.06}^{+0.08}$ & $8.8_{-0.05}^{+0.06}$ & $0.67_{-0.14}^{+0.04}$ & $0.07_{-0.07}^{+0.10}$ & $0.23_{-0.06}^{+0.12}$ & $4.4_{-0.38}^{+0.38}$ & $-$ & $-$ & $-$ & -0.6\\
5070$^e$ & 53.092006 & -27.903137 & $4.41_{-0.07}^{+0.07}$ & $10.0_{-0.03}^{+0.04}$ & $1.17_{-0.03}^{+0.03}$ & $0.7_{-0.07}^{+0.09}$ & $1.03_{-0.08}^{+0.06}$ & $13.2_{-1.96}^{+1.96}$ & $-$ & $-$ & $-$ & -1.3\\
65559$^e$ &53.041051 & -27.854478  & $4.57_{-0.08}^{+0.08}$ & $9.7_{-0.02}^{+0.03}$ & $0.99_{-0.02}^{+0.02}$ & $0.34_{-0.04}^{+0.05}$ & $0.95_{-0.07}^{+0.05}$ & $12.6_{-1.34}^{+1.34}$ & $-$ & $-$ & $-$ & -1.4\\
\\
\hline
\multicolumn{13}{p{1.8\columnwidth}}{Notes: $^a$ The quenching timescale, $\Delta t_{\rm q}$, is the difference between the time at which a galaxy quenched ($t_{\rm q}$) and its formation time ($t_{\rm f}$). $\tau_{\rm q}\equiv \Delta t_{\rm q}/t_{\rm q}$ is the normalized quenching timescale.  See Section~\ref{sec:timescales}. $^b$Member of the $z\sim3.4$ overdensity.
$^c$Members of the Cosmic Rose. $^d$Member of the $z\sim3.7$ overdensity (see Sections~\ref{sec:rose}, Appendix~\ref{app:overdensity} for further details on the overdensities). $^e$Sources are poorly fit with our model ($\chi^2_{\nu}\sim20$).}
\end{tabular}
\end{scriptsize}
\end{table*}

\begin{figure*}[tbh!]
    \centering
    \includegraphics[width=1.7\columnwidth]{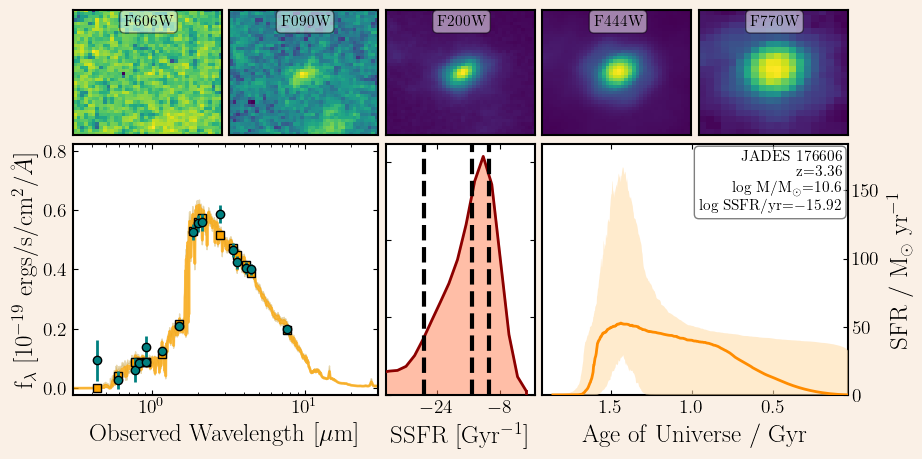}
    \includegraphics[width=1.7\columnwidth]{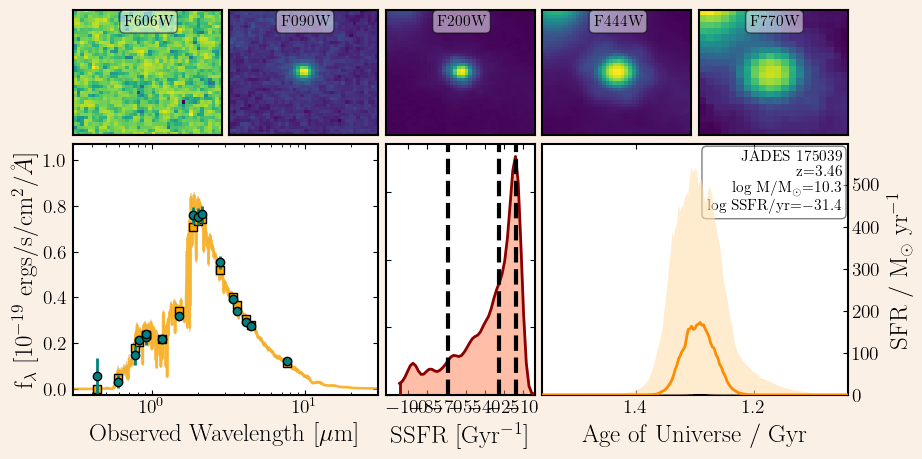}
    \includegraphics[width=1.7\columnwidth]{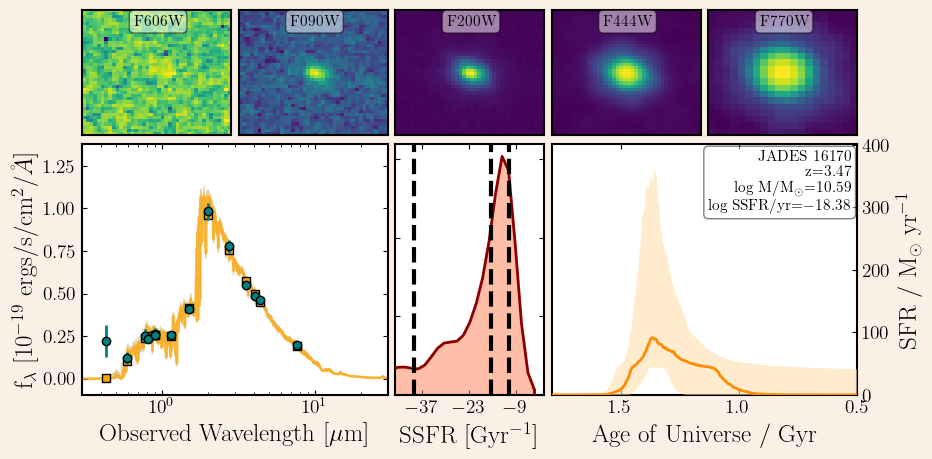}
    \caption{Properties of the UVJ-selected quiescent galaxies. \emph{Top row for each source:} F606W, F090W, F200W, F444W, F770W cutouts, $1.2\arcsec$ on a side.  \emph{Bottom row for each source:} The SEDs (left), SSFR posterior distributions (middle), and SFHs (right).  Robust candidates are highlighted with tan backgrounds.}
    \label{fig:robust}
\end{figure*}

\begin{figure*}[tbh!]\ContinuedFloat
    \centering
    \includegraphics[width=1.7\columnwidth]{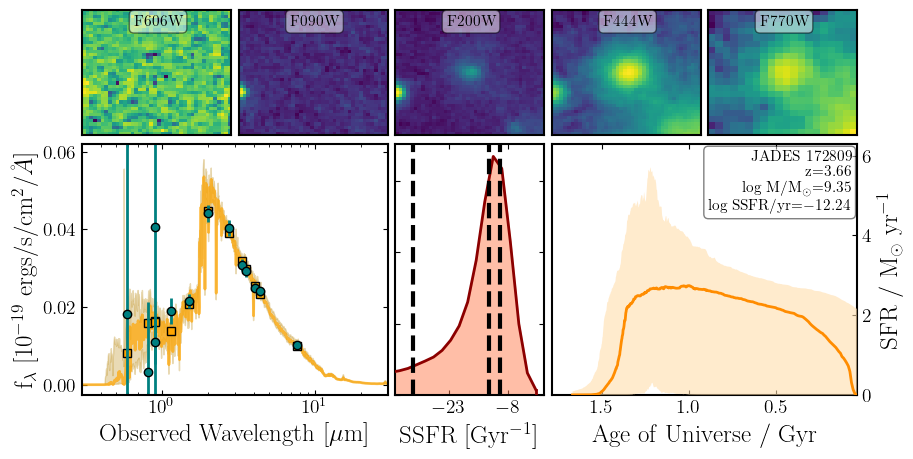}
    \includegraphics[width=1.7\columnwidth]{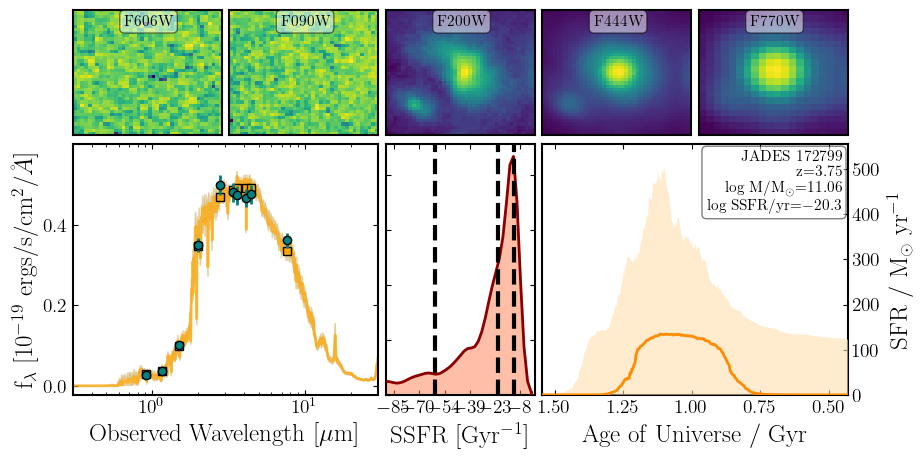}
     \includegraphics[width=1.7\columnwidth]
     {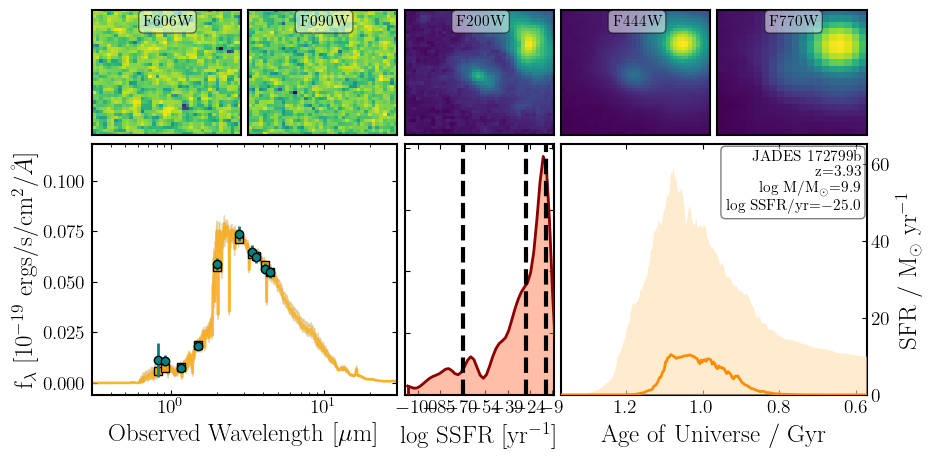}
     \caption{Continued.}
\end{figure*}

\begin{figure*}[tbh!]\ContinuedFloat
    \centering
    \includegraphics[width=1.7\columnwidth]{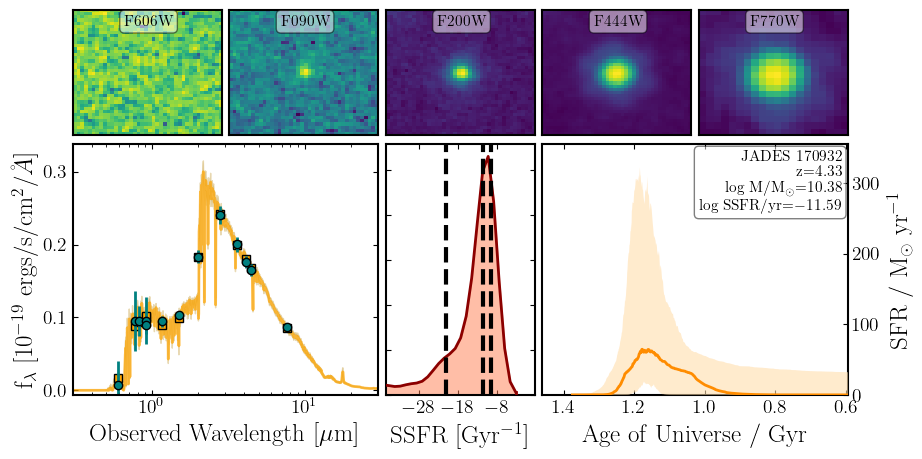}
    \caption{Continued}
    \includegraphics[width=1.7\columnwidth]{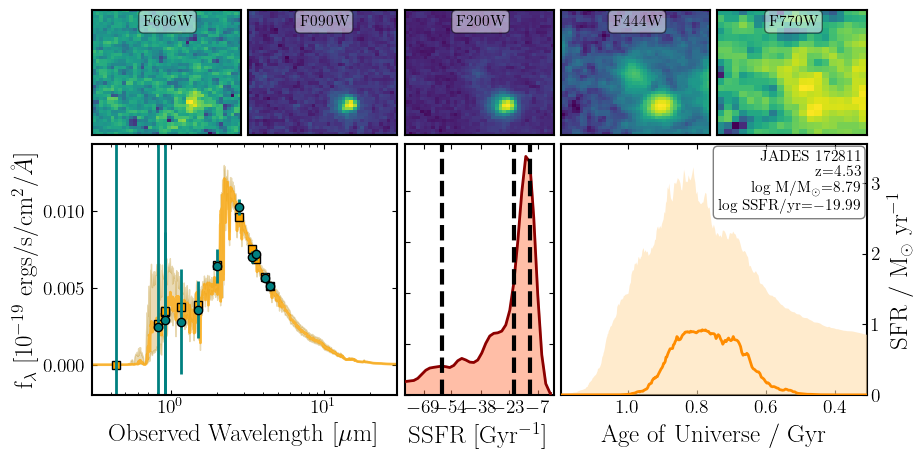}
    \includegraphics[width=1.7\columnwidth]{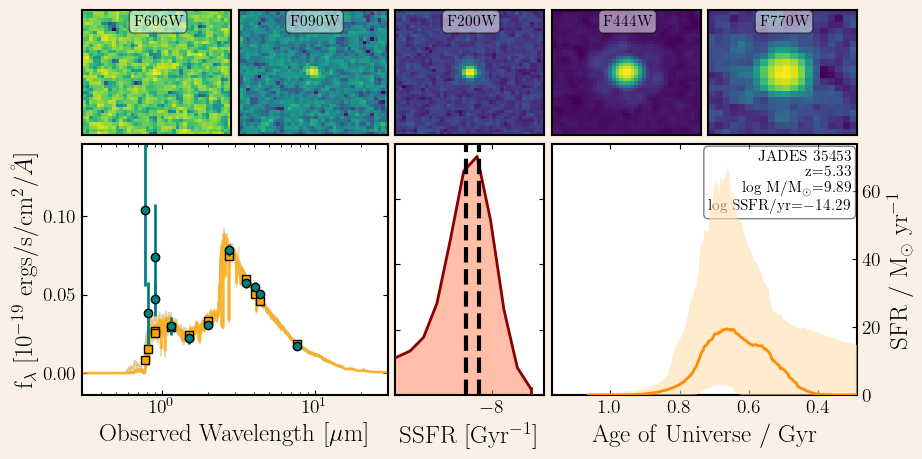}
    \caption{Continued.}
\end{figure*}

Initial quiescent galaxy selection is done using the UVJ selection (purple lines) from \citet{antwi-danso2023}, empirically derived using pre-JWST observations at $3<z<4$.
%and mock galaxies from {\sc JAGUAR} \citep{williams2018} at $4<z<6$. 
The purple dashed line (Figure~\ref{fig:uvj}, right) denotes an additional padded region \citep{antwi-danso2023}; such extensions of UVJ are commonly used to capture even younger passively evolving populations as we move to higher redshifts \citep{schreiber2018a, carnall2020, marsan2022}.  We supplement this with the selection proposed in \citet[][hereafter B19]{belli2019}, which removes the $U-V$ boundary entirely to identify young post-starbursts \citep[see also][]{forrest2020a, marsan2022}. % ref checked
For ease of discussion, we will hereafter refer to candidates that are UVJ-selected as quiescent galaxies and candidates that are selected via the B19 line only as post-starburst galaxies.  We note, however, that these distinctions are for convenience and most massive passively evolving galaxies at $z>3$ likely fall under classical, spectroscopy-based definitions of post-starburst \citep[i.e. spectral features that indicate A-type stars dominate;][]{deugenio2020}.
 
At $3<z<4$, we identify 5 quiescent galaxies via UVJ-selection.  At $4<z<6$, we find an additional 2 in the main UVJ selection and 3 in the padded region.
In the B19 PSB region, we find an additional 17 at $3<z<4$ and 1 at $4<z<6$. 
Image cutouts, SEDs, specific-SFR (SSFR) posteriors, and SFHs for these sources can be seen in Figures~\ref{fig:robust} and \ref{fig:psb1}-\ref{fig:psb2}. Upon inspection, incorporating the F770W into the SED modeling and therefore the color measurements results  
in only one minor change in classification at $z>4.$ JADES 172811 moves from the edge of the main UVJ region to the edge of the padded region when MIRI is added (Figure~\ref{fig:uvj} (right)).  However, visual inspection reveals that the F770W flux is blended with a close neighbor (Section~\ref{sec:rose}) and so we adopt the classification and modeling without the F770W datapoint.  We further inspect the SEDs, images, and SFHs (Figure~\ref{fig:psb2}) of the other two sources (JADES 5070 and 65559) in the UVJ padded region and find that their SEDs are not well fit by our model ($\chi^2_{\nu}\sim20$) and their UV emission and SSFRs (log SSFR/yr$\gtrsim-10$) are consistent with some residual star-formation. We add them to our tentative post-starburst sample. Hereafter we will indicate ``QG'' or ``PSB'' when using specific galaxy IDs.

\subsubsection{Determining selection robustness}\label{sec:robustness}

When selecting quiescent galaxies, UVJ and other color selections can be less sensitive to the assumptions that go into SED modeling than other methods, provided you can measure accurate rest-frame colors.  On the other hand, this selection comes with a loss of information. For example, it has been shown that UVJ colors are not correlated with SSFR below log SSFR/yr$^{-1}\sim-10.5$ \citep{leja2019}; robustly measuring SSFR requires additional far-UV or mid-IR observations.  As we have such observations, we test the robustness of our candidates by making a redshift-dependent cut on the measured SSFR  

\begin{equation}\label{eqn:ssfr}
    \mathrm{SSFR} < \frac{0.2}{t_{\mathrm{obs}}}
\end{equation}

where SSFR is measured using the SFR averaged over the last 100 Myr (SFR$_{100}$) and $t_{\rm obs}$ is the age of the Universe at the observed redshift \citep[e.g.][]{fontana2009, gallazzi2014, pacifici2016, merlin2018, carnall2018}. Though more model-dependent, this approach takes full advantage of our extensive photometry, including coverage of the rest-UV via HST F435W, F606W, F775W, and F814W.  

To label a QG candidate as robust, we require that $>97.5\%$ of its SSFR posterior (referred to as SSFR$_{97.5\%}$ hereafter) is below this evolving threshold.  Four of our QGs (176606, 175039, 16170 at $3<z<4$ and 35453 at $4<z<6$) pass this threshold.  One additional QG (172799) also passes the threshold; however, its SED and placement in the UVJ diagram suggest significant dust content (see Section~\ref{sec:dusty}). 
%which is degenerate with old stellar populations.  
We label this candidate as tentative. QG 178211, mentioned in Section~\ref{sec:selection} as being blended at F770W by a neighbor, meets the SSFR$_{97.5\%}$ threshold only when MIRI is excluded and is also labeled tentative.  
Two other QGs (172809 and 170932) are labeled as tentative as they only clear $\sim50-80\%$ of their SSFR posteriors below the threshold.   
These sources and their measured properties are listed in Table~\ref{tbl:properties}.

Of our sixteen PSB candidates, six meet a slightly relaxed threshold of SSFR$_{50\%}$; these six make up our robust PSB candidates and the remainder our tentative PSB candidates.   
Their properties are listed in Table~\ref{tbl:properties2}.  We note that slight adjustments of the B19 line in UVJ space would result in new PSBs being selected and others being de-selected.  We revisit this in Section~\ref{sec:disc}.

\subsubsection{Candidate samples summary}

Our robust sample of QGs spans log $M_{\star}/\Msun\sim8.8-10.6$ in stellar mass.  In Figure~\ref{fig:robust}, it is clear that their SEDs and cutouts have weak to no UV emission, consistent with their low SSFRs (log SSFR/yr$\,\lesssim-11.5$).  Our highest redshift candidate, QG 35453 at $z_{\rm phot}=5.3^{0.16}_{0.17}$, would be the highest redshift \emph{massive} QG known to date if spectroscopically confirmed.
None of these candidates have been previously identified.

Our four tentative QG candidates are more of a mixed bag (Figure~\ref{fig:robust}).  As stated above, QG 172799 has an unusual SED suggestive of high dust content (see Section~\ref{sec:dusty}) and is additionally part of a large, compact structure that includes a massive, dusty galaxy as well as QG 172809 and PSB 173604. QG 178211 is also spatially located near this group but at a higher redshift. This will be explored in Section~\ref{sec:rose}.  As these neighbors may have minor to moderate blending at F770W, we verify that their fits and measured properties with and without F770W are consistent within the uncertainties, with the exception of QG 178211 (see Section~\ref{sec:selection}). %\SA{Except for 178211.} 
Our last tentative candidate, QG 170932, has a more typical SED and no neighbors, and its failure to meet our SSFR$_{97.5}$ cut is likely due to our use of SFR averaged over 100 Myr rather than a more instantaneous measure.  

Our post-starburst candidates (Table~\ref{tbl:properties2}), on the other hand, range from log $M_{\star}/\Msun\sim8.6-10$.  They typically have more UV emission and blue U-V colors.  Our robust PSB subsample (Figures~\ref{fig:psb1}) are characterized by low SSFRs (log SSFR/yr$\,\lesssim-11$) and SFHs that rapidly rise and fall (see Section~\ref{sec:disc}).  Our tentative PSBs have higher SSFRs ($-10\gtrsim$ log SSFR/yr$\,\lesssim-9$) consistent with residual or ongoing star formation and more extended SFHs (Figure~\ref{fig:psb2}).  In Section~\ref{sec:disc}, we discuss whether the tentative PSBs should be considered candidates for quenched galaxies.

\subsection{Observed-frame three-band color selection with JWST}\label{sec:threeband}

With our QG and PSB candidates identified, we now look at (observed-frame) color selection of passively evolving galaxies with JWST, before turning to our sample's properties in Section~\ref{sec:disc}. At $z>3$, the rest-frame $J$-band redshifts to $>5\,\mu$m; as such, pre-JWST QG color selection was limited to the relatively shallow Spitzer/IRAC 5.8 and 8.0$\,\mu$m filters or to extrapolation via SED fitting in order to obtain this long wavelength anchor.  The effects of extrapolation on UVJ color selection are discussed in detail in \citet{antwi-danso2023}, which found that omission of rest-frame $J$-band data at $z>2$ results in up to one magnitude of scatter in the $V-J$ color, as well as some scatter in $U-V$ due to the rest V band moving past observed $H$-band.  The result is a contamination rate of false positives equal to the selection of true QGs by $z=3.5$.

\subsubsection{NIRCam only}\label{sec:ncselection}

\begin{figure*}[th!]
    \centering
    \includegraphics[width=1.8\columnwidth]{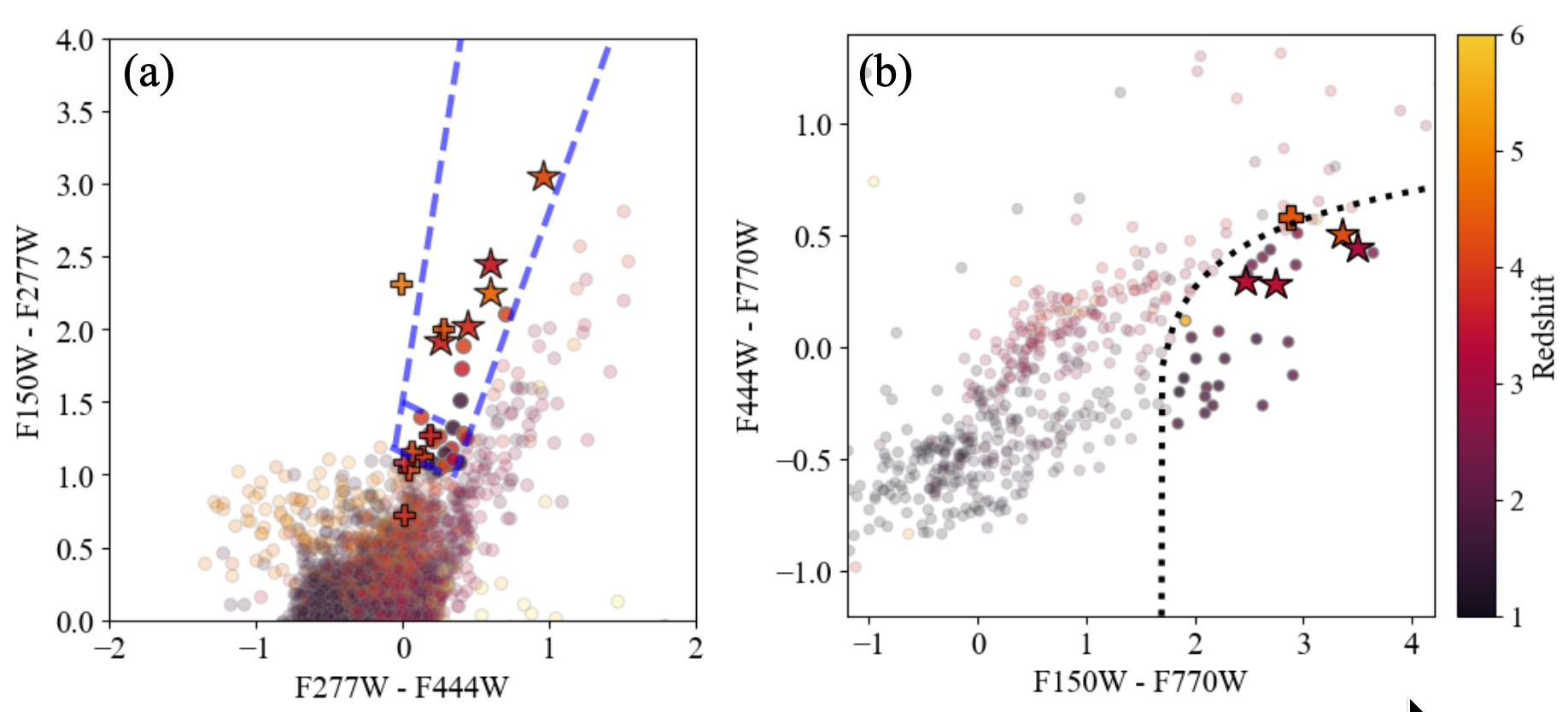}
    \caption{(a) JWST 3 band color selection using NIRCam F150W, F277W, F444W from \citet{long2023} for galaxies with F150W brighter than 28 AB.  The blue dashed lines show the QG selection criteria, with an extension toward bluer F150W-F277W colors to capture more young PSBs. (b) The 3 band color selection using F150W, F444W, F770W from \citet{lovell2023} with F150W$<28$ and F770W$<25$ AB. The black dotted line shows their suggested selection. In both panels, data points are colored by redshift and the stars and pluses denote the QG and PSB candidates in this study; circles show other galaxies. Given a rough redshift cut of $z_{\rm phot}=2.5$, we find minimum contamination rates of 60\% for the \citet{long2023} selection and 25\% for the \citet{lovell2023} selection.}
    \label{fig:longcolor}
\end{figure*}

Extrapolation is similarly necessary with a NIRCam-only UVJ selection at $z>3$, though with the distinction that NIRCam here provides sensitive imaging via $4-6$ filters longward of the 4000\AA\, break at $z=3-6$, placing stronger constraints on the continuum slope than previous $K$-band+IRAC1+IRAC2 combinations used at high redshift, minimizing for example, uncertain slopes due to strong emission lines.  In addition, our sources have photometry in 1-4 medium band filters;
(F182M, F210M, F335M, F410M) 
medium bands (and dense wavelength coverage in general) have been shown to be beneficial 
%is analogous to the success 
in reducing systematics through ground-based $J$, $H$, and $K$ medium bands at lower redshifts \citep{whitaker2010, straatman2016, marchesini2010, tomczak2014, spitler2014, esdaile2021}. 
In Figure~\ref{fig:uvj}, we show that adding the F770W to the full JADES HST+NIRCam photometry results in relatively small shifts in the measured rest-frame colors.  This results in our classifications of QGs and PSBs remaining the same with and without the F770W datapoint.

This successful extrapolation of the continuum using HST+NIRCam alone to disentangle old stellar populations from dust-reddened star formation is likely a function of our robust photometric coverage from UV-near-infrared, allowing for accurate photometric redshifts and colors. To test this, we look at the color selection presented in \citet{long2023}, which combines three NIRCam bands to bracket the 4000\AA\, break (F150W and F277W) and measure the continuum slope longward of the break (F277W and F444W).  For this test, we apply a conservative cut of F150W flux of 28 mag as suggested in \citet{long2023} to the JADES catalog within the MIRI footprint and measure the (observed) F150W - F277W and F277W - F444W colors (Figure~\ref{fig:longcolor}). We note that the typical SNR ($\gg$50) of the JADES NIRCam photometry used in this test allows us to measure very accurate observed NIRCam colors. Using the main \citet{long2023} color selection, 5/5 of our QG candidates (above the flux limit) plus one PSB candidate (65559) are selected, plus an additional 5 sources not in our candidate list.  Two are low-mass galaxies not in our parent sample at $z\sim1$. The remaining three are at $z>3$ with log SSFR/yr $\gtrsim-9$.  
This implies a minimum $60\%$ contamination rate of $z>3$ interlopers; we have not accounted for contamination due to photometric scatter that will be present in lower SNR observations.  The extended \citet{long2023} color cuts, intended to capture more of the young PSB population, select an additional 22 galaxies (16 at $z_{\rm phot}>2.5$), only three of which 
are selected by the B19 line.  

\subsubsection{NIRCam plus MIRI}\label{sec:ncmiriselection}

Can reintroducing the long baseline with MIRI reduce the contamination rate in an observed-frame three-band color selection?  Based on forward modeling of simulated quiescent galaxies in the {\sc FLARES} zoom-in simulation \citep{lovell2021, vijayan2021}, \citet{lovell2023} developed color selections for $z>5$ involving the F770W or F1280W MIRI filters combined with two NIRCam bands.  We apply their F150W - F770W vs F444W - F770W selection in Figure~\ref{fig:longcolor} (right) with the F150W magnitude cut at 28 mag as in the previous section plus a cut on F770W at 25 mag, the depth of the MIRI F770W parallels for COSMOS-Web \citep{casey2022a}.  This selection recovers 4/4 QG candidates above the flux limits and only introduces one contaminant at $z_{\rm phot}>2.5$ (and 21 low-mass contaminants at $z_{\rm phot}<2.5$ which are not in our parent sample).  With the caveat that we are working with a small area and sample, this suggests that quenched galaxy selection is more robust given dense wavelength coverage with NIRCam longward of the 4000\AA\, break \emph{or} F444W plus moderate depth MIRI imaging at F770W, given a liberal cut in redshift is possible.  We note that \citet{lovell2023} found that using F1280W as the longwave anchor produced a selection with $>80\%$ in completeness and purity.  

\section{Discussion}\label{sec:disc}

In this work, we have selected $3<z<6$ quiescent galaxies using UVJ color space and post-starburst galaxies using the B19 selection, with robustness criteria for quiescence based on a redshift-dependent threshold in SSFR.  This selection takes full advantage of 11-14 bands of high-SNR ($\gg$50) HST+NIRCam photometry plus a long-wavelength anchor at rest-frame 1-2$\mu$m provided by ultra-deep MIRI F770W photometry (though as discussed in Sections~\ref{sec:selection}, our sample is the same with and without this anchor).  In this section, we take a closer look at the properties of our sample. 

\begin{figure}[tbh!]
    \centering
    \includegraphics[width=0.9\columnwidth]{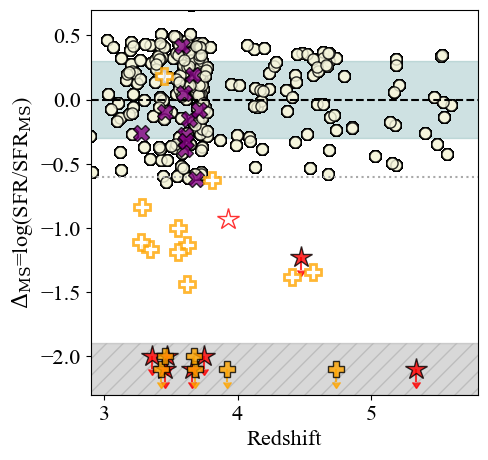}
    \caption{Our mass-limited parent sample relative to the MS from \citet{popesso2023} as a function of redshift.  Red stars (orange pluses) are our QG (PSB) candidates. The red open star is the close companion source to JADES 172799 (172799b; Section~\ref{sec:rose}). Solid plus signs indicate our primary PSB sample, while open symbols are the secondary sample.  Only one PSB is a contaminant on the MS. White circles are galaxies in our parent sample not selected as QG or PSB. Purple x's are galaxies selected by the \citet{long2023} NIRCam color selection (Section~\ref{sec:ncselection}) that are not in our sample. }

    \label{fig:ms}
\end{figure}

Figure~\ref{fig:ms} presents our QG and PSB candidates relative to the Main Sequence \citet[MS, e.g.][and references therein]{popesso2023}, i.e. $\Delta$MS = log (SFR/SFR$_{\rm MS}$), where the SFR is the median posterior of SFR$_{100}$ provided by {\sc Bagpipes} SED fitting and SFR$_{\rm MS}(z,M_{\star})$ is the SFR of a MS galaxy at a given redshift and stellar mass from \citep{popesso2023}, which covers our full mass range.  As expected from our evaluation of their SSFRs (Section~\ref{sec:selection}), our QG candidates fall well below the MS ($\gg1$ dex).  Gratifyingly, 15/16 post-starburst candidates selected in color space, ten of which are labeled tentative as they do not meet the SSFR$_{50\%}$ requirement, also fall significantly below the MS (by $\gtrsim0.6-1$ dex) and only one is an obvious contaminant on the MS, which we discard.  
This supports that extending the UVJ diagonal past traditional $U-V$ boundary B19 can select young post-starbursts, even at low stellar mass.  The $U-V$ colors of our PSBs extend down to 0.5 mag, which is consistent with the colors modeled for rapid quenching (via a top-hat SFH) of dust-free star-forming galaxies presented in \citet{merlin2018}.   Additionally, we find that galaxies in the \citet{long2023} NIRCam color selection that are not in our sample largely fall within the scatter of the MS (Figure~\ref{fig:ms}), consistent with our supposition that they are contaminants (Section~\ref{sec:ncselection}).

The remainder of our discussion will go as follows: in the next section, we dive deeper into the completeness and purity of our selection.  In Section~\ref{sec:qgdisc}, we discuss QG and PSB properties and highlight individual sources; in Section~\ref{sec:rose}, we present evidence of an association between overdense environments and our sample down to low mass in a confluence of galaxies known as the Cosmic Rose \citep{eisenstein2023} and other overdensities; and finally in Section~\ref{sec:abundance}, we examine the number density of QGs suggested by our sample.

\subsection{UVJ selection with JADES: completeness and contamination}\label{sec:completeness}

\subsubsection{Completeness}\label{sec:incompleteness}

The expected completeness (and contamination) rate of UVJ selection has varied in the literature when calibrated against other measures of quiescence.  Part of this has been shown to arise from systematic offsets in the measured rest-frame colors in different datasets \citep{kawinwanichakij2016} and part from potential quiescent populations that have colors outside of the typical UVJ selection \citep{schreiber2018a}. 

In this work, we have adopted the UVJ selection presented in \citet{antwi-danso2023}. Other common selections \citep{williams2009, whitaker2011, muzzin2013a, carnall2018} would recover the same sample within the $1\sigma$ color uncertainties.  We have tested this selection after the fact using a redshift evolving SSFR threshold (Eqn~\ref{eqn:ssfr}), finding that 6/8 (8/8) of our UVJ-selected QG candidates have $97.5\%$ ($50\%$) of their SSFR posterior distribution below this threshold.  We likewise find no candidates with low SSFR that are not UVJ or B19-selected, in agreement with some previous studies \citep{pacifici2016, carnall2018}.  

In contrast, \citet{schreiber2018a}  
$-$ using a photometric sample of 24 candidate QGs at $3<z<4$, supplemented with MOSFIRE $H$ and $K$ spectra for half the sample $-$ found that $40\%$ of candidates with low SSFRs (10x below the MS) were not found by UVJ color selection, but instead inhabited space below (labeled ``young quiescent", partially overlapping the B19 selection) and to the right (labeled ``dusty quiescent") of the traditional region. We do not find populations redder than the UVJ or B19 selection through either our SSFR criteria or comparison to the MS (Figure~\ref{fig:ms}).  \citet{carnall2023a} likewise found disagreement with the QG selection presented in the \citet{schreiber2018a} in a subsample covered by CEERS; our work supports their supposition that JWST data greatly improves the measurement of colors and SSFRs, particularly in dusty sources, reconciling the UVJ and SSFR selections.

This agreement was not necessarily expected to hold as we move into these high redshifts, however.  In this regime, galaxies are more likely to be low metallicity \citep[e.g.][]{cullen2019} and this has been predicted to slow the development of the red colors necessary for UVJ selection \citep[potentially on timescales longer than the age of the Universe at the observed redshift;][]{tacchella2018}.  Quiescent/post-starbursts galaxies observed at $z>3$ are also necessarily younger, which motivates the B19 selection for PSBs and the padded region for UVJ-selection at $z>4$ \citep{antwi-danso2023}.  Alternative color selections \citep{antwi-danso2023, gould2023, kubo2023} have also been explored to try to capture this young population. To test whether UVJ-selection is missing young QGs, we consider one additional color selection, $(ugi)_s$, presented in \citet{antwi-danso2023}. Designed to mitigate the issues arising from extrapolation to $J$-band through new, synthetic $ugi$ filters, this color selection is expected to pick out QGs 250 Myr before they enter UVJ space as it is optimized to be able to select bluer Balmer breaks. However, we find that while $(ugi)_s$ color selection is very effective at identifying our main QG candidates, it does not identify 
any new candidates for young QGs or our PSB sample. The latter is expected as $(ugi)_s$ was designed to maximize purity and minimize contamination from dusty SFGs, which makes it less sensitive to the blue PSB colors short of the break \citep{antwi-danso2023}.

From the above, we conclude that there are no obvious sources of incompleteness in our UVJ-selected sample.  Issues such as the effects of low metallicity on QG colors remain, however, and will require spectroscopic studies with JWST to resolve.

\subsubsection{Contamination}\label{sec:contamination}

Pre-JWST, the UVJ color space for quiescent galaxies was known to be contaminated at the $\sim10-30\%$ level \citep{belli2017, diaz-garcia2019, fang2018, merlin2018, leja2019}; at $z\sim3-4$, \citet{schreiber2018a} found this contamination to be dominated by dusty galaxies at low redshifts with poor photometric redshifts solutions and strong line emitters.  Now with JWST, and particularly in this work with our 11-14 bands of high SNR medium and broadband NIRCam+MIRI photometry, we can largely mitigate these contaminants by providing good sampling on the continuum longward of the 4000\AA\, break, including rest-frame $J$-band, accurately measuring photometric redshifts, identifying emission lines, and breaking the degeneracy between stellar age and dust attenuation.  
This was forecasted in \citet{merlin2018}, which measured the UVJ colors for mock catalogs based on CANDELS catalogs using the CANDELS photometric filters and then again with JWST filters. This comparison demonstrated that the addition of JWST photometry greatly reduced the scatter from measurement uncertainties and, in addition, it predicted a much cleaner separation of QGs and SFGs in UVJ space, with very few galaxies encroaching on the diagonal UVJ boundary. Given this reduction in scatter from measurement uncertainties, one might instead expect that \emph{intrinsic} scatter such as due to complex SFHs or galaxies caught in a transition phase could instead crowd the UVJ boundaries. Though our sample is small, we see evidence for a relatively clean separation along the diagonal UVJ selection, which is where the slow-quenching evolutionary tracks presented in B19 would be expected to enter. This supports that rapid quenching is the dominant mode at high redshift in massive galaxies  \citep[][B19]{whitaker2012, wild2016, rowlands2018}, as they are not spending much time in a transition phase such as the green valley at $z>3$.  This clean separation suggests that our main UVJ selection has low to no contamination.  For our PSB sample, this clean separation is not seen.  Nevertheless, we find that our PSB list captures all galaxies in our parent sample significantly below the MS, with only one contaminant on the MS.  This suggests that this selection from B19, optimized for lower redshift, higher mass galaxies, is promising for high redshift, low-mass PSBs.

\subsection{Quenched galaxies at $z=3-6$}\label{sec:qgdisc}

\subsubsection{Quenching timescales}\label{sec:timescales}

\begin{figure}[tbh!]
    \centering
    \includegraphics[width=\columnwidth]{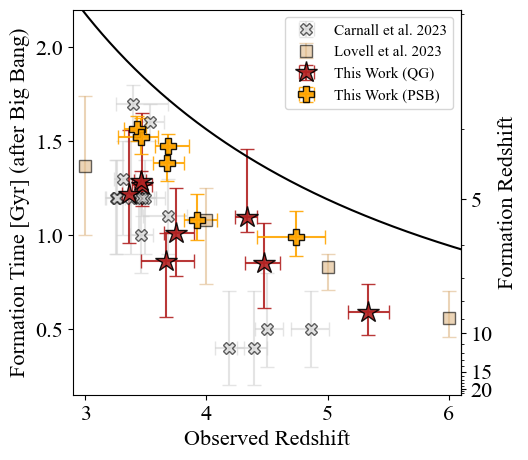}
    \caption{The formation time (since the Big Bang)  of our QG (red stars) and primary PSB (orange pluses) candidates as a function of the redshift at which they are observed.  The right y-axis shows the corresponding formation redshift.  The black line shows the age of the Universe at the observed redshift.  We compare to the CEERS sample (gray crosses) from \citet{carnall2023a} and the median timescales from the {\sc EAGLE} and {\sc FLARES} simulations \citep[tan squares;][]{lovell2023}.}

    \label{fig:formation}
\end{figure}

From the modeled SFHs, 
we can estimate properties that broadly describe the life cycles 
%lifetime 
of our QG and PSB samples, including mass-weighted age, formation time, and quenching timescales (Tables~\ref{tbl:properties}-\ref{tbl:properties2}). As in Section~\ref{sec:sedfitting}, we caution that these measurements are known to be sensitive to the choice of priors \citep{suess2022b, kaushal2023} and we limit our comparisons below to observational studies with similar prior assumptions.  Our QG and PSB candidates span $\sim200-800$ Myr in mass-weighted age, corresponding to formation redshifts of $z_{\rm f}\sim4-9$\footnote{We exclude 5070 and 65559 from consideration here due to their relatively poor fits (Section~\ref{sec:selection})}.  We show the formation time of our robust samples relative to the observed redshift in Figure~\ref{fig:formation} and compare to the CEERS sample analyzed using similar SED modeling in \citet{carnall2023a}.  Split by redshift, our QGs at $3<z<4$ have $4.5\lesssim z_{\rm f}\lesssim6.5$, comparable to CEERS.  As expected given downsizing and their low masses (log $\logM\lesssim9.5$), the majority of our robust PSBs observed at $3<z<4$ have slightly later formations times of $4\lesssim z_{\rm f}\lesssim5.5$.

Unlike the CEERS sample, however, which found that QGs observed at $4<z<5$ have formation redshifts $9<z_{\rm f}<12$, our two massive QGs at $z>4$ formed later at $5<z_{\rm f}<9$.  \citet{carnall2023a} noted that finding such high formation redshifts but no QG observed at $z>5$ was unexpected. Later formation times such as we find here are in line with predictions from the {\sc EAGLE} hydrodynamical cosmological simulation \citep{crain2015, schaye2015} and {\sc FLARES}, which predict that quiescent galaxies 
(defined as SSFR$<0.1$ Gyr$^{-1}$) with $z_{\rm f}\geq10$ should be observed at $z\geq6$ \citep{lovell2023}.  Such extreme early formation timescales as found in \citet{carnall2023a} may be linked to the extreme massive galaxy candidates being uncovered by JWST \citep[e.g.][]{glazebrook2023} and, if confirmed, could place strong constraints on mechanisms for quenching, such as the onset for feedback from supermassive black holes, which drives quenching in the {\sc EAGLE} and {\sc FLARES} simulations.  

To examine the timescales associated with quenching, we adopt $z_{\rm q}$ as defined in \citet{carnall2018} as the redshift at which the current SFR falls to $<10\%$ of the time-averaged SFR across the full SFH and calculate $\Delta t_{\rm q}\equiv t_{\rm q} - t_{\rm f}$ and $\tau_{\rm q}\equiv \Delta t_{\rm q}/t_{\rm q}$ \citep{carnall2018, tacchella2022}; the latter accounts for the difference in dynamical timescales when comparing across a wide range in redshift.  As shown in Tables~\ref{tbl:properties}-\ref{tbl:properties2}, our QGs quench on relatively rapid quenching timescales ($\Delta t_{\rm q}\sim100-600$ Myr), while our robust PSBs show a narrower range of $100-200$ Myr. The quenching timescales for the tentative PSBs are unconstrained, perhaps due to residual star formation.  The quenching of galaxies is thought to occur through multiple pathways \citep[e.g.][B19]{carnall2018, carnall2019a, tacchella2022, hamadouche2023}, which, as demonstrated in \citet{carnall2018}, is reflected in the distribution of $\tau_{\rm q}$.  That work found that $\tau_{\rm q}$ for QGs at $z<4$ forms three peaks, which was interpreted as three distinction modes of quenching:
a rapid rise in star formation followed by rapid quenching  ($\tau_{\rm q}\sim0.1$, prominent at $z>2$), a more extended SFH history followed by relatively rapid quenching ($\tau_{\rm q}\sim0.4$, dominant at $z\sim1-2$), and a slow quenching mode that appears at $z<1$ ($\tau_{\rm q}\sim0.6$).  Our QG sample spans the $\tau_{\rm q}$ values of the first and second pathways, which can be visualized in their SFHs in Figure~\ref{fig:robust}.  On the other hand, all of our robust PSBs display a relatively uniform rapid rise in star formation followed by rapid quenching (Figure~\ref{fig:psb1}).  

\subsubsection{Quenching in low-mass galaxies}\label{sec:lowmass}

Using the B19 extension of PSB selection in UVJ space, we have selected 14 PSBs that live significantly below the MS (Figure~\ref{fig:ms}), six of which are supported by SED modeling to have low SSFRs (our robust sub-sample; Table~\ref{tbl:properties2}).  Though the B19 selection was developed for intermediate-redshift, high-mass galaxies, the majority of our PSBs have log $\logM\sim8.5-9$, 1-2 orders of magnitude below the B19 sample.  This results undoubtedly in part from our smaller search volume, but nevertheless, it highlights the existence of a new population of high-redshift, low-mass passively evolving galaxies now being revealed with JWST \citep[e.g.][]{looser2023, strait2023}.  As seen in Section~\ref{fig:formation} and Figure~\ref{fig:psb1}, the SFHs of the robust PSB sub-sample are consistent with a rapid rise in star formation, followed by rapid quenching.  The mechanism behind quenching in low-mass galaxies, even at low- and intermediate-redshifts, is still undergoing intense debate.  Our PSBs are too massive to be candidates for UV-background quenching \citep{efstathiou1992}, but lower mass than the threshold for mass-driven secular quenching observed at low redshifts \citep{peng2010, geha2012}.  The latter has led low-mass quenched galaxies to be associated with environmental quenching; however, the story may be more complicated at high redshift.  Temporary (``mini'') quenching episodes driven by stochastic star formation and AGN feedback have been invoked \citep{dome2024} to explain two spectroscopically-confirmed fast-quenching, low-mass galaxies at $z\sim5-7$ \citep{looser2023, strait2023}.  This is plausible as the shallow potential wells of low-mass galaxies likely make them susceptible to gas loss through outflows and winds \citep[][]{bullock2017, mcquinn2019, gelli2023}.  Arguments have long been made that AGN, which can produce stronger feedback than stellar processes, are uncommon in dwarf galaxies \citep[e.g.][]{trebitsch2018}, however, observational \citep[e.g.][and references therein]{kaviraj2019, davis2022a} and theoretical \citep[e.g.][]{silk2017, koudmani2021} evidence is mounting to challenge this view, including at high redshift with JWST \citep{maiolino2023, harikane2023}.  However, these mini-quenching phases may be exceedingly short \citep[$\sim20-40$ Myr][]{dome2024}, which would make them difficult to catch without instantaneous SFR tracers from spectroscopy. In photometric work such as here, it is common to use tracers that probe SFR over the last $\sim100$ Myr.

On the other hand, the shallow potential wells of low-mass galaxies are also likely susceptible to environmental mechanisms such as ram pressure stripping \citep{cortese2021, boselli2022} in filamentary or group-scale overdensities \citep[e.g.][]{benitez-llambay2013, bluck2020, castignani2022a, vulcani2021} or the enhancement of internal quenching mechanisms through increased interactions or mergers \citep{vijayaraghavan2013, bahe2019}.  The effectiveness and timescales of these processes have proven difficult to assess and are thought to range from rapid $\sim100$ Myr to slow ($>1$ Gyr) depending on local conditions \citep[see][for a discussion on environmental quenching mechanisms]{alberts2022a}.  It is thus still a challenge to identify mass- vs environmental-quenching in a given low-mass galaxy.  However, we can make some progress by looking at the demographics of larger samples; in Section~\ref{sec:rose}, we examine the local environment of our low-mass PSBs. 

\subsubsection{Dusty Quiescent Galaxies}\label{sec:dusty}

The response of galaxy colors to increasing stellar age is known to cause a strong gradient in the region of UVJ space used to select QGs, with more recently quenched galaxies moving from the bluer, lower-lefthand region to the redder upper-righthand region of the selection as they age \citep[e.g.][B19]{whitaker2012}.  As argued in \citet{carnall2020}, we can expect the upper half of the UVJ selection (associated with ages $\gtrsim1$ Gyr in massive galaxies; B19) to depopulate at $z>3$ as not enough time has elapsed to produce these red colors via an aging population.  Alternatively, such colors at $z>3$ could be caused by dust\footnote{Obscured AGN can also cause red optical colors.  To rule this out, we searched the catalog from \citet{lyu2022a} for AGN within our parent sample. We find that QG 176606 is AGN, but selected in X-ray and not the mid-infrared.  This indicates the AGN is likely not luminous in the optical to mid-infrared.  QG 172799 has no indication of AGN. We don't find any other indications of AGN among our QG and PSB samples, though the \citet{lyu2022a} catalog doesn't cover about a third of the MIRI parallel.}, which reddens galaxies along a similar vector.  Two of our QG candidates, 176606 and 172799, occupy this space, and their dusty nature is supported visually in their SEDs (Figures~\ref{fig:robust}) and by their measured V-band attenuations of $A_{\rm V}=1$ and $A_{\rm V}=2$, respectively. 

While photometric disentanglement of the age dust degeneracy is notoriously difficult, our photometric sampling including 4 and 2 NIRCam medium bands for 176606 and 172799, respectively, gives us confidence that we have selected bona fide candidates for dusty QGs suitable for spectroscopic follow-up. 
%(for example, H$\beta$/[OIII] in F210M for 176606 and [OII] in F182M for 172799). 
Confirmation of quenched galaxies with residual dust has interesting implications as the typical expectation is that the cold interstellar medium (ISM) has been destroyed, evacuated, or heated in order for galaxies to halt star formation \citep[e.g.][]{dave2012, lilly2013}.  Infrared stacking of photometric samples has detected non-negligible cold \citep{gobat2018, magdis2021} and warm \citep{blanquez-sese2023} dust in quiescent populations, albeit at lower redshifts.  In addition, a handful of spectroscopically-confirmed QGs  %to lack emission lines 
have direct detections of dust in the far-infrared \citep{whitaker2021b, lee2023, morishita2022}.  These studies seem to contradict the idea of full destruction of the cold ISM; however, a number of unknowns still prevent firm conclusions, 
including whether dust can be accreted via minor merging \citep[see e.g.][]{caliendo2021, woodrum2022}, what dust destruction mechanisms dominate and on what timescales \citep[e.g.][]{whitaker2021}, and how our assumptions about dust temperatures in quiescent galaxies bias our interpretation \citep[e.g.][]{cochrane2022}.  Spectroscopic confirmation of the nature of the $z>3$ dusty QG candidates presented here would provide valuable constraints given the early epoch.

%While the excellent wavelength sampling and long-wavelength coverage of our current data provide tantalizing hints of surviving dust in quenched galaxies, future spectroscopic follow-up is needed to confirm the exact nature of these sources.  

\subsection{Environmental quenching of low-mass galaxies at high redshift: the Cosmic Rose}\label{sec:rose}

\begin{figure*}[tbh!]
    \centering
    \includegraphics[width=2.2\columnwidth]{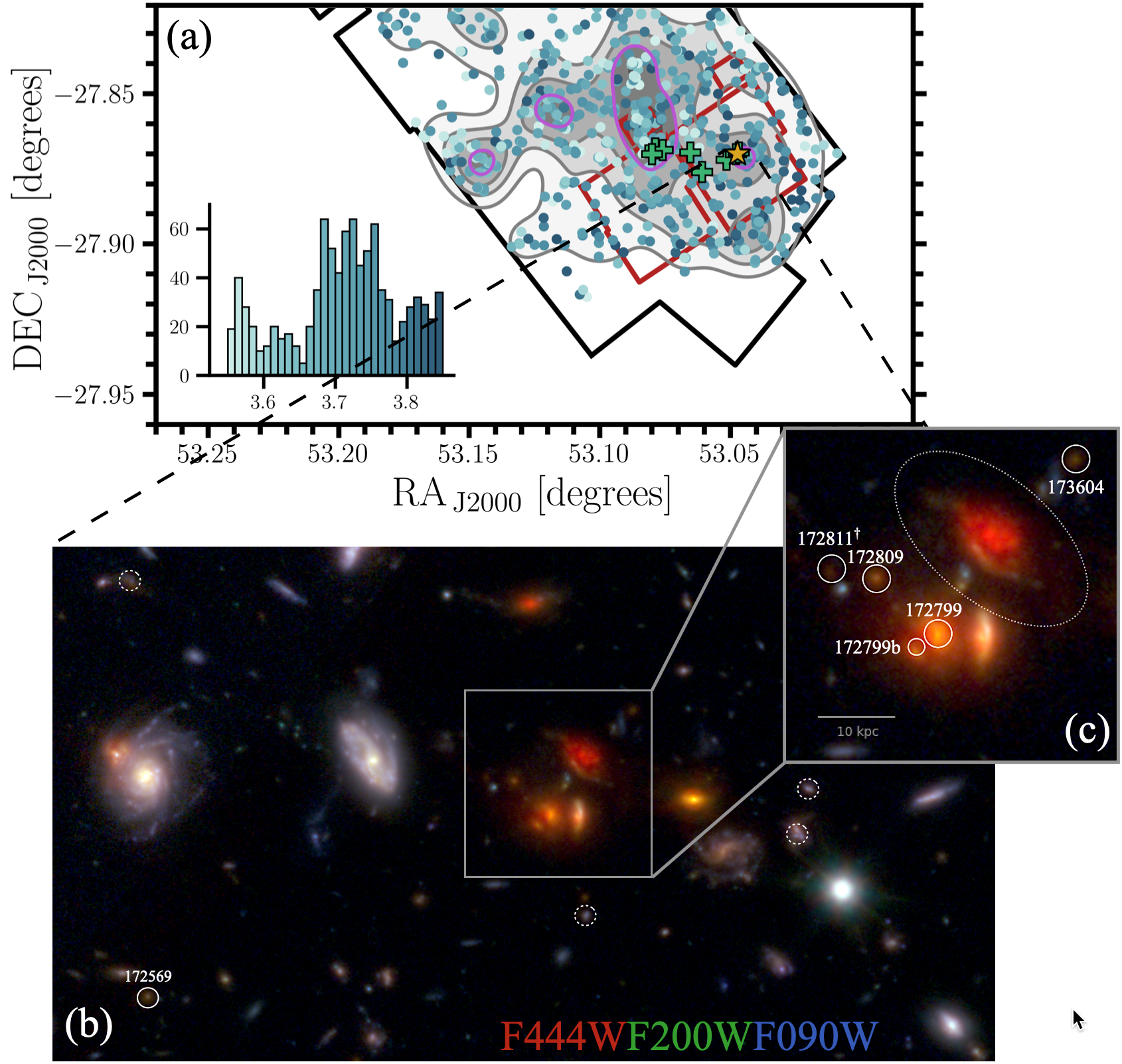}
    %\vspace{-5pt}
    \caption{(a) Overdensities of galaxies at $3.55<z<3.85$ in the JADES NIRCam (black outline) and MIRI parallel (red outline) field-of-views.  Contours increment by $1\sigma$, with the purple contour outlining $4\sigma$ peaks. The yellow star indicates the Cosmic Rose.  The green pluses show low-mass PSBs with redshifts consistent with $z=3.7$. The inset histogram shows the redshift distribution of galaxies in the overdensity. (b) RGB (F444W, F200W, F090W) image of $\sim150$ kpc around the Cosmic Rose. (c) Zoom-in of the Cosmic Rose. Quiescent and PSB (SFG) galaxies in the $z\sim3.7$ overdensity are circled with solid  (dotted) lines ($^{\dag}$172811 is at $z=4.53$, its proximity to the Rose is a projection effect). QGs and PSBs are labeled. % change to the v0.8.1 outline!
    }

    \label{fig:rose}
\end{figure*}

Beyond its dusty, yet old nature, QG 172799 is the base of the so-called ``Cosmic Rose" \citep{eisenstein2023}, a visually striking structure composed of two massive red galaxies, QG candidate 172799, and a log $\logM=11$ dusty SFG (JADES 172813\footnote{Also known as ALESS009.1, spectroscopically confirmed via CO(4-3) at $z_{\rm spec}=3.694$
\citep[][]{birkin2021}.}) with $\mathrm{SFR}\sim730\,\Msun$ yr$^{-1}$ and $A_{\rm V}\sim3.5$\footnote{Due to its size, the properties reported for JADES 172813 were measured here using a Kron aperture.}.  This pair, reminiscent of the Jekyll and Hyde galaxies observed at similar redshifts \citep{glazebrook2017, schreiber2018b, kokorev2023a}, are surrounded by lower mass quenched galaxies with similar photometric redshifts (Table~\ref{tbl:properties}-\ref{tbl:properties2}).  In Figure~\ref{fig:rose}, we show the F444W, F200W, F090W RGB image of the Rose, which reveals that 172799 has similar colors as 172809, which we have identified as a log $\logM=9.4$, relatively old ($\mathrm{age_{\rm MW}}\sim0.6$ Gyr) QG. Both QG candidates are fit with $z_{\rm phot}\sim3.7$, which would put them at a physical separation of 11 kpc, significantly less than the distance from the Milky Way to the Magellanic Clouds \citep[$\sim50$ kpc;][]{fitzpatrick2002}. PSB 173604 sits just northwest of the Rose. The image further reveals a potential third companion not identified by the JADES catalog deblending algorithm.  Using a smaller aperture ($d=0.3\arcsec$) to isolate this companion, we measure its HST+NIRCam photometry and perform SED fitting.  We do not repeat the fit with F770W as the small separation means this companion is likely significantly blended with 172799 in the MIRI photometry.  The companion, which we dub 172799b, has a photometric redshift within $1\sigma$ of 172799, a stellar mass of log $\logM=9.9$, and meets the criteria for a QG candidate based on UVJ and SSFR$_{\rm 97.5}$ (Tables~\ref{tbl:properties}, Figure~\ref{fig:uvj}, \ref{fig:robust}, \ref{fig:ms}).  

The close proximity of these four quenched galaxies strongly suggests they are related and that their quenching may be environmentally driven.  For massive galaxies, environmental quenching is notoriously difficult to disentangle from secular quenching, even in obvious cases of visible signatures such as ram pressure stripped tails or morphological disturbance from major mergers \citep{alberts2022a}.  For low-mass galaxies, on the other hand, secular quenching mechanisms may act over long timescales and quenching of dwarf galaxies is known to correlate with overdense environments locally \citep[e.g.,][]{peng2010, peng2012, bluck2014, bluck2016} and at $z\sim2$ \citep{ji2018}.  As discussed in Section~\ref{sec:lowmass}, however, this is an area of intense debate, particularly at high redshift.

Seven of our low-mass PSBs have redshifts consistent with $z=3.7$ within $2\sigma$.  To explore whether this is a coincidence, we look for a larger structure around the Cosmic Rose following the procedure in \citet{sandles2023} (see Appendix~\ref{app:overdensity} for details).  And indeed, we find that there is a $>4\sigma$ overdensity (Figure~\ref{fig:rose}) that peaks just north of the MIRI parallel ($\alpha, \delta=53.08324339, -27.85463419$) surrounded by secondary $\sim4\sigma$ peaks; the Cosmic Rose is located at the edge of a $4.3\sigma$ secondary peak with a $\sim11\arcsec$ (80 kpc) radius.  This overdensity has been previously identified and was recently confirmed using optical spectroscopy \citep[][and references therein]{shah2023}. Overplotted are the locations of the 7 low-mass PSBs which all coincide with regions that are overdense at the $3-4\sigma$ level.  
%This suggests that environmental quenching is involved.  

Among our remaining low-mass PSBs, 4 have redshifts consistent with $z\sim3.4$, so we repeat this large-scale structure analysis (Appendix~\ref{app:overdensity}) and find that these four are also associated with an overdensity at the $3-4\sigma$ level \citep[see also][and references therein]{shah2023}.  This means that, out of 12 log $\logM=8.5-9.5$ PSBs, we find that only one (PSB 79086 at $z\sim4.7$) is not associated with an overdensity. A similar association was found for a spectroscopically-confirmed low-mass (log $\logM=8.97$) quiescent galaxy at $z=2.34$ in JADES \citep{sandles2023}. These preliminary findings are consistent with a recent result from the CEERS team \citep{bluck2023}, which examined quenching as a function of the stellar potential (stellar mass divided by half-light radius).  They found that stellar potential is the best predictor of quiescence in \emph{massive} galaxies, which they attributed to a tight correlation between stellar potential and black hole mass, making it a tracer of the integrated effects of AGN feedback \citep{deugenio2023}.  Conversely, they found stellar potential was not a predictor of quenching in low-mass (log $\logM=9-10$) galaxies, which they interpreted as ruling out quenching mechanisms that scale with stellar mass and indirectly supporting environmental quenching.  Though we also do not directly demonstrate environmental quenching in this work, the overwhelming association between low-mass PSBs and overdense environments in our sample at $z>3$ strongly supports that we focus on establishing or refuting a causal link with environmental quenching in future follow-up.
%Our work here, the first to examine a large sample of quenched low-mass galaxies at high redshift, does not directly demonstrate environmental quenching but the association between low-mass quenching and overdense environments.  Whether this link is causal remains to be seen.

\subsection{The Abundance of Quiescent Galaxies at $3<z<6$}\label{sec:abundance}

Understanding the quenching pathways that dominate galaxy evolution across cosmic time requires tracing the abundances of quiescent galaxies from their first emergence to later times. Ground-based and HST studies have revealed that QGs at $z\sim2-3$ are abundant with relatively old stellar ages \citep[up to $1-2$ Gyr;][B19]{carnall2020}, implying a substantial population already in place at earlier epochs.  Pre-JWST estimates of the massive QG number density at $3<z<4$, however, vary by over an order of magnitude \citep{valentino2023}, changing dramatically with the number of filters/wavelength sampling and depth, which highlights the strong dependence of selection functions on measured abundances. In particular, deeper, well-sampled surveys (e.g. space-based or medium-band surveys) tend to identify larger abundances \citep[although these studies are typically limited to smaller areas;][]{straatman2014, straatman2015, tomczak2014, schreiber2018a, shahidi2020} compared to sparsely-sampled ground-based (but wide area) surveys \citep{muzzin2013a,weaver2022,valentino2020,davidzon2017}. 
Simulations have similarly discrepant results at early times \citep{valentino2023, cecchi2019, merlin2019, girelli2019} but this is reflective of the overall poor empirical constraints informing feedback and evolutionary models at $z>3$ compared to lower redshifts. 

\begin{figure}[tbh!]
    \centering
    \includegraphics[width=\columnwidth]{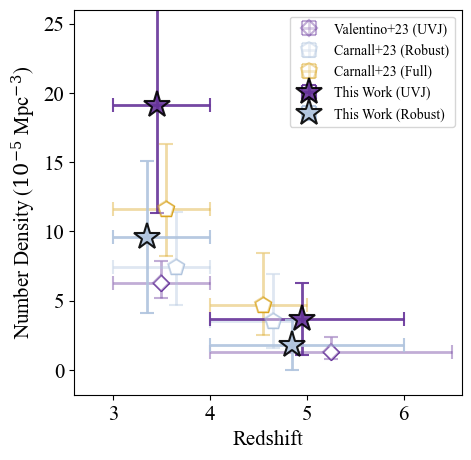}
    \caption{The number density of photometrically-selected, massive (log $\logM\gtrsim9.5$) quiescent galaxies at $z\sim3-6$ from our robust (purple stars) and full UVJ-selected (pale blue stars) samples. Points are staggered slightly along the redshift axis for visual clarity.  
    We compare to the massive (log $\logM\gtrsim9.5$) robust and full samples from \citet{carnall2023a} (pentagons) and the UVJ-selected sample from \citet{valentino2023} (diamonds).  We find that the abundance of quiescent galaxies in the MIRI parallel agrees with the higher estimates from the deep CEERS survey \citet{carnall2023a}; however, we caution that our field contains known overdensities (Sections~\ref{sec:rose}, \ref{app:overdensity}).}

    \label{fig:number}
\end{figure}

In this work, we have the disadvantage of examining a relatively small area (8.8 sq. arcmin) but the advantage of exquisitely deep data over observed 0.4-7.7$\mu$m, allowing us to identify a robust, complete photometric sample of QGs with low contamination (Section~\ref{sec:completeness}).  In our two redshift bins, we find number densities of
%$9.6\pm5.5-15.9\pm7.1\e{-5}$ and $1.8\pm1.8-5.5\pm3.2\e{-5}$ 
$10^{-3.7\pm0.2}-10^{-4.0\pm0.3}$ Mpc$^{-3}$ at $3<z<4$ and $10^{-4.4\pm0.3}-10^{-4.7\pm0.4}$ Mpc$^{-3}$ at $4<z<6$ for our log $\logM>9.4$  full and robust  UVJ-selected QG samples (Figure~\ref{fig:number}), in good agreement with the recent results from CEERS \citep{carnall2023} and a factor of 2-3x higher than the similarly selected UVJ numbers derived from multiple JWST fields, most of which are shallower and have more sparse wavelength coverage \citep{valentino2023}.  This discrepancy could be due to cosmic variance; \citet{valentino2023} found a factor of $2-3$x in field-to-field variation in the number densities in 11 fields ranging from $2-35$ sq arcmin in size.  Part of this may be driven by the (often unaccounted for) presence of overdensities, and indeed, as discussed in Section~\ref{sec:rose}, we have found that the JADES MIRI parallel contains multiple 3-4$\sigma$ galaxy overdensities \citep[see also][]{shah2023}. 
Nevertheless, we suggest that further investigation is needed to test whether these results from small but deep fields indicate that wide, shallow JWST fields are limited in their ability to constrain quiescent galaxy abundances, given a lack of wavelength coverage (no HST or MIRI, sparse NIRCam) and lower significance detections which increase the uncertainties on measured parameters such as colors or SSFR.  

\section{Conclusions}\label{sec:conclusions}

In this work, we evaluate the selection and properties of quiescent and post-starburst galaxies at $3<z<6$ in a mass-limited sample (log $\logM\geq8.5$) using 13-16 bands of HST+NIRCam photometry combined with ultra-deep JADES MIRI F770W imaging, which constrains the rest-frame $J$-band anchor commonly used in color selection of quiescent galaxies.  SED fitting is done using a double power-law SFH through the {\sc Bagpipes} fitting code from which we measure rest-frame colors and galaxy properties.  Our main conclusions are as follows:

\begin{itemize}
    \item The derivation of photometric redshifts and stellar masses yields consistent results with and without the inclusion of the F770W data point (probing rest-frame $1-2\,\mu$m) in the SED fitting (Section~\ref{sec:masses}). This is likely the result of the dense wavelength coverage in JADES (8-11 NIRCam bands including 1-4 medium band filters per source), which places robust constraints on the continuum longward of the 4000\AA\, break.
    
    \item Selection of quiescent and post-starburst galaxies is done using the standard UVJ color diagram plus the extended PSB selection from B19.  The robustness of our candidates is evaluated using a redshift-evolving SSFR cut that takes full advantage of our dataset. Over our 8.8 sq arcmin area, we determine a final sample of 4 (5) robust (tentative) UVJ-selected QGs (Table~\ref{tbl:properties}, Figure~\ref{fig:robust}).  This corresponds to number densities of massive QGs at $3<z<6$ (see Section~\ref{sec:abundance}) in good agreement with results from JWST surveys to similar depths \citep{carnall2023}, with the caveat that our 8.8 sq arcmin area contains known overdensities (Section~\ref{sec:rose}, Appendix~\ref{app:overdensity}).  We additionally identify 6 (9) robust (tentative) B19-selected PSBs. We identify only one Main Sequence galaxy contaminant in our B19 selection.

    \item As with the stellar masses, our sample of QG and PSB candidates is selected equally well with and without the F770W data point, again pointing to the constraining power of the JADES dataset.  For similar surveys, QG-selection therefore doesn't require MIRI data at $z>3$. For fields with sparser JWST wavelength coverage, we test (observed) 3-band color selections presented in the literature \citep{long2023, lovell2023} and find that NIRCam 3-band (F150W, F277W, F444W) selection suffers from high $\gtrsim60\%$ contamination.  A 3-band selection including relatively shallow MIRI F770W reduces this contamination rate.
   
    \item Our full QG and PSB galaxies have a range in mass-weighted ages ($\sim200-800$ Myr), corresponding to formation redshifts of $z_{\rm f}\sim4-9$.  We do not find examples of extremely early ($z=9-12$) formation times \citep[e.g.][]{carnall2023}, despite identifying potentially the highest redshift massive QG at $z_{\rm phot}=5.3$.  
    
    \item The range in quenching timescales for our sample ($\tau_{\rm q}\sim100-600$ Myr) is consistent with rapid quenching pathways.  B19-selected robust PSBs have uniformly short quenching timescales ($100-200$ Myr).

    \item Through the B19 selection, we identify a substantial new population of low-mass (log $\logM=8.5-9.5$) post-starbursts with SFHs that are characterized by a rapid rise in star formation following by rapid quenching (Figure~\ref{fig:psb1}-\ref{fig:psb2}) and/or living significantly below the star-forming Main Sequence (Figure~\ref{fig:ms}). This demonstrates that UVJ-based PSB selection can be extended to low masses.
    
    \item We characterize the nature of the Cosmic Rose (Figure~\ref{fig:rose}), a complex of galaxies dominated by a massive, dusty QG and a massive, dusty SFG at $z\sim3.7$ $-$ a so-called Jeykell and Hyde pair.  Three lower mass QGs are within $\sim20$ kpc of the complex center, indicating likely efficient environmental quenching within this system.  
    
    \item An investigation of a larger area reveals that the Cosmic Rose is part of a larger overdensity at $z\sim3.7$ \citep[see also][]{shah2023} that encompasses fully half of our low-mass PSBs.  The other half (save one) is located in an overdensity at $z\sim3.4$.  This provides compelling evidence that quenching of log $\logM\sim8.5-9.5$ is associated with overdense environments and potentially driven by environmental-quenching mechanisms.
\end{itemize}

The launch of JWST has opened up new opportunities to trace the evolution of quenching back to the emergence of the first quiescent galaxies and down to the low-mass regime for which little is known beyond the low-redshift Universe. To take full advantage of this opportunity, we will need both detailed spectroscopic analysis and robust selection of statistical samples from deep JWST surveys.

%% Also note that the akcnowlodgment environment does not support long amounts of text. If you have a lot of people and institutions to acknowledge, do not use this command. Instead, create a new \section{Acknowledgments}.

\section*{Acknowledgments} The author thanks Andras Gaspar for his work on constructing empirical MIRI PSFs and Adam Carnall for discussions about {\sc Bagpipes}. The JADES Collaboration thanks the Instrument Development
Teams and the JWST instrument teams at the European
Space Agency and the Space Telescope Science Institute for the support that made this program possible.
The authors acknowledge use of the lux supercomputer at UC Santa Cruz, funded by NSF MRI grant
AST 1828315. This work was performed in part at Aspen Center for Physics, which is supported by National Science Foundation grant PHY-2210452.
SA, JL, IS, GHR acknowledge support from the JWST Mid-Infrared Instrument (MIRI) Science Team Lead, grant 80NSSC18K0555, from NASA Goddard Space Flight Center to the University of Arizona and SA, CCW, JMH, ZJ, KH, DJE, BDJ, MR, BR, CNAW acknowledge the JWST/NIRCam contract to the University of Arizona NAS5-02015.  The research of CCW is supported by NOIRLab, which is managed by the Association of Universities for Research in Astronomy (AURA) under a cooperative agreement with the National Science Foundation.
WB, FDE acknowledge support by the Science and Technology Facilities Council (STFC), by the ERC through Advanced Grant 695671 ``QUENCH'', and by the UKRI Frontier Research grant RISEandFALL.
S. Arribas acknowledges support from Grant PID2021-127718NB-I00 funded by the Spanish Ministry of Science and Innovation/State Agency of Research (MICIN/AEI/ 10.13039/501100011033). NB acknowledges the Cosmic Dawn Center (DAWN), funded by the Danish National Research Foundation under grant no.140. AJB acknowledges funding from the "First Galaxies" Advanced Grant from the European Research Council (ERC) under the European Union’s Horizon 2020 research and innovation programme (Grant agreement No. 789056). SC acknowledges support by European Union's HE ERC Starting Grant No. 101040227 - WINGS. ECL acknowledges support of an STFC Webb Fellowship (ST/W001438/1).


% \vspace{5mm}
% \facilities{Facilities go here.}


\software{{\sc astropy} \citep{astropycollaboration2022}, {\sc Bagpipes} \citep{carnall2018}, {\sc Multinest} \citep{feroz2019}, {\sc PyMultinest} \citep{buchner2014}}


\clearpage

\appendix
\renewcommand\thefigure{\thesection.\arabic{figure}}    

% \section{Classification of sources}
% \setcounter{figure}{0}

% Here is a table with the classification of all sources mentioned in this work.

% \begin{table*}[htb!]
% \begin{footnotesize}
% \caption{Classification Summary  \SA{Add Ra, Dec.} \label{tbl:classification}}\begin{tabular}{lccccccc}
% \hline\hline
% ID & UVJ & PSB & SSFR$_{97.5}$ & SSFR$_{50}$ & NIRCam Colors & FLARES & ugi  \\
%  &  &  &  &  & \citet{long2023} & \citet{lovell2023} & \citet{antwi-danso2023}   \\
%  \hline
% 176606 & $\checkmark$ & $\checkmark$ & $\checkmark$ & $\checkmark$ & $\checkmark$ & $\checkmark$ & $\checkmark$ \\

% \hline
% \end{tabular}
% \end{footnotesize}
% \end{table*}

\section{Post-Starburst Candidate Cutouts and SEDs}
\setcounter{figure}{0}

\begin{figure*}[!h]
    \centering
    \includegraphics[width=1.7\columnwidth]{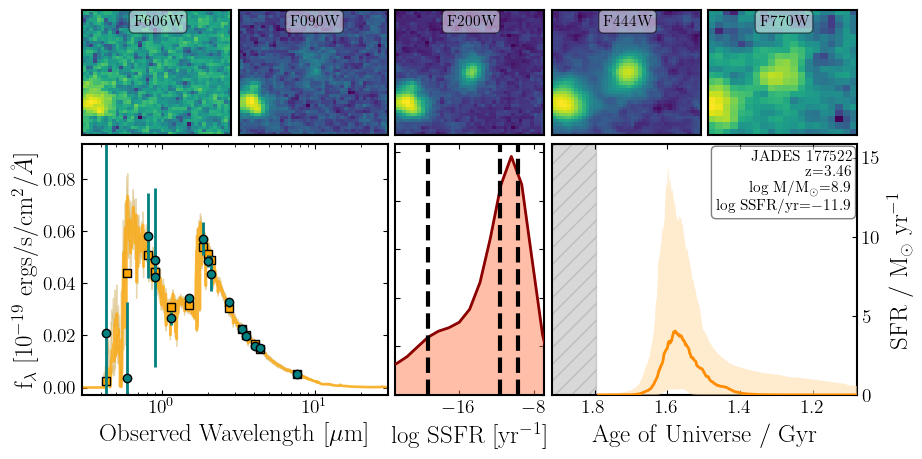}
    \includegraphics[width=1.7\columnwidth]{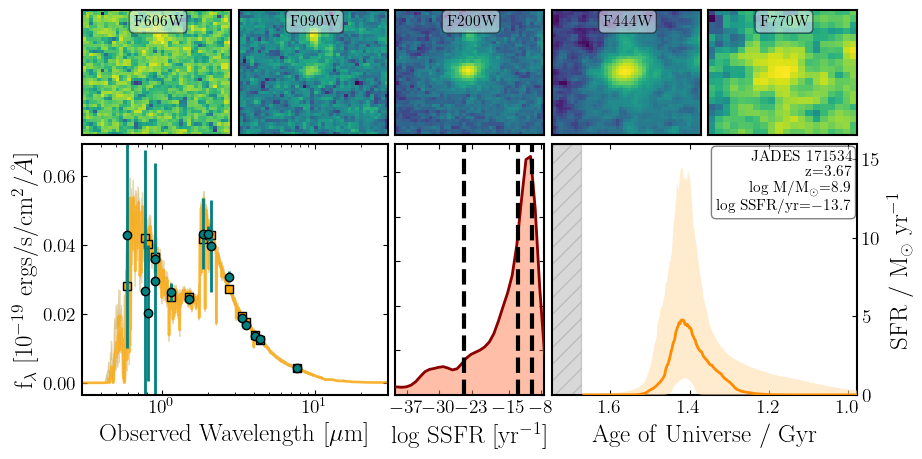}
    \caption{Properties of robust post-starburst galaxies selected via B19. \emph{Top rows for each source:} F606W, F090W, F200W, F444W, F770W cutouts, $1.2\arcsec$ on a side.  \emph{Bottom rows for each source:} The SEDs (left), SSFR posterior distributions (middle), and SFHs (right).  }
    \label{fig:psb1}
\end{figure*}

\begin{figure*}[tbh!]\ContinuedFloat
    \centering
    \includegraphics[width=1.7\columnwidth]{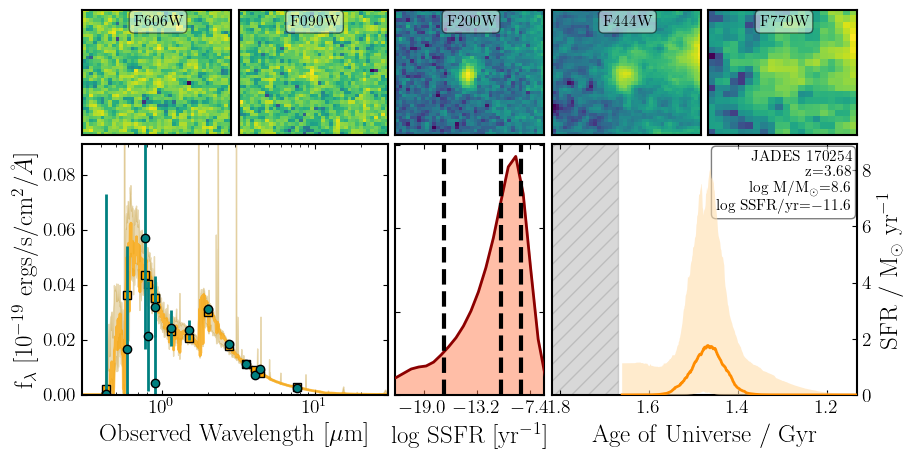}
    \includegraphics[width=1.7\columnwidth]{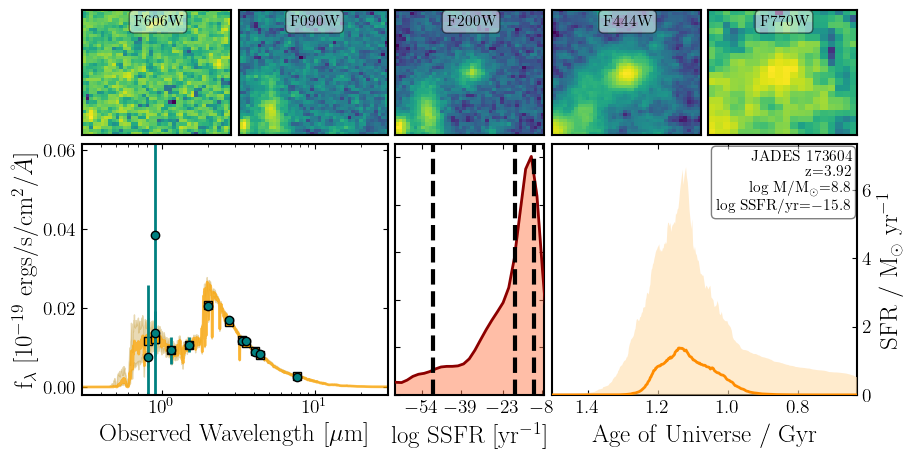}
     \includegraphics[width=1.7\columnwidth]{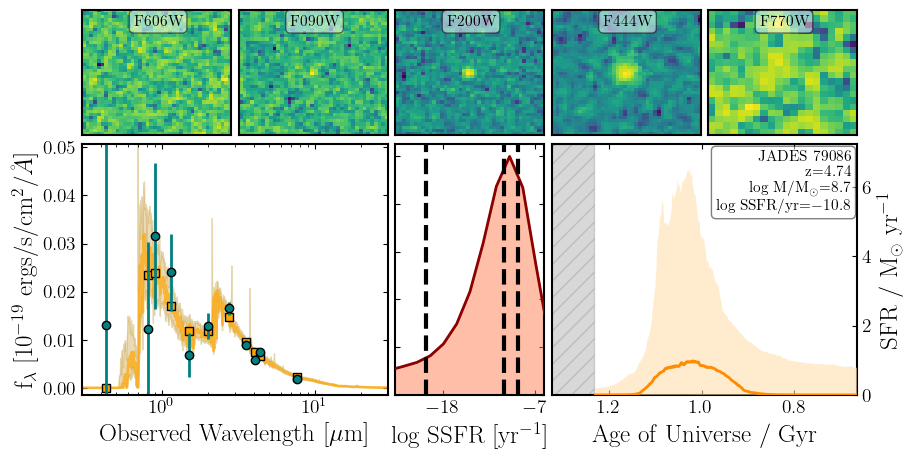}
    \caption{Continued}
\end{figure*}

\begin{figure*}[tbh!]
    \centering
    \includegraphics[width=1.7\columnwidth]{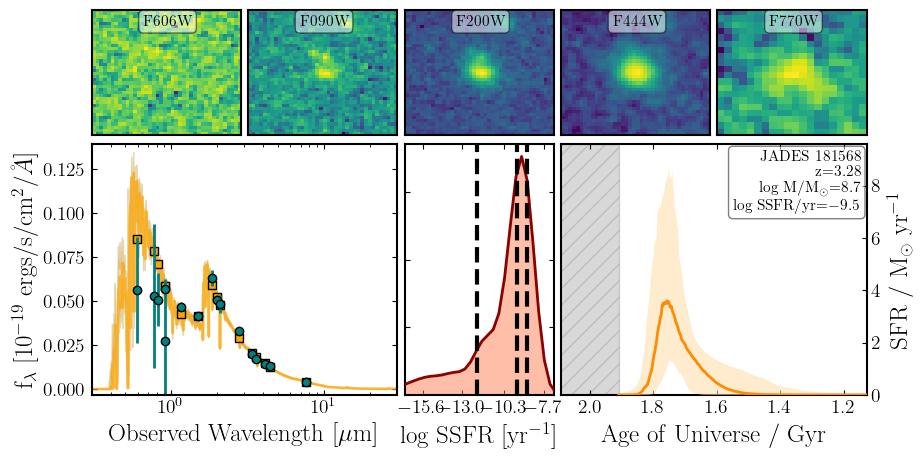}
    \includegraphics[width=1.7\columnwidth]{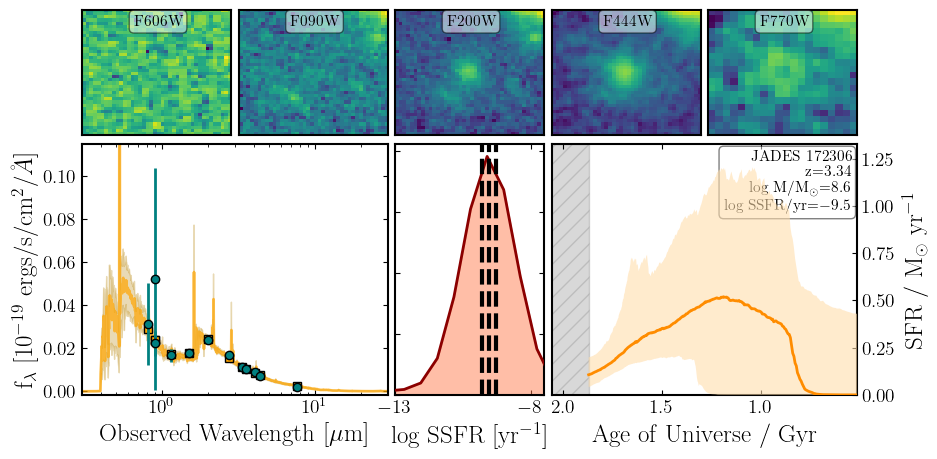}
    \includegraphics[width=1.7\columnwidth]{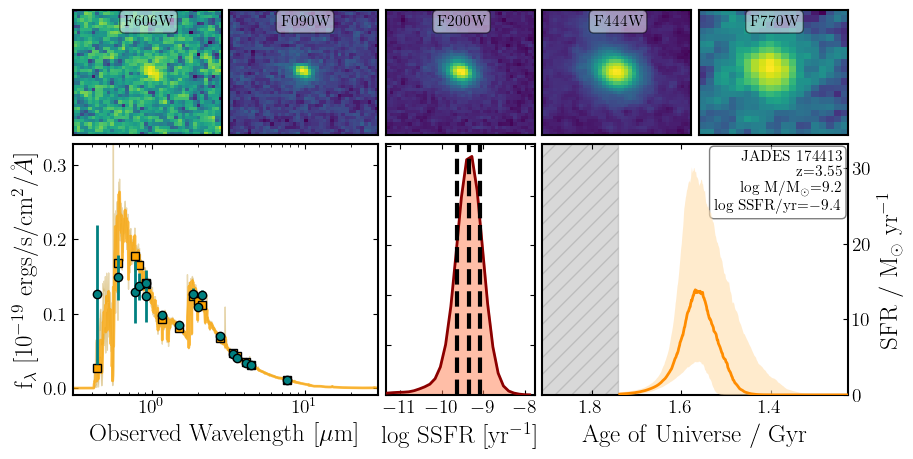}
    \caption{Properties of the tentative post-starburst galaxies selected via B19. \emph{Top rows for each source:} F606W, F090W, F200W, F444W, F770W cutouts, $1.2\arcsec$ on a side.  \emph{Bottom rows for each source:} The SEDs (left), SSFR posterior distributions (middle), and SFHs (right).  }
    \label{fig:psb2}
\end{figure*}

\begin{figure*}[tbh!]\ContinuedFloat
    \centering
    \includegraphics[width=1.7\columnwidth]{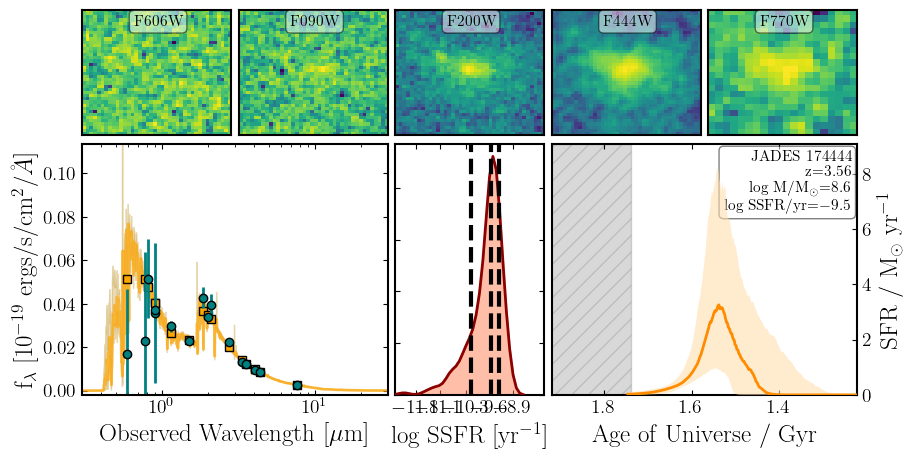}
    \includegraphics[width=1.7\columnwidth]{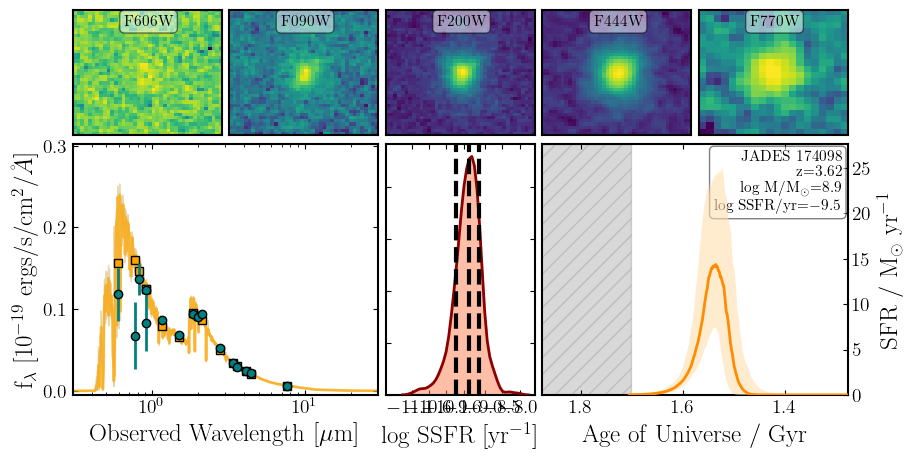}
     \includegraphics[width=1.7\columnwidth]{79086_sed_w_sfh.png}
    \caption{Continued}
\end{figure*}

\begin{figure*}[tbh!]\ContinuedFloat
    \centering
    \includegraphics[width=1.7\columnwidth]{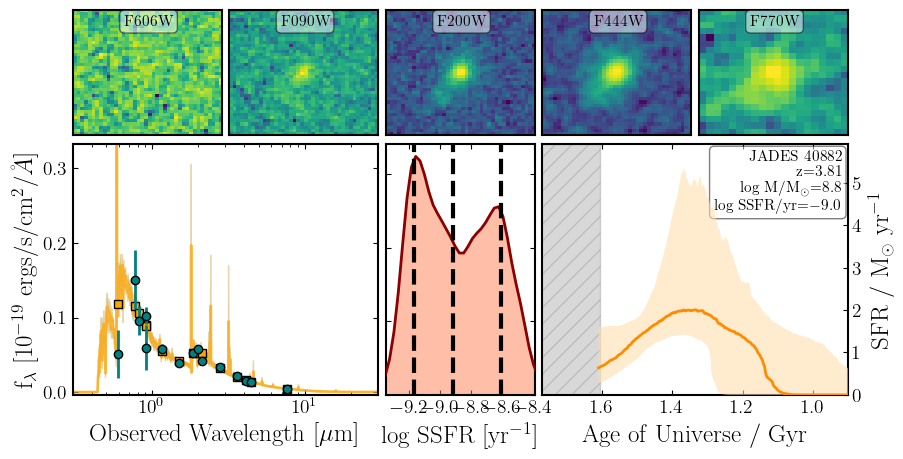}
    \includegraphics[width=1.7\columnwidth]{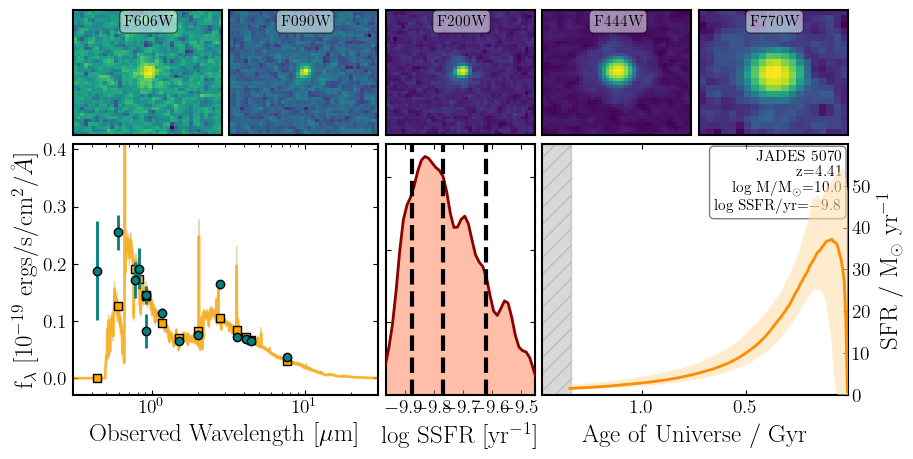}
     \includegraphics[width=1.7\columnwidth]{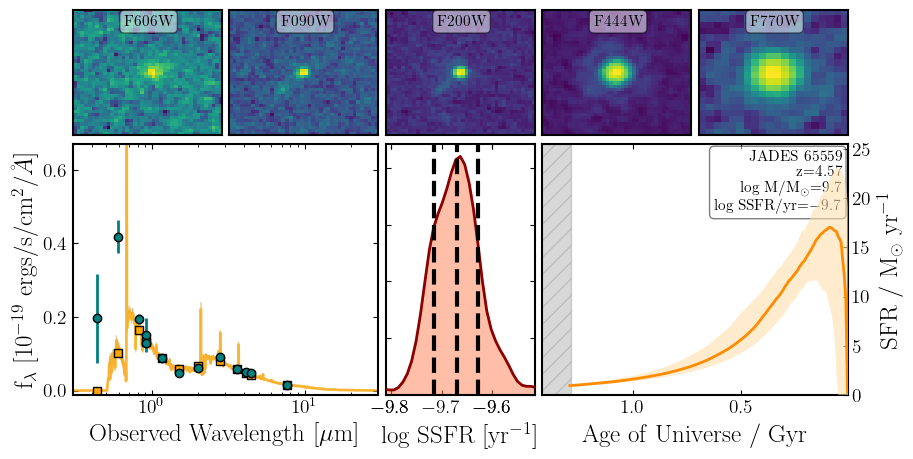}
    \caption{Continued}
\end{figure*}

% \begin{figure}[tbh!]
%     \centering
%     \includegraphics[width=\columnwidth]{174413_sed.png}
%     \includegraphics[width=\columnwidth]{184872_sed.png}
%     \includegraphics[width=\columnwidth]{5070_sed.png}
%     \includegraphics[width=\columnwidth]{65559_sed.png}
%     \caption{The SEDs (left), SSFR posterior distributions (middle), and F606W, F090W, F444W,F770W cutouts (right) of the post-starburst galaxy sample.}

%     \label{fig:psb}
% \end{figure}
\clearpage

\section{Overdensities at $z=3-4$ in JADES GOODS-S}\label{app:overdensity}
\setcounter{figure}{0}

In this section, we outline the identification of the galaxy overdensities discussed in Section~\ref{sec:rose}.  For an overdensity associated with the Cosmic Rose (Figure~\ref{fig:rose}), we searched the v0.8.1 JADES GOODS-S photometric catalog \citep{rieke2023} for relatively bright ($<28$ mag in F200W, F277W, F356W, and F444W) galaxies at $3.55<z_{\rm phot}<3.85$ using PSF-matched Kron photometry and EAZY photometric redshifts \citep{hainline2023}. 
%In order to explore this further, we first quantify the significance of a photometric overdensity at $z \sim 3.7$ in the vicinity of the Cosmic Rose. We searched the publicly available JADES GOODS-S Data Release 2 \citep{Eisenstein2023b:JOF} catalog for relatively bright galaxies in the vicinity of the Cosmic Rose by requiring objects to be brighter than $28$ AB mag in F200W, F277W, F356W, and F444W using PSF-matched Kron photometry. 
%This filter set was chosen since these are the four broad-band filters that are stacked to create the NIRCam detection image, which was constructed in order not to bias against short wavelength (SW) dropouts (e.g., F090W, F115W, or F150W). We further select galaxies in the vicinity of the Cosmic Rose by requiring objects to have photometric redshifts $3.55 < z_{\,\mathrm{phot}} < 3.85$, adopting the values that correspond to the fit where the EAZY likelihood was maximized ($\chi^{2}$ was minimized). 
We choose to not make any cut on the photometric redshift uncertainty
%, quantified as the difference between the 16th and 84th percentiles of the photometric redshift posterior distribution, 
since this would bias against the quiescent and/or post-starburst galaxy candidates identified here. Following these selections, we are left with a photometric sample of $N = 940$ sources.
%that are potentially associated with the Cosmic Rose. 

Following the methodology briefly described in \citet{sandles2023}, we then utilize a kernel density estimator (KDE) assuming Gaussian kernels to estimate the underlying density field of our photometric sample. The assumed bandwidth (or smoothing scale) is optimized by maximizing the likelihood cross-validation (LCV) quantity \citep{chartab2020}, which provides an optimal trade-off between under- and over-smoothing. Our optimized bandwidth is $0.86$ cMpc, which roughly corresponds to $0.42$ arcmin at $z = 3.7$. The KDE method identifies $N = 4$ spatially distinct galaxy overdensities which have peak significance levels larger than $4\sigma$, where $\sigma$ corresponds to the standard deviation of the density values across the entire JADES GOODS-S field. One of these identified galaxy overdensities spatially and kinematically encompass the Cosmic Rose, containing $N = 16$ potential members with a peak significance level of roughly $4.3\sigma$ at $\langle z_{\,\mathrm{phot}} \rangle = 3.759$. The maximum separation between these potential constituent members is $0.79$ cMpc, or roughly $0.37$ arcmin.  The full $z\sim3.7$ overdensity identified here is coincident with a structure (``Sparsh'' at $<z>=3.696$) independently identified using spectroscopic and photometric optical catalogs and Voronoi tessellation Monte Carlo (VMC) mapping in \citet{shah2023}.

\begin{figure}[tbh!]
    \centering
    \includegraphics[width=\columnwidth]{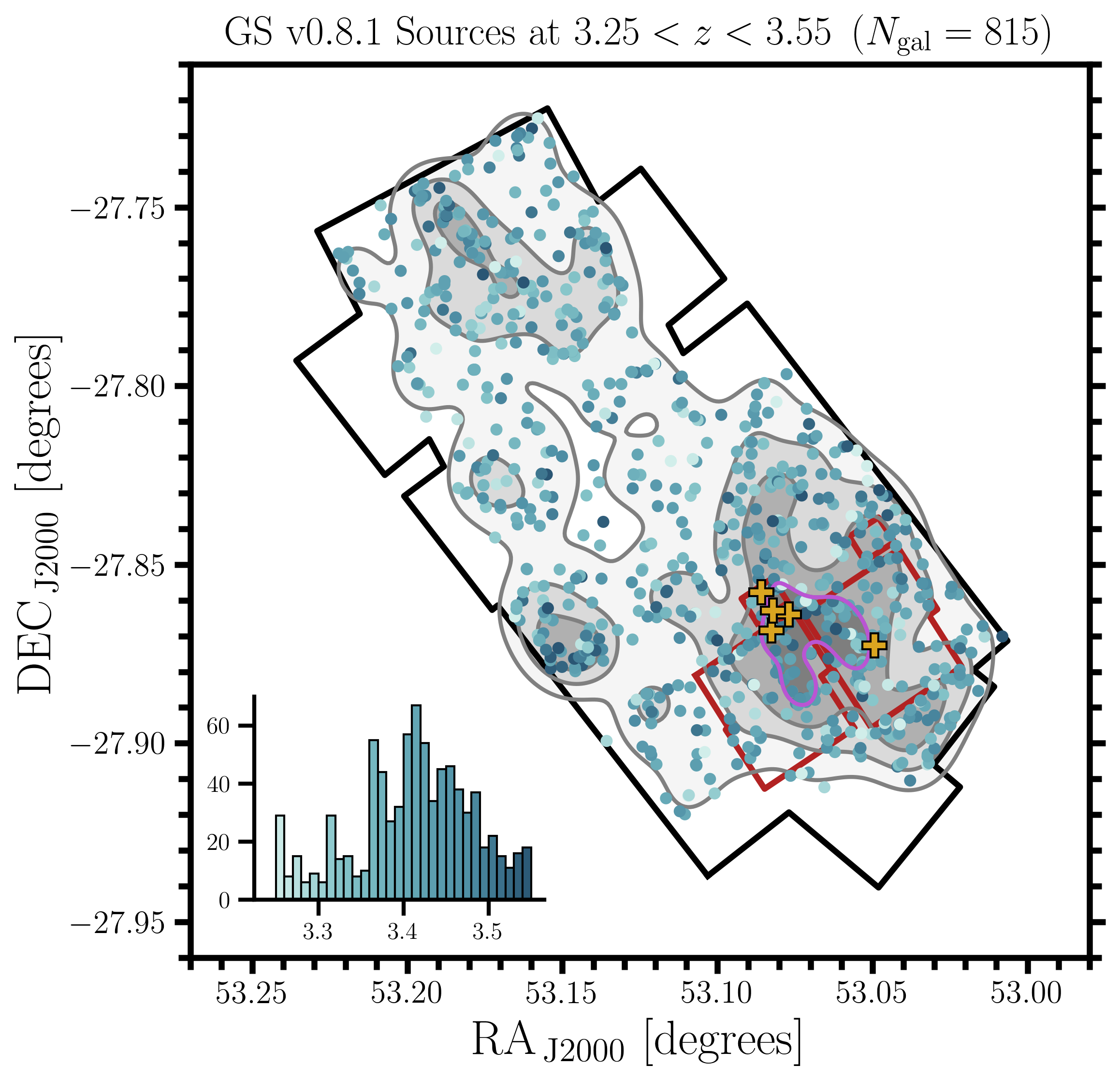}
    \caption{Overdensities of galaxies at $3.25<z<3.55$ in the JADES NIRCam (black outline) and MIRI parallel (red outline) field-of-views. Contours increment by $1\sigma$, with the purple
    contour outlining $4\sigma$ peaks. The yellow pluses show low-mass PSBs that are part of the $z\sim3.4$ overdensity %(see Section~\ref{sec:rose}).
    }

    \label{fig:overdensity}
\end{figure}

We repeat this procedure at $3.25<z<3.55$, finding $N_{\rm gal}=8.15$ galaxies.  We estimate the underlying density field using an optimized bandwidth is $0.9$ cMpc and find a $>4\sigma$ peak at $z\sim3.4$ (Figure~\ref{fig:overdensity}).  This overdensity is coincident with ``Smruti'' at $<z>=3.466$ identified in \citep{shah2023}.

\newpage

\bibliography{main}{}

\begin{thebibliography}{}
\expandafter\ifx\csname natexlab\endcsname\relax\def\natexlab#1{#1}\fi
\providecommand{\url}[1]{\href{#1}{#1}}
\providecommand{\dodoi}[1]{doi:~\href{http://doi.org/#1}{\nolinkurl{#1}}}
\providecommand{\doeprint}[1]{\href{http://ascl.net/#1}{\nolinkurl{http://ascl.net/#1}}}
\providecommand{\doarXiv}[1]{\href{https://arxiv.org/abs/#1}{\nolinkurl{https://arxiv.org/abs/#1}}}

\bibitem[{Akhshik {et~al.}(2021)Akhshik, Whitaker, Leja, Mahler, Sharon, Brammer, Toft, Bezanson, Man, Nelson, Pacifici, Wellons, \& Williams}]{akhshik2021}
Akhshik, M., Whitaker, K.~E., Leja, J., {et~al.} 2021, ApJ, 907, L8, \dodoi{10.3847/2041-8213/abd416}

\bibitem[{Akhshik {et~al.}(2023)Akhshik, Whitaker, Leja, Richard, Spilker, Song, Brammer, Bezanson, Ebeling, Gallazzi, Mahler, Mowla, Nelson, Pacifici, Sharon, Toft, Williams, Wright, \& Zabl}]{akhshik2023}
---. 2023, ApJ, 943, 179, \dodoi{10.3847/1538-4357/aca677}

\bibitem[{Alatalo {et~al.}(2016)Alatalo, Lisenfeld, Lanz, Appleton, Ardila, Cales, Kewley, Lacy, Medling, Nyland, Rich, \& Urry}]{alatalo2016}
Alatalo, K., Lisenfeld, U., Lanz, L., {et~al.} 2016, ApJ, 827, 106, \dodoi{10.3847/0004-637X/827/2/106}

\bibitem[{Alberts \& Noble(2022)}]{alberts2022a}
Alberts, S., \& Noble, A. 2022, Universe, 8, 554, \dodoi{10.3390/universe8110554}

\bibitem[{{Antwi-Danso} {et~al.}(2023{\natexlab{a}}){Antwi-Danso}, Papovich, Esdaile, Nanayakkara, Glazebrook, Hutchison, Whitaker, Marsan, Diaz, Marchesini, Muzzin, Tran, Setton, Kaushal, Speagle, \& Cole}]{antwi-danso2023a}
{Antwi-Danso}, J., Papovich, C., Esdaile, J., {et~al.} 2023{\natexlab{a}}, The {{FENIKS Survey}}: {{Spectroscopic Confirmation}} of {{Massive Quiescent Galaxies}} at z {\textasciitilde} 3-5, \dodoi{10.48550/arXiv.2307.09590}

\bibitem[{{Antwi-Danso} {et~al.}(2023{\natexlab{b}}){Antwi-Danso}, Papovich, Leja, Marchesini, Marsan, Martis, Labb{\'e}, Muzzin, Glazebrook, Straatman, \& Tran}]{antwi-danso2023}
{Antwi-Danso}, J., Papovich, C., Leja, J., {et~al.} 2023{\natexlab{b}}, ApJ, 943, 166, \dodoi{10.3847/1538-4357/aca294}

\bibitem[{Arrabal~Haro {et~al.}(2023)Arrabal~Haro, Dickinson, Finkelstein, Kartaltepe, Donnan, Burgarella, Carnall, Cullen, Dunlop, Fern{\'a}ndez, Fujimoto, Jung, Krips, Larson, Papovich, {P{\'e}rez-Gonz{\'a}lez}, Amor{\'i}n, Bagley, Buat, Casey, Chworowsky, Cohen, Ferguson, Giavalisco, {Huertas-Company}, Hutchison, Kocevski, Koekemoer, Lucas, McLeod, McLure, Pirzkal, Seill{\'e}, Trump, Weiner, Wilkins, \& Zavala}]{arrabalharo2023}
Arrabal~Haro, P., Dickinson, M., Finkelstein, S.~L., {et~al.} 2023, Nature, 622, 707, \dodoi{10.1038/s41586-023-06521-7}

\bibitem[{{Astropy Collaboration} {et~al.}(2022){Astropy Collaboration}, {Price-Whelan}, Lim, Earl, Starkman, Bradley, Shupe, Patil, Corrales, Brasseur, N{\"o}the, Donath, Tollerud, Morris, Ginsburg, Vaher, Weaver, Tocknell, Jamieson, {van Kerkwijk}, Robitaille, Merry, Bachetti, G{\"u}nther, Aldcroft, {Alvarado-Montes}, Archibald, B{\'o}di, Bapat, Barentsen, Baz{\'a}n, Biswas, Boquien, Burke, Cara, Cara, Conroy, Conseil, Craig, Cross, Cruz, D'Eugenio, Dencheva, Devillepoix, Dietrich, Eigenbrot, Erben, Ferreira, {Foreman-Mackey}, Fox, Freij, Garg, Geda, Glattly, Gondhalekar, Gordon, Grant, Greenfield, Groener, Guest, Gurovich, Handberg, Hart, {Hatfield-Dodds}, Homeier, Hosseinzadeh, Jenness, Jones, Joseph, Kalmbach, Karamehmetoglu, Ka{\l}uszy{\'n}ski, Kelley, Kern, Kerzendorf, Koch, Kulumani, Lee, Ly, Ma, MacBride, Maljaars, Muna, Murphy, Norman, O'Steen, Oman, Pacifici, Pascual, {Pascual-Granado}, Patil, Perren, Pickering, Rastogi, Roulston, Ryan, Rykoff, Sabater, Sakurikar, Salgado, Sanghi, Saunders,
  Savchenko, Schwardt, {Seifert-Eckert}, Shih, Jain, Shukla, Sick, Simpson, Singanamalla, Singer, Singhal, Sinha, Sip{\H o}cz, Spitler, Stansby, Streicher, {\v S}umak, Swinbank, Taranu, Tewary, Tremblay, {de Val-Borro}, Van~Kooten, Vasovi{\'c}, Verma, {de Miranda Cardoso}, Williams, Wilson, Winkel, {Wood-Vasey}, Xue, Yoachim, Zhang, Zonca, \& {Astropy Project Contributors}}]{astropycollaboration2022}
{Astropy Collaboration}, {Price-Whelan}, A.~M., Lim, P.~L., {et~al.} 2022, ApJ, 935, 167, \dodoi{10.3847/1538-4357/ac7c74}

\bibitem[{Bah{\'e} {et~al.}(2019)Bah{\'e}, Schaye, Barnes, Dalla~Vecchia, Kay, Bower, Hoekstra, McGee, \& Theuns}]{bahe2019}
Bah{\'e}, Y.~M., Schaye, J., Barnes, D.~J., {et~al.} 2019, MNRAS, 485, 2287, \dodoi{10.1093/mnras/stz361}

\bibitem[{Baker {et~al.}(2023{\natexlab{a}})Baker, Maiolino, Bluck, Belfiore, Curti, D'Eugenio, Piotrowska, Tacchella, \& Trussler}]{baker2023}
Baker, W.~M., Maiolino, R., Bluck, A. F.~L., {et~al.} 2023{\natexlab{a}}, The Black Hole Mass Metallicity Relation and Insights into Galaxy Quenching, \dodoi{10.48550/arXiv.2309.00670}

\bibitem[{Baker {et~al.}(2023{\natexlab{b}})Baker, Tacchella, Johnson, Nelson, Suess, D'Eugenio, Curti, {de Graaff}, Ji, Maiolino, Robertson, Scholtz, Alberts, Arribas, Boyett, Bunker, Carniani, Charlot, Chen, Chevallard, {Curtis-Lake}, Danhaive, DeCoursey, Egami, Eisenstein, Endsley, Hausen, Helton, Kumari, Looser, Maseda, Pusk{\'a}s, Rieke, Sandles, Sun, {\"U}bler, Williams, Willmer, \& Witstok}]{baker2023a}
Baker, W.~M., Tacchella, S., Johnson, B.~D., {et~al.} 2023{\natexlab{b}}, Inside-out Growth in the Early {{Universe}}: A Core in a Vigorously Star-Forming Disc, \dodoi{10.48550/arXiv.2306.02472}

\bibitem[{Baldry {et~al.}(2004)Baldry, Glazebrook, Brinkmann, Ivezi{\'c}, Lupton, Nichol, \& Szalay}]{baldry2004}
Baldry, I.~K., Glazebrook, K., Brinkmann, J., {et~al.} 2004, ApJ, 600, 681, \dodoi{10.1086/380092}

\bibitem[{Balogh {et~al.}(2000)Balogh, Navarro, \& Morris}]{balogh2000}
Balogh, M.~L., Navarro, J.~F., \& Morris, S.~L. 2000, ApJ, 540, 113, \dodoi{10.1086/309323}

\bibitem[{Beckwith {et~al.}(2006)Beckwith, Stiavelli, Koekemoer, Caldwell, Ferguson, Hook, Lucas, Bergeron, Corbin, Jogee, Panagia, Robberto, Royle, Somerville, \& Sosey}]{beckwith2006}
Beckwith, S. V.~W., Stiavelli, M., Koekemoer, A.~M., {et~al.} 2006, AJ, 132, 1729, \dodoi{10.1086/507302}

\bibitem[{Belli {et~al.}(2019)Belli, Newman, \& Ellis}]{belli2019}
Belli, S., Newman, A.~B., \& Ellis, R.~S. 2019, ApJ, 874, 17, \dodoi{10.3847/1538-4357/ab07af}

\bibitem[{Belli {et~al.}(2017)Belli, Genzel, F{\"o}rster~Schreiber, Wisnioski, Wilman, Wuyts, Mendel, Beifiori, Bender, Brammer, Burkert, Chan, Davies, Davies, Fabricius, Fossati, Galametz, Lang, Lutz, Momcheva, Nelson, Saglia, Tacconi, Tadaki, {\"U}bler, \& {van Dokkum}}]{belli2017}
Belli, S., Genzel, R., F{\"o}rster~Schreiber, N.~M., {et~al.} 2017, ApJ, 841, L6, \dodoi{10.3847/2041-8213/aa70e5}

\bibitem[{Belli {et~al.}(2021)Belli, Contursi, Genzel, Tacconi, {F{\"o}rster-Schreiber}, Lutz, Combes, Neri, {Garc{\'i}a-Burillo}, Schuster, {Herrera-Camus}, Tadaki, Davies, Davies, Johnson, Lee, Leja, Nelson, Price, Shangguan, Shimizu, Tacchella, \& {\"U}bler}]{belli2021}
Belli, S., Contursi, A., Genzel, R., {et~al.} 2021, ApJ, 909, L11, \dodoi{10.3847/2041-8213/abe6a6}

\bibitem[{{Ben{\'i}tez-Llambay} {et~al.}(2013){Ben{\'i}tez-Llambay}, Navarro, Abadi, Gottl{\"o}ber, Yepes, Hoffman, \& Steinmetz}]{benitez-llambay2013}
{Ben{\'i}tez-Llambay}, A., Navarro, J.~F., Abadi, M.~G., {et~al.} 2013, ApJ, 763, L41, \dodoi{10.1088/2041-8205/763/2/L41}

\bibitem[{Bezanson {et~al.}(2019)Bezanson, Spilker, Williams, Whitaker, Narayanan, Weiner, \& Franx}]{bezanson2019}
Bezanson, R., Spilker, J., Williams, C.~C., {et~al.} 2019, ApJ, 873, L19, \dodoi{10.3847/2041-8213/ab0c9c}

\bibitem[{Birkin {et~al.}(2021)Birkin, Weiss, Wardlow, Smail, Swinbank, Dudzevi{\v c}i{\=u}t{\.e}, An, Ao, Chapman, Chen, {da Cunha}, Dannerbauer, Gullberg, Hodge, Ikarashi, Ivison, Matsuda, Stach, Walter, Wang, \& {van der Werf}}]{birkin2021}
Birkin, J.~E., Weiss, A., Wardlow, J.~L., {et~al.} 2021, MNRAS, 501, 3926, \dodoi{10.1093/mnras/staa3862}

\bibitem[{{Bl{\'a}nquez-Ses{\'e}} {et~al.}(2023){Bl{\'a}nquez-Ses{\'e}}, Magdis, {G{\'o}mez-Guijarro}, Shuntov, Kokorev, Brammer, Valentino, {D{\'i}az-Santos}, Paspaliaris, Rigopoulou, Hjorth, Langeroodi, Gobat, Jin, Sillassen, Gillman, Greve, \& Lee}]{blanquez-sese2023}
{Bl{\'a}nquez-Ses{\'e}}, D., Magdis, G.~E., {G{\'o}mez-Guijarro}, C., {et~al.} 2023, Uncovering the {{MIR}} Emission of Quiescent Galaxies with \${{JWST}}\$,  {arXiv}, \dodoi{10.48550/arXiv.2310.01601}

\bibitem[{Bluck {et~al.}(2022)Bluck, Maiolino, Brownson, Conselice, Ellison, Piotrowska, \& Thorp}]{bluck2022}
Bluck, A. F.~L., Maiolino, R., Brownson, S., {et~al.} 2022, A\&A, 659, A160, \dodoi{10.1051/0004-6361/202142643}

\bibitem[{Bluck {et~al.}(2014)Bluck, Mendel, Ellison, Moreno, Simard, Patton, \& Starkenburg}]{bluck2014}
Bluck, A. F.~L., Mendel, J.~T., Ellison, S.~L., {et~al.} 2014, MNRAS, 441, 599, \dodoi{10.1093/mnras/stu594}

\bibitem[{Bluck {et~al.}(2023{\natexlab{a}})Bluck, Piotrowska, \& Maiolino}]{bluck2023a}
Bluck, A. F.~L., Piotrowska, J.~M., \& Maiolino, R. 2023{\natexlab{a}}, ApJ, 944, 108, \dodoi{10.3847/1538-4357/acac7c}

\bibitem[{Bluck {et~al.}(2016)Bluck, Mendel, Ellison, Patton, Simard, Henriques, Torrey, Teimoorinia, Moreno, \& Starkenburg}]{bluck2016}
Bluck, A. F.~L., Mendel, J.~T., Ellison, S.~L., {et~al.} 2016, MNRAS, 462, 2559, \dodoi{10.1093/mnras/stw1665}

\bibitem[{Bluck {et~al.}(2020)Bluck, Maiolino, Piotrowska, Trussler, Ellison, S{\'a}nchez, Thorp, Teimoorinia, Moreno, \& Conselice}]{bluck2020}
Bluck, A. F.~L., Maiolino, R., Piotrowska, J.~M., {et~al.} 2020, MNRAS, 499, 230, \dodoi{10.1093/mnras/staa2806}

\bibitem[{Bluck {et~al.}(2023{\natexlab{b}})Bluck, Conselice, Ormerod, Piotrowska, Adams, Austin, Caruana, Duncan, Ferreira, Goubert, Harvey, Trussler, \& Maiolino}]{bluck2023}
Bluck, A. F.~L., Conselice, C.~J., Ormerod, K., {et~al.} 2023{\natexlab{b}}, Galaxy Quenching at the High Redshift Frontier: {{A}} Fundamental Test of Cosmological Models in the Early Universe with {{JWST-CEERS}},  {arXiv}, \dodoi{10.48550/arXiv.2311.02526}

\bibitem[{Boselli {et~al.}(2022)Boselli, Fossati, \& Sun}]{boselli2022}
Boselli, A., Fossati, M., \& Sun, M. 2022, Astron. Astrophys. Rev. Vol. 30 Issue 1 Artic. Id3, 30, 3, \dodoi{10.1007/s00159-022-00140-3}

\bibitem[{Brammer {et~al.}(2008)Brammer, {van Dokkum}, \& Coppi}]{brammer2008}
Brammer, G.~B., {van Dokkum}, P.~G., \& Coppi, P. 2008, ApJ, 686, 1503, \dodoi{10.1086/591786}

\bibitem[{Bruzual \& Charlot(2003)}]{bruzual2003}
Bruzual, G., \& Charlot, S. 2003, MNRAS, 344, 1000, \dodoi{10.1046/j.1365-8711.2003.06897.x}

\bibitem[{Buchner {et~al.}(2014)Buchner, Georgakakis, Nandra, Hsu, Rangel, Brightman, Merloni, Salvato, Donley, \& Kocevski}]{buchner2014}
Buchner, J., Georgakakis, A., Nandra, K., {et~al.} 2014, Astron. Astrophys., 564, A125, \dodoi{10.1051/0004-6361/201322971}

\bibitem[{Bullock \& {Boylan-Kolchin}(2017)}]{bullock2017}
Bullock, J.~S., \& {Boylan-Kolchin}, M. 2017, Annu. Rev. Astron. Astrophys. Vol 55 Issue 1 Pp 343-387, 55, 343, \dodoi{10.1146/annurev-astro-091916-055313}

\bibitem[{Butcher \& Oemler(1978)}]{butcher1978}
Butcher, H., \& Oemler, Jr., A. 1978, ApJ, 226, 559, \dodoi{10.1086/156640}

\bibitem[{Byler {et~al.}(2017)Byler, Dalcanton, Conroy, \& Johnson}]{byler2017}
Byler, N., Dalcanton, J.~J., Conroy, C., \& Johnson, B.~D. 2017, ApJ, 840, 44, \dodoi{10.3847/1538-4357/aa6c66}

\bibitem[{Caliendo {et~al.}(2021)Caliendo, Whitaker, Akhshik, Wilson, Williams, Spilker, Mahler, Pope, Sharon, Aguilar, Bezanson, Chavez~Dagostino, {G{\'o}mez-Ruiz}, Monta{\~n}a, Toft, {Velazquez de la Rosa}, \& Zeballos}]{caliendo2021}
Caliendo, J.~N., Whitaker, K.~E., Akhshik, M., {et~al.} 2021, ApJ, 910, L7, \dodoi{10.3847/2041-8213/abe132}

\bibitem[{Calzetti {et~al.}(2000)Calzetti, Armus, Bohlin, Kinney, Koornneef, \& {Storchi-Bergmann}}]{calzetti2000}
Calzetti, D., Armus, L., Bohlin, R.~C., {et~al.} 2000, ApJ, 533, 682, \dodoi{10.1086/308692}

\bibitem[{Carnall {et~al.}(2019{\natexlab{a}})Carnall, Leja, Johnson, McLure, Dunlop, \& Conroy}]{carnall2019a}
Carnall, A.~C., Leja, J., Johnson, B.~D., {et~al.} 2019{\natexlab{a}}, ApJ, 873, 44, \dodoi{10.3847/1538-4357/ab04a2}

\bibitem[{Carnall {et~al.}(2018)Carnall, McLure, Dunlop, \& Dav{\'e}}]{carnall2018}
Carnall, A.~C., McLure, R.~J., Dunlop, J.~S., \& Dav{\'e}, R. 2018, MNRAS, 480, 4379, \dodoi{10.1093/mnras/sty2169}

\bibitem[{Carnall {et~al.}(2019{\natexlab{b}})Carnall, McLure, Dunlop, Cullen, McLeod, Wild, Johnson, Appleby, Dav{\'e}, Amorin, Bolzonella, Castellano, Cimatti, Cucciati, Gargiulo, Garilli, Marchi, Pentericci, Pozzetti, Schreiber, Talia, \& Zamorani}]{carnall2019}
Carnall, A.~C., McLure, R.~J., Dunlop, J.~S., {et~al.} 2019{\natexlab{b}}, MNRAS, 490, 417, \dodoi{10.1093/mnras/stz2544}

\bibitem[{Carnall {et~al.}(2020)Carnall, Walker, McLure, Dunlop, McLeod, Cullen, Wild, Amorin, Bolzonella, Castellano, Cimatti, Cucciati, Fontana, Gargiulo, Garilli, Jarvis, Pentericci, Pozzetti, Zamorani, Calabro, Hathi, \& Koekemoer}]{carnall2020}
Carnall, A.~C., Walker, S., McLure, R.~J., {et~al.} 2020, MNRAS, 496, 695, \dodoi{10.1093/mnras/staa1535}

\bibitem[{Carnall {et~al.}(2023{\natexlab{a}})Carnall, McLeod, McLure, Dunlop, Begley, Cullen, Donnan, Hamadouche, Jewell, Jones, Pollock, \& Wild}]{carnall2023a}
Carnall, A.~C., McLeod, D.~J., McLure, R.~J., {et~al.} 2023{\natexlab{a}}, MNRAS, 520, 3974, \dodoi{10.1093/mnras/stad369}

\bibitem[{Carnall {et~al.}(2023{\natexlab{b}})Carnall, McLure, Dunlop, McLeod, Wild, Cullen, Magee, Begley, Cimatti, Donnan, Hamadouche, Jewell, \& Walker}]{carnall2023}
Carnall, A.~C., McLure, R.~J., Dunlop, J.~S., {et~al.} 2023{\natexlab{b}}, A Massive Quiescent Galaxy at Redshift 4.658

\bibitem[{Casey {et~al.}(2022)Casey, Kartaltepe, Drakos, Franco, Ilbert, Rose, Cox, Nightingale, Robertson, Silverman, Koekemoer, Massey, McCracken, Rhodes, Akins, Amvrosiadis, {Arango-Toro}, Bagley, Capak, Champagne, Chartab, Chavez~Ortiz, Cooke, Cooper, Darvish, Ding, Faisst, Finkelstein, Fujimoto, Gentile, Gillman, Gould, Gozaliasl, Harish, Hayward, He, Hemmati, Hirschmann, Jin, Khostovan, Kokorev, Lambrides, Laigle, Leung, Liu, Liaudat, Long, Magdis, Mahler, Mainieri, Manning, Maraston, Martin, McCleary, McKinney, McPartland, Mobasher, Pattnaik, Renzini, Rich, Sanders, Sattari, Scognamiglio, Scoville, Sheth, Shuntov, Sparre, Suzuki, Talia, Toft, Trakhtenbrot, Urry, Valentino, Vanderhoof, Vardoulaki, Weaver, Whitaker, Wilkins, Yang, \& Zavala}]{casey2022a}
Casey, C.~M., Kartaltepe, J.~S., Drakos, N.~E., {et~al.} 2022, {{COSMOS-Web}}: {{An Overview}} of the {{JWST Cosmic Origins Survey}}

\bibitem[{Castignani {et~al.}(2022)Castignani, Combes, Jablonka, Finn, Rudnick, Vulcani, Desai, Zaritsky, \& Salom{\'e}}]{castignani2022a}
Castignani, G., Combes, F., Jablonka, P., {et~al.} 2022, Astron. Astrophys., 657, A9, \dodoi{10.1051/0004-6361/202040141}

\bibitem[{Cecchi {et~al.}(2019)Cecchi, Bolzonella, Cimatti, \& Girelli}]{cecchi2019}
Cecchi, R., Bolzonella, M., Cimatti, A., \& Girelli, G. 2019, ApJ, 880, L14, \dodoi{10.3847/2041-8213/ab2c80}

\bibitem[{Chartab {et~al.}(2020)Chartab, Mobasher, Darvish, Finkelstein, Guo, Kodra, Lee, Newman, Pacifici, Papovich, Sattari, Shahidi, Dickinson, Faber, Ferguson, Giavalisco, \& Jafariyazani}]{chartab2020}
Chartab, N., Mobasher, B., Darvish, B., {et~al.} 2020, ApJ, 890, 7, \dodoi{10.3847/1538-4357/ab61fd}

\bibitem[{Chevallard \& Charlot(2016)}]{chevallard2016}
Chevallard, J., \& Charlot, S. 2016, MNRAS, 462, 1415, \dodoi{10.1093/mnras/stw1756}

\bibitem[{Cochrane {et~al.}(2022)Cochrane, Hayward, \& {Angl{\'e}s-Alc{\'a}zar}}]{cochrane2022}
Cochrane, R.~K., Hayward, C.~C., \& {Angl{\'e}s-Alc{\'a}zar}, D. 2022, ApJ, 939, L27, \dodoi{10.3847/2041-8213/ac951d}

\bibitem[{Cortese {et~al.}(2021/ed)Cortese, Catinella, \& Smith}]{cortese2021}
Cortese, L., Catinella, B., \& Smith, R. 2021/ed, Publ. Astron. Soc. Aust., 38, \dodoi{10.1017/pasa.2021.18}

\bibitem[{Crain {et~al.}(2015)Crain, Schaye, Bower, Furlong, Schaller, Theuns, Dalla~Vecchia, Frenk, McCarthy, Helly, Jenkins, {Rosas-Guevara}, White, \& Trayford}]{crain2015}
Crain, R.~A., Schaye, J., Bower, R.~G., {et~al.} 2015, MNRAS, 450, 1937, \dodoi{10.1093/mnras/stv725}

\bibitem[{Cullen {et~al.}(2019)Cullen, McLure, Dunlop, Khochfar, Dav{\'e}, Amor{\'i}n, Bolzonella, Carnall, Castellano, Cimatti, Cirasuolo, Cresci, Fynbo, Fontanot, Gargiulo, Garilli, Guaita, Hathi, Hibon, Mannucci, Marchi, McLeod, Pentericci, Pozzetti, Shapley, Talia, \& Zamorani}]{cullen2019}
Cullen, F., McLure, R.~J., Dunlop, J.~S., {et~al.} 2019, MNRAS, 487, 2038, \dodoi{10.1093/mnras/stz1402}

\bibitem[{Dav{\'e} {et~al.}(2012)Dav{\'e}, Finlator, \& Oppenheimer}]{dave2012}
Dav{\'e}, R., Finlator, K., \& Oppenheimer, B.~D. 2012, MNRAS, 421, 98, \dodoi{10.1111/j.1365-2966.2011.20148.x}

\bibitem[{Davidzon {et~al.}(2017)Davidzon, Ilbert, Laigle, Coupon, McCracken, Delvecchio, Masters, Capak, Hsieh, Le~F{\`e}vre, Tresse, Bethermin, Chang, Faisst, Le~Floc'h, Steinhardt, Toft, Aussel, Dubois, Hasinger, Salvato, Sanders, Scoville, \& Silverman}]{davidzon2017}
Davidzon, I., Ilbert, O., Laigle, C., {et~al.} 2017, Astron. Astrophys., 605, A70, \dodoi{10.1051/0004-6361/201730419}

\bibitem[{Davis {et~al.}(2022)Davis, Kaviraj, Hardcastle, Martin, Jackson, Kraljic, Malek, Peirani, Smith, Volonteri, \& Wang}]{davis2022a}
Davis, F., Kaviraj, S., Hardcastle, M.~J., {et~al.} 2022, MNRAS, 511, 4109, \dodoi{10.1093/mnras/stac068}

\bibitem[{Desprez {et~al.}(2023)Desprez, Martis, Asada, Sawicki, Willott, Muzzin, Abraham, Brada{\v c}, Brammer, {Estrada-Carpenter}, Iyer, Matharu, Mowla, Noirot, Sarrouh, Strait, Gledhill, \& Rihtar{\v s}i{\v c}}]{desprez2023}
Desprez, G., Martis, N.~S., Asada, Y., {et~al.} 2023, \${\textbackslash}{{Lambda}}\${{CDM}} Not Dead yet: Massive High-z {{Balmer}} Break Galaxies Are Less Common than Previously Reported,  {arXiv}, \dodoi{10.48550/arXiv.2310.03063}

\bibitem[{D'Eugenio {et~al.}(2020)D'Eugenio, Daddi, Gobat, Strazzullo, Lustig, Delvecchio, Jin, Puglisi, Calabr{\'o}, Mancini, Dickinson, Cimatti, \& Onodera}]{deugenio2020}
D'Eugenio, C., Daddi, E., Gobat, R., {et~al.} 2020, ApJ, 892, L2, \dodoi{10.3847/2041-8213/ab7a96}

\bibitem[{D'Eugenio {et~al.}(2023)D'Eugenio, {Perez-Gonzalez}, Maiolino, Scholtz, Perna, Circosta, Uebler, Arribas, Boeker, Bunker, Carniani, Charlot, Chevallard, Cresci, {Curtis-Lake}, Jones, Kumari, Lamperti, Looser, Parlanti, Rix, Robertson, Rodriguez Del~Pino, Tacchella, Venturi, \& Willott}]{deugenio2023}
D'Eugenio, F., {Perez-Gonzalez}, P., Maiolino, R., {et~al.} 2023, A Fast-Rotator Post-Starburst Galaxy Quenched by Supermassive Black-Hole Feedback at Z=3, \dodoi{10.48550/arXiv.2308.06317}

\bibitem[{{D{\'i}az-Garc{\'i}a} {et~al.}(2019){D{\'i}az-Garc{\'i}a}, Cenarro, {L{\'o}pez-Sanjuan}, Ferreras, Cervi{\~n}o, {Fern{\'a}ndez-Soto}, Gonz{\'a}lez~Delgado, M{\'a}rquez, Povi{\'c}, San~Roman, Viironen, Moles, {Crist{\'o}bal-Hornillos}, {L{\'o}pez-Comazzi}, Alfaro, {Aparicio-Villegas}, Ben{\'i}tez, Broadhurst, {Cabrera-Ca{\~n}o}, Castander, Cepa, Husillos, Infante, Aguerri, Mart{\'i}nez, Masegosa, Molino, {del Olmo}, Perea, Prada, \& Quintana}]{diaz-garcia2019}
{D{\'i}az-Garc{\'i}a}, L.~A., Cenarro, A.~J., {L{\'o}pez-Sanjuan}, C., {et~al.} 2019, Astron. Amp Astrophys. Vol. 631 IdA156 NUMPAGES37NUMPAGES Pp, 631, A156, \dodoi{10.1051/0004-6361/201832788}

\bibitem[{Dome {et~al.}(2024)Dome, Tacchella, Fialkov, Ceverino, Dekel, Ginzburg, Lapiner, \& Looser}]{dome2024}
Dome, T., Tacchella, S., Fialkov, A., {et~al.} 2024, MNRAS, 527, 2139, \dodoi{10.1093/mnras/stad3239}

\bibitem[{Efstathiou(1992)}]{efstathiou1992}
Efstathiou, G. 1992, MNRAS, 256, 43P, \dodoi{10.1093/mnras/256.1.43P}

\bibitem[{Eisenstein {et~al.}(2023)Eisenstein, Willott, Alberts, Arribas, Bonaventura, Bunker, Cameron, Carniani, Charlot, {Curtis-Lake}, D'Eugenio, Endsley, Ferruit, Giardino, Hainline, Hausen, Jakobsen, Johnson, Maiolino, Rieke, Rieke, Rix, Robertson, Stark, Tacchella, Williams, Willmer, Baker, Baum, Bhatawdekar, Boyett, Chen, Chevallard, Circosta, Curti, Danhaive, DeCoursey, {de Graaff}, Dressler, Egami, Helton, Hviding, Ji, Jones, Kumari, L{\"u}tzgendorf, Laseter, Looser, Lyu, Maseda, Nelson, Parlanti, Perna, Pusk{\'a}s, Rawle, Rodr{\'i}guez Del~Pino, Sandles, Saxena, Scholtz, Sharpe, Shivaei, Silcock, Simmonds, Skarbinski, Smit, Stone, Suess, Sun, Tang, Topping, {\"U}bler, Villanueva, Wallace, Whitler, Witstok, \& Woodrum}]{eisenstein2023}
Eisenstein, D.~J., Willott, C., Alberts, S., {et~al.} 2023, Overview of the {{JWST Advanced Deep Extragalactic Survey}} ({{JADES}}), \dodoi{10.48550/arXiv.2306.02465}

\bibitem[{Endsley {et~al.}(2023)Endsley, Stark, Whitler, Topping, Chen, Plat, Chisholm, \& Charlot}]{endsley2023}
Endsley, R., Stark, D.~P., Whitler, L., {et~al.} 2023, MNRAS, 524, 2312, \dodoi{10.1093/mnras/stad1919}

\bibitem[{Esdaile {et~al.}(2021)Esdaile, Labb{\'e}, Glazebrook, {Antwi-Danso}, Papovich, Taylor, Marsan, Muzzin, Straatman, Marchesini, Diaz, Spitler, Tran, \& Goodsell}]{esdaile2021}
Esdaile, J., Labb{\'e}, I., Glazebrook, K., {et~al.} 2021, AJ, 162, 225, \dodoi{10.3847/1538-3881/ac2148}

\bibitem[{Fang {et~al.}(2018)Fang, Faber, Koo, {Rodr{\'i}guez-Puebla}, Guo, Barro, Behroozi, Brammer, Chen, Dekel, Ferguson, Gawiser, Giavalisco, Kartaltepe, Kocevski, Koekemoer, McGrath, McIntosh, Newman, Pacifici, Pandya, {P{\'e}rez-Gonz{\'a}lez}, Primack, Salmon, Trump, Weiner, Willner, Acquaviva, Dahlen, Finkelstein, Finlator, Fontana, Galametz, Grogin, Gruetzbauch, Johnson, Mobasher, Papovich, Pforr, Salvato, Santini, {van der Wel}, Wiklind, \& Wuyts}]{fang2018}
Fang, J.~J., Faber, S.~M., Koo, D.~C., {et~al.} 2018, ApJ, 858, 100, \dodoi{10.3847/1538-4357/aabcba}

\bibitem[{Ferland {et~al.}(2013)Ferland, Porter, {van Hoof}, Williams, Abel, Lykins, Shaw, Henney, \& Stancil}]{ferland2013}
Ferland, G.~J., Porter, R.~L., {van Hoof}, P. A.~M., {et~al.} 2013, Rev. Mex. Astron. Astrofisica, 49, 137, \dodoi{10.48550/arXiv.1302.4485}

\bibitem[{Feroz {et~al.}(2019)Feroz, Hobson, Cameron, \& Pettitt}]{feroz2019}
Feroz, F., Hobson, M.~P., Cameron, E., \& Pettitt, A.~N. 2019, Open J. Astrophys., 2, 10, \dodoi{10.21105/astro.1306.2144}

\bibitem[{Fitzpatrick {et~al.}(2002)Fitzpatrick, Ribas, Guinan, DeWarf, Maloney, \& Massa}]{fitzpatrick2002}
Fitzpatrick, E.~L., Ribas, I., Guinan, E.~F., {et~al.} 2002, ApJ, 564, 260, \dodoi{10.1086/324184}

\bibitem[{Fontana {et~al.}(2009)Fontana, Santini, Grazian, Pentericci, Fiore, Castellano, Giallongo, Menci, Salimbeni, Cristiani, Nonino, \& Vanzella}]{fontana2009}
Fontana, A., Santini, P., Grazian, A., {et~al.} 2009, Astron. Astrophys., 501, 15, \dodoi{10.1051/0004-6361/200911650}

\bibitem[{Forrest {et~al.}(2020)Forrest, Marsan, Annunziatella, Wilson, Muzzin, Marchesini, Cooper, Chan, McConachie, Gomez, {Kado-Fong}, La~Barbera, {Lange-Vagle}, Nantais, Nonino, Saracco, Stefanon, \& {van der Burg}}]{forrest2020a}
Forrest, B., Marsan, Z.~C., Annunziatella, M., {et~al.} 2020, ApJ, 903, 47, \dodoi{10.3847/1538-4357/abb819}

\bibitem[{French {et~al.}(2015)French, Yang, Zabludoff, Narayanan, Shirley, Walter, Smith, \& Tremonti}]{french2015}
French, K.~D., Yang, Y., Zabludoff, A., {et~al.} 2015, ApJ, 801, 1, \dodoi{10.1088/0004-637X/801/1/1}

\bibitem[{{Gaia Collaboration} {et~al.}(2018){Gaia Collaboration}, Brown, Vallenari, Prusti, {de Bruijne}, Babusiaux, {Bailer-Jones}, Biermann, Evans, Eyer, Jansen, Jordi, Klioner, Lammers, Lindegren, Luri, Mignard, Panem, Pourbaix, Randich, Sartoretti, Siddiqui, Soubiran, {van Leeuwen}, Walton, Arenou, Bastian, Cropper, Drimmel, Katz, Lattanzi, Bakker, Cacciari, Casta{\~n}eda, Chaoul, Cheek, De~Angeli, Fabricius, Guerra, Holl, Masana, Messineo, Mowlavi, Nienartowicz, Panuzzo, Portell, Riello, Seabroke, Tanga, Th{\'e}venin, {Gracia-Abril}, Comoretto, {Garcia-Reinaldos}, Teyssier, Altmann, Andrae, Audard, {Bellas-Velidis}, Benson, Berthier, Blomme, Burgess, Busso, Carry, Cellino, Clementini, Clotet, Creevey, Davidson, De~Ridder, Delchambre, Dell'Oro, Ducourant, {Fern{\'a}ndez-Hern{\'a}ndez}, Fouesneau, Fr{\'e}mat, Galluccio, {Garc{\'i}a-Torres}, {Gonz{\'a}lez-N{\'u}{\~n}ez}, {Gonz{\'a}lez-Vidal}, Gosset, Guy, Halbwachs, Hambly, Harrison, Hern{\'a}ndez, Hestroffer, Hodgkin, Hutton, Jasniewicz,
  {Jean-Antoine-Piccolo}, Jordan, Korn, {Krone-Martins}, Lanzafame, Lebzelter, L{\"o}ffler, Manteiga, Marrese, {Mart{\'i}n-Fleitas}, Moitinho, Mora, Muinonen, Osinde, Pancino, Pauwels, Petit, {Recio-Blanco}, Richards, Rimoldini, Robin, Sarro, Siopis, Smith, Sozzetti, S{\"u}veges, Torra, {van Reeven}, Abbas, Abreu~Aramburu, Accart, Aerts, Altavilla, {\'A}lvarez, Alvarez, Alves, Anderson, Andrei, Anglada~Varela, Antiche, Antoja, Arcay, Astraatmadja, Bach, Baker, {Balaguer-N{\'u}{\~n}ez}, Balm, Barache, Barata, Barbato, Barblan, Barklem, Barrado, Barros, Barstow, Bartholom{\'e}~Mu{\~n}oz, Bassilana, Becciani, Bellazzini, Berihuete, Bertone, Bianchi, Bienaym{\'e}, {Blanco-Cuaresma}, Boch, Boeche, Bombrun, Borrachero, Bossini, Bouquillon, Bourda, Bragaglia, Bramante, Breddels, Bressan, Brouillet, Br{\"u}semeister, Brugaletta, Bucciarelli, Burlacu, Busonero, Butkevich, Buzzi, Caffau, Cancelliere, Cannizzaro, {Cantat-Gaudin}, Carballo, Carlucci, Carrasco, Casamiquela, Castellani, {Castro-Ginard}, Charlot, Chemin,
  Chiavassa, Cocozza, Costigan, Cowell, Crifo, Crosta, Crowley, Cuypers, Dafonte, Damerdji, Dapergolas, David, David, {de Laverny}, De~Luise, De~March, {de Martino}, {de Souza}, {de Torres}, Debosscher, {del Pozo}, Delbo, Delgado, Delgado, Di~Matteo, Diakite, Diener, Distefano, Dolding, Drazinos, Dur{\'a}n, Edvardsson, Enke, Eriksson, Esquej, Eynard~Bontemps, Fabre, Fabrizio, Faigler, Falc{\~a}o, Farr{\`a}s~Casas, Federici, Fedorets, Fernique, Figueras, Filippi, Findeisen, Fonti, Fraile, Fraser, Fr{\'e}zouls, Gai, Galleti, Garabato, {Garc{\'i}a-Sedano}, Garofalo, Garralda, Gavel, Gavras, Gerssen, Geyer, Giacobbe, Gilmore, Girona, Giuffrida, Glass, Gomes, Granvik, Gueguen, Guerrier, Guiraud, {Guti{\'e}rrez-S{\'a}nchez}, Haigron, Hatzidimitriou, Hauser, Haywood, Heiter, Helmi, Heu, Hilger, Hobbs, Hofmann, Holland, Huckle, Hypki, Icardi, Jan{\ss}en, {Jevardat de Fombelle}, Jonker, Juh{\'a}sz, Julbe, Karampelas, Kewley, Klar, Kochoska, Kohley, Kolenberg, Kontizas, Kontizas, Koposov, Kordopatis,
  {Kostrzewa-Rutkowska}, Koubsky, Lambert, Lanza, Lasne, Lavigne, Le~Fustec, {Le Poncin-Lafitte}, Lebreton, Leccia, Leclerc, {Lecoeur-Taibi}, Lenhardt, Leroux, Liao, Licata, Lindstr{\o}m, Lister, Livanou, Lobel, L{\'o}pez, Managau, Mann, Mantelet, Marchal, Marchant, Marconi, Marinoni, Marschalk{\'o}, Marshall, Martino, Marton, Mary, Massari, Matijevi{\v c}, Mazeh, McMillan, Messina, Michalik, Millar, Molina, Molinaro, Moln{\'a}r, Montegriffo, Mor, Morbidelli, Morel, Morris, Mulone, Muraveva, Musella, Nelemans, Nicastro, Noval, O'Mullane, Ord{\'e}novic, {Ord{\'o}{\~n}ez-Blanco}, Osborne, Pagani, Pagano, Pailler, Palacin, Palaversa, Panahi, Pawlak, Piersimoni, Pineau, Plachy, Plum, Poggio, Poujoulet, Pr{\v s}a, Pulone, Racero, Ragaini, Rambaux, {Ramos-Lerate}, Regibo, Reyl{\'e}, Riclet, Ripepi, Riva, Rivard, Rixon, Roegiers, Roelens, {Romero-G{\'o}mez}, Rowell, Royer, {Ruiz-Dern}, Sadowski, Sagrist{\`a}~Sell{\'e}s, Sahlmann, Salgado, Salguero, Sanna, {Santana-Ros}, Sarasso, Savietto, Schultheis, Sciacca, Segol,
  Segovia, S{\'e}gransan, Shih, Siltala, Silva, Smart, Smith, Solano, Solitro, Sordo, Soria~Nieto, Souchay, Spagna, Spoto, Stampa, Steele, Steidelm{\"u}ller, Stephenson, Stoev, Suess, Surdej, Szabados, {Szegedi-Elek}, Tapiador, Taris, Tauran, Taylor, Teixeira, Terrett, Teyssandier, Thuillot, Titarenko, Torra~Clotet, Turon, Ulla, Utrilla, Uzzi, Vaillant, Valentini, Valette, {van Elteren}, Van~Hemelryck, {van Leeuwen}, Vaschetto, Vecchiato, Veljanoski, Viala, Vicente, Vogt, {von Essen}, Voss, Votruba, Voutsinas, Walmsley, Weiler, Wertz, Wevers, Wyrzykowski, Yoldas, {\v Z}erjal, Ziaeepour, Zorec, Zschocke, Zucker, Zurbach, \& Zwitter}]{gaiacollaboration2018}
{Gaia Collaboration}, Brown, A. G.~A., Vallenari, A., {et~al.} 2018, Astron. Astrophys., 616, A1, \dodoi{10.1051/0004-6361/201833051}

\bibitem[{Gallazzi {et~al.}(2014)Gallazzi, Bell, Zibetti, Brinchmann, \& Kelson}]{gallazzi2014}
Gallazzi, A., Bell, E.~F., Zibetti, S., Brinchmann, J., \& Kelson, D.~D. 2014, ApJ, 788, 72, \dodoi{10.1088/0004-637X/788/1/72}

\bibitem[{G{\'a}sp{\'a}r {et~al.}(2021)G{\'a}sp{\'a}r, Rieke, Guillard, Dicken, Gastaud, Alberts, Morrison, Ressler, Argyriou, \& Glasse}]{gaspar2021}
G{\'a}sp{\'a}r, A., Rieke, G.~H., Guillard, P., {et~al.} 2021, Publ. Astron. Soc. Pac., 133, 014504, \dodoi{10.1088/1538-3873/abcd04}

\bibitem[{Geha {et~al.}(2012)Geha, Blanton, Yan, \& Tinker}]{geha2012}
Geha, M., Blanton, M.~R., Yan, R., \& Tinker, J.~L. 2012, ApJ, 757, 85, \dodoi{10.1088/0004-637X/757/1/85}

\bibitem[{Gelli {et~al.}(2023)Gelli, Salvadori, Ferrara, \& Pallottini}]{gelli2023}
Gelli, V., Salvadori, S., Ferrara, A., \& Pallottini, A. 2023, Can Supernovae Quench Star Formation in High-\$z\$ Galaxies?, \dodoi{10.48550/arXiv.2310.03065}

\bibitem[{{Gim{\'e}nez-Arteaga} {et~al.}(2023){Gim{\'e}nez-Arteaga}, Oesch, Brammer, Valentino, Mason, Weibel, Barrufet, Fujimoto, Heintz, Nelson, Strait, Suess, \& Gibson}]{gimenez-arteaga2023}
{Gim{\'e}nez-Arteaga}, C., Oesch, P.~A., Brammer, G.~B., {et~al.} 2023, ApJ, 948, 126, \dodoi{10.3847/1538-4357/acc5ea}

\bibitem[{Girelli {et~al.}(2019)Girelli, Bolzonella, \& Cimatti}]{girelli2019}
Girelli, G., Bolzonella, M., \& Cimatti, A. 2019, Astron. Astrophys., 632, A80, \dodoi{10.1051/0004-6361/201834547}

\bibitem[{Glazebrook {et~al.}(2017)Glazebrook, Schreiber, Labb{\'e}, Nanayakkara, Kacprzak, Oesch, Papovich, Spitler, Straatman, Tran, \& Yuan}]{glazebrook2017}
Glazebrook, K., Schreiber, C., Labb{\'e}, I., {et~al.} 2017, Nature, 544, 71, \dodoi{10.1038/nature21680}

\bibitem[{Glazebrook {et~al.}(2023)Glazebrook, Nanayakkara, Jacobs, Leethochawalit, Calabr{\`o}, Bonchi, Castellano, Fontana, Mason, Merlin, Morishita, Paris, Trenti, Treu, Santini, Wang, Boyett, Bradac, Brammer, Jones, Marchesini, Nonino, \& Vulcani}]{glazebrook2023}
Glazebrook, K., Nanayakkara, T., Jacobs, C., {et~al.} 2023, ApJ, 947, L25, \dodoi{10.3847/2041-8213/acba8b}

\bibitem[{Gobat {et~al.}(2018)Gobat, Daddi, Magdis, Bournaud, Sargent, Martig, Jin, Finoguenov, B{\'e}thermin, Hwang, Renzini, Wilson, Aretxaga, Yun, Strazzullo, \& Valentino}]{gobat2018}
Gobat, R., Daddi, E., Magdis, G., {et~al.} 2018, Nat. Astron., 2, 239, \dodoi{10.1038/s41550-017-0352-5}

\bibitem[{Gould {et~al.}(2023)Gould, Brammer, Valentino, Whitaker, Weaver, Lagos, Rizzo, Franco, Hseih, Ilbert, Jin, Magdis, McCracken, Mobasher, Shuntov, Steinhardt, Strait, \& Toft}]{gould2023}
Gould, K. M.~L., Brammer, G., Valentino, F., {et~al.} 2023, {{COSMOS2020}}: {{Exploring}} the Dawn of Quenching for Massive Galaxies at 3 {$<$} z {$<$} 5 with a New Colour Selection Method,  {arXiv}, \dodoi{10.48550/arXiv.2302.10934}

\bibitem[{Hainline {et~al.}(2023)Hainline, Johnson, Robertson, Tacchella, Helton, Sun, Eisenstein, Simmonds, Topping, Whitler, Willmer, Rieke, Suess, Hviding, Cameron, Alberts, Baker, Bhatawdekar, Boyett, Bunker, Carniani, Charlot, Chen, Curti, {Curtis-Lake}, D'Eugenio, Egami, Endsley, Hausen, Ji, Looser, Lyu, Maiolino, Nelson, Puskas, Rawle, Sandles, Saxena, Smit, Stark, Williams, Willott, \& Witstok}]{hainline2023}
Hainline, K.~N., Johnson, B.~D., Robertson, B., {et~al.} 2023, The {{Cosmos}} in Its {{Infancy}}: {{JADES Galaxy Candidates}} at z {$>$} 8 in {{GOODS-S}} and {{GOODS-N}}, \dodoi{10.48550/arXiv.2306.02468}

\bibitem[{Hamadouche {et~al.}(2022)Hamadouche, Carnall, McLure, Dunlop, McLeod, Cullen, Begley, Bolzonella, Buitrago, Castellano, Cucciati, Fontana, Gargiulo, Moresco, Pozzetti, \& Zamorani}]{hamadouche2022}
Hamadouche, M.~L., Carnall, A.~C., McLure, R.~J., {et~al.} 2022, MNRAS, 512, 1262, \dodoi{10.1093/mnras/stac535}

\bibitem[{Hamadouche {et~al.}(2023)Hamadouche, Carnall, McLure, Dunlop, Begley, Cullen, McLeod, Donnan, \& Stanton}]{hamadouche2023}
---. 2023, MNRAS, 521, 5400, \dodoi{10.1093/mnras/stad773}

\bibitem[{Harikane {et~al.}(2023)Harikane, Zhang, Nakajima, Ouchi, Isobe, Ono, Hatano, Xu, \& Umeda}]{harikane2023}
Harikane, Y., Zhang, Y., Nakajima, K., {et~al.} 2023, ApJ, 959, 39, \dodoi{10.3847/1538-4357/ad029e}

\bibitem[{Hubble(1926)}]{hubble1926}
Hubble, E.~P. 1926, ApJ, 64, 321, \dodoi{10.1086/143018}

\bibitem[{Illingworth {et~al.}(2016)Illingworth, Magee, Bouwens, Oesch, Labbe, {van Dokkum}, Whitaker, Holden, Franx, \& Gonzalez}]{illingworth2016}
Illingworth, G., Magee, D., Bouwens, R., {et~al.} 2016, The {{Hubble Legacy Fields}} ({{HLF-GOODS-S}}) v1.5 {{Data Products}}: {{Combining}} 2442 {{Orbits}} of {{GOODS-S}}/{{CDF-S Region ACS}} and {{WFC3}}/{{IR Images}}, \dodoi{10.48550/arXiv.1606.00841}

\bibitem[{Ji {et~al.}(2018)Ji, Giavalisco, Williams, Faber, Ferguson, Guo, Liu, \& Lee}]{ji2018}
Ji, Z., Giavalisco, M., Williams, C.~C., {et~al.} 2018, ApJ, 862, 135, \dodoi{10.3847/1538-4357/aacc2c}

\bibitem[{Johnson {et~al.}(2021)Johnson, Leja, Conroy, \& Speagle}]{johnson2021}
Johnson, B.~D., Leja, J., Conroy, C., \& Speagle, J.~S. 2021, ApJS, 254, 22, \dodoi{10.3847/1538-4365/abef67}

\bibitem[{Kauffmann {et~al.}(2003)Kauffmann, Heckman, White, Charlot, Tremonti, Brinchmann, Bruzual, Peng, Seibert, Bernardi, Blanton, Brinkmann, Castander, Cs{\'a}bai, Fukugita, Ivezic, Munn, Nichol, Padmanabhan, Thakar, Weinberg, \& York}]{kauffmann2003}
Kauffmann, G., Heckman, T.~M., White, S. D.~M., {et~al.} 2003, MNRAS, 341, 33, \dodoi{10.1046/j.1365-8711.2003.06291.x}

\bibitem[{Kaushal {et~al.}(2023)Kaushal, Nersesian, Bezanson, {van der Wel}, Leja, Carnall, Zibetti, Khullar, Franx, Muzzin, De~Graaff, Pacifici, Whitaker, Bell, \& Martorano}]{kaushal2023}
Kaushal, Y., Nersesian, A., Bezanson, R., {et~al.} 2023, A Census of Star Formation Histories of Massive Galaxies at 0.6 {$<$} z {$<$} 1 from Spectro-Photometric Modeling Using {{Bagpipes}} and {{Prospector}}, \dodoi{10.48550/arXiv.2307.03725}

\bibitem[{Kaviraj {et~al.}(2019)Kaviraj, Martin, \& Silk}]{kaviraj2019}
Kaviraj, S., Martin, G., \& Silk, J. 2019, MNRAS, 489, L12, \dodoi{10.1093/mnrasl/slz102}

\bibitem[{Kawinwanichakij {et~al.}(2016)Kawinwanichakij, Quadri, Papovich, Kacprzak, Labb{\'e}, Spitler, Straatman, Tran, Allen, Behroozi, Cowley, Dekel, Glazebrook, Hartley, Kelson, Koo, Lee, Lu, Nanayakkara, Persson, Primack, Tilvi, Tomczak, \& {van Dokkum}}]{kawinwanichakij2016}
Kawinwanichakij, L., Quadri, R.~F., Papovich, C., {et~al.} 2016, ApJ, 817, 9, \dodoi{10.3847/0004-637X/817/1/9}

\bibitem[{Kokorev {et~al.}(2023)Kokorev, Jin, Magdis, Caputi, Valentino, Dayal, Trebitsch, Brammer, Fujimoto, Bauer, Iani, Kohno, Bl{\'a}nquez~Ses{\'e}, {G{\'o}mez-Guijarro}, Rinaldi, \& {Navarro-Carrera}}]{kokorev2023a}
Kokorev, V., Jin, S., Magdis, G.~E., {et~al.} 2023, ApJ, 945, L25, \dodoi{10.3847/2041-8213/acbd9d}

\bibitem[{Koudmani {et~al.}(2021)Koudmani, Henden, \& Sijacki}]{koudmani2021}
Koudmani, S., Henden, N.~A., \& Sijacki, D. 2021, MNRAS, 503, 3568, \dodoi{10.1093/mnras/stab677}

\bibitem[{Kriek {et~al.}(2016)Kriek, Conroy, {van Dokkum}, Shapley, Choi, Reddy, Siana, {van de Voort}, Coil, \& Mobasher}]{kriek2016}
Kriek, M., Conroy, C., {van Dokkum}, P.~G., {et~al.} 2016, Nature, 540, 248, \dodoi{10.1038/nature20570}

\bibitem[{Kriek {et~al.}(2023)Kriek, Beverage, Price, Suess, Barro, Bezanson, Conroy, Cutler, Franx, Lin, Lorenz, Ma, Momcheva, Mowla, Pasha, {van Dokkum}, \& Whitaker}]{kriek2023}
Kriek, M., Beverage, A.~G., Price, S.~H., {et~al.} 2023, The {{Heavy Metal Survey}}: {{Star Formation Constraints}} and {{Dynamical Masses}} of 21 {{Massive Quiescent Galaxies}} at Z{\textasciitilde}1.4-2.2, \dodoi{10.48550/arXiv.2311.16232}

\bibitem[{Kroupa(2001)}]{kroupa2001}
Kroupa, P. 2001, MNRAS, 322, 231, \dodoi{10.1046/j.1365-8711.2001.04022.x}

\bibitem[{Kubo {et~al.}(2023)Kubo, Nagao, Uchiyama, Yamashita, Toba, Kajisawa, \& Yamamoto}]{kubo2023}
Kubo, M., Nagao, T., Uchiyama, H., {et~al.} 2023, New Technique to Select Recent Fast-Quenching Galaxies at \$z{\textbackslash}sim2\$ Using the Optical Colors, \dodoi{10.48550/arXiv.2310.09703}

\bibitem[{Labb{\'e} {et~al.}(2005)Labb{\'e}, Huang, Franx, Rudnick, Barmby, Daddi, {van Dokkum}, Fazio, F{\"o}rster~Schreiber, Moorwood, Rix, R{\"o}ttgering, Trujillo, \& {van der Werf}}]{labbe2005}
Labb{\'e}, I., Huang, J., Franx, M., {et~al.} 2005, ApJ, 624, L81, \dodoi{10.1086/430700}

\bibitem[{Labb{\'e} {et~al.}(2023)Labb{\'e}, {van Dokkum}, Nelson, Bezanson, Suess, Leja, Brammer, Whitaker, Mathews, Stefanon, \& Wang}]{labbe2023a}
Labb{\'e}, I., {van Dokkum}, P., Nelson, E., {et~al.} 2023, Nature, 616, 266, \dodoi{10.1038/s41586-023-05786-2}

\bibitem[{Larson {et~al.}(1980)Larson, Tinsley, \& Caldwell}]{larson1980}
Larson, R.~B., Tinsley, B.~M., \& Caldwell, C.~N. 1980, ApJ, 237, 692, \dodoi{10.1086/157917}

\bibitem[{Lee {et~al.}(2023)Lee, Steidel, Brammer, {F{\"o}rster-Schreiber}, Renzini, Liu, {Herrera-Camus}, Naab, Price, {\"U}bler, Arriagada, \& Magdis}]{lee2023}
Lee, M.~M., Steidel, C.~C., Brammer, G., {et~al.} 2023, arXiv e-prints, arXiv:2311.00023, \dodoi{10.48550/arXiv.2311.00023}

\bibitem[{Leja {et~al.}(2019{\natexlab{a}})Leja, Carnall, Johnson, Conroy, \& Speagle}]{leja2019b}
Leja, J., Carnall, A.~C., Johnson, B.~D., Conroy, C., \& Speagle, J.~S. 2019{\natexlab{a}}, ApJ, 876, 3, \dodoi{10.3847/1538-4357/ab133c}

\bibitem[{Leja {et~al.}(2019{\natexlab{b}})Leja, Tacchella, \& Conroy}]{leja2019}
Leja, J., Tacchella, S., \& Conroy, C. 2019{\natexlab{b}}, ApJ, 880, L9, \dodoi{10.3847/2041-8213/ab2f8c}

\bibitem[{Leung {et~al.}(2023)Leung, Wild, Papathomas, Carnall, Zheng, Boardman, Wang, \& Johansson}]{leung2023}
Leung, H.-H., Wild, V., Papathomas, M., {et~al.} 2023, Chemical Evolution of Local Post-Starburst Galaxies: {{Implications}} for the Mass-Metallicity Relation, \dodoi{10.48550/arXiv.2309.16626}

\bibitem[{Lilly {et~al.}(2013)Lilly, Carollo, Pipino, Renzini, \& Peng}]{lilly2013}
Lilly, S.~J., Carollo, C.~M., Pipino, A., Renzini, A., \& Peng, Y. 2013, ApJ, 772, 119, \dodoi{10.1088/0004-637X/772/2/119}

\bibitem[{Long {et~al.}(2023)Long, {Antwi-Danso}, Lambrides, Lovell, {de la Vega}, Valentino, Zavala, Casey, Wilkins, Yung, Haro, Bagley, Bisigello, Chworowsky, Cooper, Cooper, Cooray, Croton, Dickinson, Finkelstein, Franco, Gould, Hirschmann, Hutchison, Kartaltepe, Kocevski, Koekemoer, Lucas, McKinney, Papovich, {Perez-Gonzalez}, Pirzkal, \& Santini}]{long2023}
Long, A.~S., {Antwi-Danso}, J., Lambrides, E.~L., {et~al.} 2023, Efficient {{NIRCam Selection}} of {{Quiescent Galaxies}} at 3 {$<$} z {$<$} 6 in {{CEERS}},  {arXiv}, \dodoi{10.48550/arXiv.2305.04662}

\bibitem[{Looser {et~al.}(2023)Looser, D'Eugenio, Maiolino, Witstok, Sandles, {Curtis-Lake}, Chevallard, Tacchella, Johnson, Baker, Suess, Carniani, Ferruit, Arribas, Bonaventura, Bunker, Cameron, Charlot, Curti, {de Graaff}, Maseda, Rawle, Rix, Rodriguez Del~Pino, Smit, {\"U}bler, Willott, Alberts, Egami, Eisenstein, Endsley, Hausen, Rieke, Robertson, Shivaei, Williams, Boyett, Chen, Ji, Jones, Kumari, Nelson, Perna, Saxena, \& Scholtz}]{looser2023}
Looser, T.~J., D'Eugenio, F., Maiolino, R., {et~al.} 2023, Discovery of a Quiescent Galaxy at Z=7.3, \dodoi{10.48550/arXiv.2302.14155}

\bibitem[{Lovell {et~al.}(2021)Lovell, Vijayan, Thomas, Wilkins, Barnes, Irodotou, \& Roper}]{lovell2021}
Lovell, C.~C., Vijayan, A.~P., Thomas, P.~A., {et~al.} 2021, MNRAS, 500, 2127, \dodoi{10.1093/mnras/staa3360}

\bibitem[{Lovell {et~al.}(2023)Lovell, Roper, Vijayan, Seeyave, Irodotou, Wilkins, Conselice, Fortuni, Kuusisto, Merlin, Santini, \& Thomas}]{lovell2023}
Lovell, C.~C., Roper, W., Vijayan, A.~P., {et~al.} 2023, MNRAS, 525, 5520, \dodoi{10.1093/mnras/stad2550}

\bibitem[{Lyu {et~al.}(2022)Lyu, Alberts, Rieke, \& Rujopakarn}]{lyu2022a}
Lyu, J., Alberts, S., Rieke, G.~H., \& Rujopakarn, W. 2022, {{AGN Selection}} and {{Demographics}} in {{GOODS-S}}/{{HUDF}} from {{X-ray}} to {{Radio}}

\bibitem[{Lyu {et~al.}(2023)Lyu, Alberts, Rieke, Shivaei, {Perez-Gonzalez}, Sun, Hainline, Baum, Bonaventura, Bunker, Egami, Eisenstein, Florian, Ji, Johnson, Morrison, Rieke, Robertson, Rujopakarn, Tacchella, Scholtz, \& Willmer}]{lyu2023}
Lyu, J., Alberts, S., Rieke, G.~H., {et~al.} 2023, {{AGN Selection}} and {{Demographics}}: {{A New Age}} with {{JWST}}/{{MIRI}}, \dodoi{10.48550/arXiv.2310.12330}

\bibitem[{Magdis {et~al.}(2021)Magdis, Gobat, Valentino, Daddi, Zanella, Kokorev, Toft, Jin, \& Whitaker}]{magdis2021}
Magdis, G.~E., Gobat, R., Valentino, F., {et~al.} 2021, Astron. Astrophys., 647, A33, \dodoi{10.1051/0004-6361/202039280}

\bibitem[{Maiolino \& Mannucci(2019)}]{maiolino2019}
Maiolino, R., \& Mannucci, F. 2019, A\&AR, 27, 3, \dodoi{10.1007/s00159-018-0112-2}

\bibitem[{Maiolino {et~al.}(2023)Maiolino, Scholtz, {Curtis-Lake}, Carniani, Baker, {de Graaff}, Tacchella, {\"U}bler, D'Eugenio, Witstok, Curti, Arribas, Bunker, Charlot, Chevallard, Eisenstein, Egami, Ji, Jones, Lyu, Rawle, Robertson, Rujopakarn, Perna, Sun, Venturi, Williams, \& Willott}]{maiolino2023}
Maiolino, R., Scholtz, J., {Curtis-Lake}, E., {et~al.} 2023, {{JADES}}. {{The}} Diverse Population of Infant {{Black Holes}} at 4, \dodoi{10.48550/arXiv.2308.01230}

\bibitem[{Man \& Belli(2018)}]{man2018}
Man, A., \& Belli, S. 2018, Nat. Astron., 2, 695, \dodoi{10.1038/s41550-018-0558-1}

\bibitem[{Marchesini {et~al.}(2010)Marchesini, Whitaker, Brammer, {van Dokkum}, Labb{\'e}, Muzzin, Quadri, Kriek, Lee, Rudnick, Franx, Illingworth, \& Wake}]{marchesini2010}
Marchesini, D., Whitaker, K.~E., Brammer, G., {et~al.} 2010, ApJ, 725, 1277, \dodoi{10.1088/0004-637X/725/1/1277}

\bibitem[{Marchesini {et~al.}(2023)Marchesini, Brammer, Morishita, Bergamini, Wang, Bradac, {Roberts-Borsani}, Strait, Treu, Fontana, Jones, Santini, Vulcani, Acebron, Calabr{\`o}, Castellano, Glazebrook, Grillo, Mercurio, Nanayakkara, Rosati, Tubthong, \& Vanzella}]{marchesini2023}
Marchesini, D., Brammer, G., Morishita, T., {et~al.} 2023, ApJ, 942, L25, \dodoi{10.3847/2041-8213/acaaac}

\bibitem[{Marsan {et~al.}(2022)Marsan, Muzzin, Marchesini, Stefanon, Martis, Annunziatella, Chan, Cooper, Forrest, Gomez, McConachie, \& Wilson}]{marsan2022}
Marsan, Z.~C., Muzzin, A., Marchesini, D., {et~al.} 2022, ApJ, 924, 25, \dodoi{10.3847/1538-4357/ac312a}

\bibitem[{McQuinn {et~al.}(2019)McQuinn, {van Zee}, \& Skillman}]{mcquinn2019}
McQuinn, {\relax Kristen}. B.~W., {van Zee}, L., \& Skillman, E.~D. 2019, ApJ, 886, 74, \dodoi{10.3847/1538-4357/ab4c37}

\bibitem[{Merlin {et~al.}(2018)Merlin, Fontana, Castellano, Santini, Torelli, Boutsia, Wang, Grazian, Pentericci, Schreiber, Ciesla, McLure, Derriere, Dunlop, \& Elbaz}]{merlin2018}
Merlin, E., Fontana, A., Castellano, M., {et~al.} 2018, MNRAS, 473, 2098, \dodoi{10.1093/mnras/stx2385}

\bibitem[{Merlin {et~al.}(2019)Merlin, Fortuni, Torelli, Santini, Castellano, Fontana, Grazian, Pentericci, Pilo, \& Schmidt}]{merlin2019}
Merlin, E., Fortuni, F., Torelli, M., {et~al.} 2019, MNRAS, 490, 3309, \dodoi{10.1093/mnras/stz2615}

\bibitem[{Morishita {et~al.}(2022)Morishita, {Abdurro'uf}, Hirashita, Newman, Stiavelli, \& Chiaberge}]{morishita2022}
Morishita, T., {Abdurro'uf}, Hirashita, H., {et~al.} 2022, ApJ, 938, 144, \dodoi{10.3847/1538-4357/ac9055}

\bibitem[{Muzzin {et~al.}(2013)Muzzin, Marchesini, Stefanon, Franx, McCracken, {Milvang-Jensen}, Dunlop, Fynbo, Brammer, Labb{\'e}, \& {van Dokkum}}]{muzzin2013a}
Muzzin, A., Marchesini, D., Stefanon, M., {et~al.} 2013, ApJ, 777, 18, \dodoi{10.1088/0004-637X/777/1/18}

\bibitem[{Nanayakkara {et~al.}(2022)Nanayakkara, Glazebrook, Jacobs, Kawinwanichakij, Schreiber, Brammer, Esdaile, Kacprzak, Labbe, Lagos, Marchesini, Marsan, Oesch, Papovich, Remus, \& Tran}]{nanayakkara2022}
Nanayakkara, T., Glazebrook, K., Jacobs, C., {et~al.} 2022, A Population of Faint, Old, and Massive Quiescent Galaxies at 3 {$<$} z {$<$} 4 Revealed by {{JWST NIRSpec Spectroscopy}}, \dodoi{10.48550/arXiv.2212.11638}

\bibitem[{Nersesian {et~al.}(2023)Nersesian, {van der Wel}, Gallazzi, Leja, Bezanson, Bell, D'Eugenio, {de Graaff}, Kaushal, Martorano, Maseda, \& Zibetti}]{nersesian2023}
Nersesian, A., {van der Wel}, A., Gallazzi, A., {et~al.} 2023, Less Is Less: Photometry Alone Cannot Predict the Observed Spectral Indices of \$z{\textbackslash}sim1\$ Galaxies from the {{LEGA-C}} Spectroscopic Survey, \dodoi{10.48550/arXiv.2310.18000}

\bibitem[{Newman {et~al.}(2018)Newman, Belli, Ellis, \& Patel}]{newman2018}
Newman, A.~B., Belli, S., Ellis, R.~S., \& Patel, S.~G. 2018, ApJ, 862, 125, \dodoi{10.3847/1538-4357/aacd4d}

\bibitem[{Noll {et~al.}(2009)Noll, Burgarella, Giovannoli, Buat, Marcillac, \& {Mu{\~n}oz-Mateos}}]{noll2009}
Noll, S., Burgarella, D., Giovannoli, E., {et~al.} 2009, Astron. Astrophys., 507, 1793, \dodoi{10.1051/0004-6361/200912497}

\bibitem[{Oesch {et~al.}(2023)Oesch, Brammer, Naidu, Bouwens, Chisholm, Illingworth, Matthee, Nelson, Qin, Reddy, Shapley, Shivaei, {van Dokkum}, Weibel, Whitaker, Wuyts, {Covelo-Paz}, Endsley, Fudamoto, Giovinazzo, {Herard-Demanche}, Kerutt, Kramarenko, Labbe, Leonova, Lin, Magee, Marchesini, Maseda, Mason, Matharu, Meyer, Neufeld, Lyon, Schaerer, Sharma, Shuntov, Smit, Stefanon, Wyithe, \& Xiao}]{oesch2023}
Oesch, P.~A., Brammer, G., Naidu, R.~P., {et~al.} 2023, The {{JWST FRESCO Survey}}: {{Legacy NIRCam}}/{{Grism Spectroscopy}} and {{Imaging}} in the Two {{GOODS Fields}},  {arXiv}.
\newblock \doeprint{2304.02026}

\bibitem[{Oke \& Gunn(1983)}]{oke1983}
Oke, J.~B., \& Gunn, J.~E. 1983, ApJ, 266, 713, \dodoi{10.1086/160817}

\bibitem[{Ormerod {et~al.}(2023)Ormerod, Conselice, Adams, Harvey, Austin, Trussler, Ferreira, Caruana, Lucatelli, Li, \& Roper}]{ormerod2023a}
Ormerod, K., Conselice, C.~J., Adams, N.~J., {et~al.} 2023, MNRAS, \dodoi{10.1093/mnras/stad3597}

\bibitem[{Pacifici {et~al.}(2016)Pacifici, Kassin, Weiner, Holden, Gardner, Faber, Ferguson, Koo, Primack, Bell, Dekel, Gawiser, Giavalisco, Rafelski, Simons, Barro, Croton, Dav{\'e}, Fontana, Grogin, Koekemoer, Lee, Salmon, Somerville, \& Behroozi}]{pacifici2016}
Pacifici, C., Kassin, S.~A., Weiner, B.~J., {et~al.} 2016, ApJ, 832, 79, \dodoi{10.3847/0004-637X/832/1/79}

\bibitem[{Papovich {et~al.}(2023)Papovich, Cole, Yang, Finkelstein, Barro, Buat, Burgarella, {P{\'e}rez-Gonz{\'a}lez}, Santini, Seill{\'e}, Shen, Arrabal~Haro, Bagley, Bell, Bisigello, Calabr{\`o}, Casey, Castellano, Chworowsky, Cleri, Costantin, Cooper, Dickinson, Ferguson, Fontana, Giavalisco, Grazian, Grogin, Hathi, Holwerda, Hutchison, Kartaltepe, Kewley, Kirkpatrick, Kocevski, Koekemoer, Larson, Long, Lucas, Pentericci, Pirzkal, Ravindranath, Somerville, Trump, Urbano~Stawinski, Weiner, Wilkins, Yung, \& Zavala}]{papovich2023}
Papovich, C., Cole, J.~W., Yang, G., {et~al.} 2023, ApJ, 949, L18, \dodoi{10.3847/2041-8213/acc948}

\bibitem[{Park {et~al.}(2023)Park, Belli, Conroy, Tacchella, Leja, Cutler, Johnson, Nelson, \& Emami}]{park2023}
Park, M., Belli, S., Conroy, C., {et~al.} 2023, ApJ, 953, 119, \dodoi{10.3847/1538-4357/acd54a}

\bibitem[{Peng {et~al.}(2015)Peng, Maiolino, \& Cochrane}]{peng2015}
Peng, Y., Maiolino, R., \& Cochrane, R. 2015, Nature, 521, 192, \dodoi{10.1038/nature14439}

\bibitem[{Peng {et~al.}(2012)Peng, Lilly, Renzini, \& Carollo}]{peng2012}
Peng, Y.-j., Lilly, S.~J., Renzini, A., \& Carollo, M. 2012, ApJ, 757, 4, \dodoi{10.1088/0004-637X/757/1/4}

\bibitem[{Peng {et~al.}(2010)Peng, Lilly, Kova{\v c}, Bolzonella, Pozzetti, Renzini, Zamorani, Ilbert, Knobel, Iovino, Maier, Cucciati, Tasca, Carollo, Silverman, Kampczyk, {de Ravel}, Sanders, Scoville, Contini, Mainieri, Scodeggio, Kneib, Le~F{\`e}vre, Bardelli, Bongiorno, Caputi, Coppa, {de la Torre}, Franzetti, Garilli, Lamareille, Le~Borgne, Le~Brun, Mignoli, Montero, Pello, Ricciardelli, Tanaka, Tresse, Vergani, Welikala, Zucca, Oesch, Abbas, Barnes, Bordoloi, Bottini, Cappi, Cassata, Cimatti, Fumana, Hasinger, Koekemoer, Leauthaud, Maccagni, Marinoni, McCracken, Memeo, Meneux, Nair, Porciani, Presotto, \& Scaramella}]{peng2010}
Peng, Y.-j., Lilly, S.~J., Kova{\v c}, K., {et~al.} 2010, ApJ, 721, 193, \dodoi{10.1088/0004-637X/721/1/193}

\bibitem[{{P{\'e}rez-Gonz{\'a}lez} {et~al.}(2023){P{\'e}rez-Gonz{\'a}lez}, Barro, Annunziatella, Costantin, {Garc{\'i}a-Argum{\'a}nez}, McGrath, M{\'e}rida, Zavala, Haro, Bagley, Backhaus, Behroozi, Bell, Bisigello, Buat, Calabr{\`o}, Casey, Cleri, Coogan, Cooper, Cooray, Dekel, Dickinson, Elbaz, Ferguson, Finkelstein, Fontana, Franco, Gardner, Giavalisco, {G{\'o}mez-Guijarro}, Grazian, Grogin, Guo, {Huertas-Company}, Jogee, Kartaltepe, Kewley, Kirkpatrick, Kocevski, Koekemoer, Long, Lotz, Lucas, Papovich, Pirzkal, Ravindranath, Somerville, Tacchella, Trump, Wang, Wilkins, Wuyts, Yang, \& Yung}]{perez-gonzalez2023}
{P{\'e}rez-Gonz{\'a}lez}, P.~G., Barro, G., Annunziatella, M., {et~al.} 2023, ApJL, 946, L16, \dodoi{10.3847/2041-8213/acb3a5}

\bibitem[{Piotrowska {et~al.}(2022)Piotrowska, Bluck, Maiolino, \& Peng}]{piotrowska2022}
Piotrowska, J.~M., Bluck, A. F.~L., Maiolino, R., \& Peng, Y. 2022, MNRAS, 512, 1052, \dodoi{10.1093/mnras/stab3673}

\bibitem[{Popesso {et~al.}(2023)Popesso, Concas, Cresci, Belli, Rodighiero, Inami, Dickinson, Ilbert, Pannella, \& Elbaz}]{popesso2023}
Popesso, P., Concas, A., Cresci, G., {et~al.} 2023, MNRAS, 519, 1526, \dodoi{10.1093/mnras/stac3214}

\bibitem[{{Rieke} {et~al.}(2023){Rieke}, {Robertson}, {Tacchella}, {Hainline}, {Johnson}, {Hausen}, {Ji}, {Willmer}, {Eisenstein}, {Pusk{\'a}s}, {Alberts}, {Arribas}, {Baker}, {Baum}, {Bhatawdekar}, {Bonaventura}, {Boyett}, {Bunker}, {Cameron}, {Carniani}, {Charlot}, {Chevallard}, {Chen}, {Curti}, {Curtis-Lake}, {Danhaive}, {DeCoursey}, {Dressler}, {Egami}, {Endsley}, {Helton}, {Hviding}, {Kumari}, {Looser}, {Lyu}, {Maiolino}, {Maseda}, {Nelson}, {Rieke}, {Rix}, {Sandles}, {Saxena}, {Sharpe}, {Shivaei}, {Skarbinski}, {Smit}, {Stark}, {Stone}, {Suess}, {Sun}, {Topping}, {{\"U}bler}, {Villanueva}, {Wallace}, {Williams}, {Willott}, {Whitler}, {Witstok}, \& {Woodrum}}]{rieke2023}
{Rieke}, M.~J., {Robertson}, B., {Tacchella}, S., {et~al.} 2023, \apjs, 269, 16, \dodoi{10.3847/1538-4365/acf44d}

\bibitem[{Rowlands {et~al.}(2015)Rowlands, Wild, Nesvadba, Sibthorpe, Mortier, Lehnert, \& {da Cunha}}]{rowlands2015}
Rowlands, K., Wild, V., Nesvadba, N., {et~al.} 2015, MNRAS, 448, 258, \dodoi{10.1093/mnras/stu2714}

\bibitem[{Rowlands {et~al.}(2018)Rowlands, Wild, Bourne, Bremer, Brough, Driver, Hopkins, Owers, Phillipps, Pimbblet, Sansom, Wang, Alpaslan, {Bland-Hawthorn}, Colless, Holwerda, \& Taylor}]{rowlands2018}
Rowlands, K., Wild, V., Bourne, N., {et~al.} 2018, MNRAS, 473, 1168, \dodoi{10.1093/mnras/stx1903}

\bibitem[{Salim {et~al.}(2018)Salim, Boquien, \& Lee}]{salim2018}
Salim, S., Boquien, M., \& Lee, J.~C. 2018, ApJ, 859, 11, \dodoi{10.3847/1538-4357/aabf3c}

\bibitem[{Sanders {et~al.}(2021)Sanders, Shapley, Jones, Reddy, Kriek, Siana, Coil, Mobasher, Shivaei, Dav{\'e}, Azadi, Price, Leung, Freeman, Fetherolf, {de Groot}, Zick, \& Barro}]{sanders2021}
Sanders, R.~L., Shapley, A.~E., Jones, T., {et~al.} 2021, ApJ, 914, 19, \dodoi{10.3847/1538-4357/abf4c1}

\bibitem[{Sandles {et~al.}(2023)Sandles, D'Eugenio, Helton, Maiolino, Hainline, Baker, Williams, Alberts, Bunker, Carniani, Charlot, Chevallard, Curti, {Curtis-Lake}, Eisenstein, Ji, Johnson, Looser, Rawle, Robertson, Rodr{\'i}guez Del~Pino, Tacchella, {\"U}bler, Willmer, \& Willott}]{sandles2023}
Sandles, L., D'Eugenio, F., Helton, J.~M., {et~al.} 2023, {{JADES}}: Deep Spectroscopy of a Low-Mass Galaxy at Redshift 2.3 Quenched by Environment, \dodoi{10.48550/arXiv.2307.08633}

\bibitem[{Schaye {et~al.}(2015)Schaye, Crain, Bower, Furlong, Schaller, Theuns, Dalla~Vecchia, Frenk, McCarthy, Helly, Jenkins, {Rosas-Guevara}, White, Baes, Booth, Camps, Navarro, Qu, Rahmati, Sawala, Thomas, \& Trayford}]{schaye2015}
Schaye, J., Crain, R.~A., Bower, R.~G., {et~al.} 2015, MNRAS, 446, 521, \dodoi{10.1093/mnras/stu2058}

\bibitem[{Schreiber {et~al.}(2018{\natexlab{a}})Schreiber, Glazebrook, Nanayakkara, Kacprzak, Labb{\'e}, Oesch, Yuan, Tran, Papovich, Spitler, \& Straatman}]{schreiber2018a}
Schreiber, C., Glazebrook, K., Nanayakkara, T., {et~al.} 2018{\natexlab{a}}, Astron. Astrophys., 618, A85, \dodoi{10.1051/0004-6361/201833070}

\bibitem[{Schreiber {et~al.}(2018{\natexlab{b}})Schreiber, Labb{\'e}, Glazebrook, Bekiaris, Papovich, Costa, Elbaz, Kacprzak, Nanayakkara, Oesch, Pannella, Spitler, Straatman, Tran, \& Wang}]{schreiber2018b}
Schreiber, C., Labb{\'e}, I., Glazebrook, K., {et~al.} 2018{\natexlab{b}}, Astron. Astrophys., 611, A22, \dodoi{10.1051/0004-6361/201731917}

\bibitem[{Setton {et~al.}(2023)Setton, Dey, Khullar, Bezanson, Newman, Aguilar, Ahlen, Andrews, Brooks, {de la Macorra}, Dey, Eftekharzadeh, {Font-Ribera}, A~Gontcho, Kremin, Juneau, Landriau, Meisner, Miquel, Moustakas, Pearl, Prada, Tarl{\'e}, Siudek, Weaver, Zhou, \& Zou}]{setton2023}
Setton, D.~J., Dey, B., Khullar, G., {et~al.} 2023, ApJ, 947, L31, \dodoi{10.3847/2041-8213/acc9b5}

\bibitem[{Shah {et~al.}(2023)Shah, Lemaux, Forrest, Cucciati, Hung, Staab, Hathi, Lubin, Gal, Shen, Zamorani, Giddings, Bardelli, Pasqua~Cassara, Cassata, Contini, {Golden-Marx}, Guaita, Gururajan, Koekemoer, McLeod, Tasca, Tresse, Vergani, \& Zucca}]{shah2023}
Shah, E.~A., Lemaux, B., Forrest, B., {et~al.} 2023, Identification and {{Characterization}} of {{Six Spectroscopically Confirmed Massive Protostructures}} at \$2.5, \dodoi{10.48550/arXiv.2312.04634}

\bibitem[{Shahidi {et~al.}(2020)Shahidi, Mobasher, Nayyeri, Hemmati, Wiklind, Chartab, Dickinson, Finkelstein, Pacifici, Papovich, Ferguson, Fontana, Giavalisco, Koekemoer, Newman, Sattari, \& Somerville}]{shahidi2020}
Shahidi, A., Mobasher, B., Nayyeri, H., {et~al.} 2020, ApJ, 897, 44, \dodoi{10.3847/1538-4357/ab96c5}

\bibitem[{Shibuya {et~al.}(2015)Shibuya, Ouchi, \& Harikane}]{shibuya2015}
Shibuya, T., Ouchi, M., \& Harikane, Y. 2015, ApJ Suppl. Ser., 219, 15, \dodoi{10.1088/0067-0049/219/2/15}

\bibitem[{Silk(2017)}]{silk2017}
Silk, J. 2017, ApJ, 839, L13, \dodoi{10.3847/2041-8213/aa67da}

\bibitem[{Somerville \& Dav{\'e}(2015)}]{somerville2015a}
Somerville, R.~S., \& Dav{\'e}, R. 2015, Annu. Rev. Astron. Astrophys., 53, 51, \dodoi{10.1146/annurev-astro-082812-140951}

\bibitem[{Spilker {et~al.}(2022)Spilker, Suess, Setton, Bezanson, Feldmann, Greene, Kriek, Lower, Narayanan, \& Verrico}]{spilker2022}
Spilker, J.~S., Suess, K.~A., Setton, D.~J., {et~al.} 2022, ApJ, 936, L11, \dodoi{10.3847/2041-8213/ac75ea}

\bibitem[{Spitler {et~al.}(2014)Spitler, Straatman, Labb{\'e}, Glazebrook, Tran, Kacprzak, Quadri, Papovich, Persson, {van Dokkum}, Allen, Kawinwanichakij, Kelson, McCarthy, Mehrtens, Monson, Nanayakkara, Rees, Tilvi, \& Tomczak}]{spitler2014}
Spitler, L.~R., Straatman, C. M.~S., Labb{\'e}, I., {et~al.} 2014, ApJ, 787, L36, \dodoi{10.1088/2041-8205/787/2/L36}

\bibitem[{Stanway {et~al.}(2005)Stanway, McMahon, \& Bunker}]{stanway2005}
Stanway, E.~R., McMahon, R.~G., \& Bunker, A.~J. 2005, MNRAS, 359, 1184, \dodoi{10.1111/j.1365-2966.2005.08977.x}

\bibitem[{Straatman {et~al.}(2014)Straatman, Labb{\'e}, Spitler, Allen, Altieri, Brammer, Dickinson, {van Dokkum}, Inami, Glazebrook, Kacprzak, Kawinwanichakij, Kelson, McCarthy, Mehrtens, Monson, Murphy, Papovich, Persson, Quadri, Rees, Tomczak, Tran, \& Tilvi}]{straatman2014}
Straatman, C. M.~S., Labb{\'e}, I., Spitler, L.~R., {et~al.} 2014, ApJ, 783, L14, \dodoi{10.1088/2041-8205/783/1/L14}

\bibitem[{Straatman {et~al.}(2015)Straatman, Labb{\'e}, Spitler, Glazebrook, Tomczak, Allen, Brammer, Cowley, {van Dokkum}, Kacprzak, Kawinwanichakij, Mehrtens, Nanayakkara, Papovich, Persson, Quadri, Rees, Tilvi, Tran, \& Whitaker}]{straatman2015}
---. 2015, ApJ, 808, L29, \dodoi{10.1088/2041-8205/808/1/L29}

\bibitem[{Straatman {et~al.}(2016)Straatman, Spitler, Quadri, Labb{\'e}, Glazebrook, Persson, Papovich, Tran, Brammer, Cowley, Tomczak, Nanayakkara, Alcorn, Allen, Broussard, {van Dokkum}, Forrest, {van Houdt}, Kacprzak, Kawinwanichakij, Kelson, Lee, McCarthy, Mehrtens, Monson, Murphy, Rees, Tilvi, \& Whitaker}]{straatman2016}
Straatman, C. M.~S., Spitler, L.~R., Quadri, R.~F., {et~al.} 2016, ApJ, 830, 51, \dodoi{10.3847/0004-637X/830/1/51}

\bibitem[{Strait {et~al.}(2023)Strait, Brammer, Muzzin, Desprez, Asada, Abraham, Brada{\v c}, Iyer, Martis, Mowla, Noirot, Sarrouh, Sawicki, Willott, Gould, Grindlay, Matharu, \& Rihtar{\v s}i{\v c}}]{strait2023}
Strait, V., Brammer, G., Muzzin, A., {et~al.} 2023, ApJ, 949, L23, \dodoi{10.3847/2041-8213/acd457}

\bibitem[{Suess {et~al.}(2017)Suess, Bezanson, Spilker, Kriek, Greene, Feldmann, Hunt, \& Narayanan}]{suess2017}
Suess, K.~A., Bezanson, R., Spilker, J.~S., {et~al.} 2017, ApJ, 846, L14, \dodoi{10.3847/2041-8213/aa85dc}

\bibitem[{Suess {et~al.}(2022{\natexlab{a}})Suess, Kriek, Bezanson, Greene, Setton, Spilker, Feldmann, Goulding, Johnson, Leja, Narayanan, {Hall-Hooper}, Hunt, Lower, \& Verrico}]{suess2022}
Suess, K.~A., Kriek, M., Bezanson, R., {et~al.} 2022{\natexlab{a}}, ApJ, 926, 89, \dodoi{10.3847/1538-4357/ac404a}

\bibitem[{Suess {et~al.}(2022{\natexlab{b}})Suess, Leja, Johnson, Bezanson, Greene, Kriek, Lower, Narayanan, Setton, \& Spilker}]{suess2022b}
Suess, K.~A., Leja, J., Johnson, B.~D., {et~al.} 2022{\natexlab{b}}, ApJ, 935, 146, \dodoi{10.3847/1538-4357/ac82b0}

\bibitem[{Suzuki {et~al.}(2022)Suzuki, Glazebrook, Schreiber, Kodama, Kacprzak, Leiton, Nanayakkara, Oesch, Papovich, Spitler, Straatman, Tran, \& Wang}]{suzuki2022}
Suzuki, T.~L., Glazebrook, K., Schreiber, C., {et~al.} 2022, ApJ, 936, 61, \dodoi{10.3847/1538-4357/ac7ce3}

\bibitem[{Tacchella {et~al.}(2018)Tacchella, Bose, Conroy, Eisenstein, \& Johnson}]{tacchella2018}
Tacchella, S., Bose, S., Conroy, C., Eisenstein, D.~J., \& Johnson, B.~D. 2018, ApJ, 868, 92, \dodoi{10.3847/1538-4357/aae8e0}

\bibitem[{Tacchella {et~al.}(2022{\natexlab{a}})Tacchella, Finkelstein, Bagley, Dickinson, Ferguson, Giavalisco, Graziani, Grogin, Hathi, Hutchison, Jung, Koekemoer, Larson, Papovich, Pirzkal, {Rojas-Ruiz}, Song, Schneider, Somerville, Wilkins, \& Yung}]{tacchella2022a}
Tacchella, S., Finkelstein, S.~L., Bagley, M., {et~al.} 2022{\natexlab{a}}, ApJ, 927, 170, \dodoi{10.3847/1538-4357/ac4cad}

\bibitem[{Tacchella {et~al.}(2022{\natexlab{b}})Tacchella, Conroy, Faber, Johnson, Leja, Barro, Cunningham, Deason, Guhathakurta, Guo, Hernquist, Koo, McKinnon, Rockosi, Speagle, {van Dokkum}, \& Yesuf}]{tacchella2022}
Tacchella, S., Conroy, C., Faber, S.~M., {et~al.} 2022{\natexlab{b}}, ApJ, 926, 134, \dodoi{10.3847/1538-4357/ac449b}

\bibitem[{Tacchella {et~al.}(2023)Tacchella, Johnson, Robertson, Carniani, D'Eugenio, Kumari, Maiolino, Nelson, Suess, {\"U}bler, Williams, Adebusola, Alberts, Arribas, Bhatawdekar, Bonaventura, Bowler, Bunker, Cameron, Curti, Egami, Eisenstein, Frye, Hainline, Helton, Ji, Looser, Lyu, Perna, Rawle, Rieke, Rieke, Saxena, Sandles, Shivaei, Simmonds, Sun, Willmer, Willott, \& Witstok}]{tacchella2023}
Tacchella, S., Johnson, B.~D., Robertson, B.~E., {et~al.} 2023, MNRAS, 522, 6236, \dodoi{10.1093/mnras/stad1408}

\bibitem[{Tomczak {et~al.}(2014)Tomczak, Quadri, Tran, Labb{\'e}, Straatman, Papovich, Glazebrook, Allen, Brammer, Kacprzak, Kawinwanichakij, Kelson, McCarthy, Mehrtens, Monson, Persson, Spitler, Tilvi, \& {van Dokkum}}]{tomczak2014}
Tomczak, A.~R., Quadri, R.~F., Tran, K.-V.~H., {et~al.} 2014, ApJ, 783, 85, \dodoi{10.1088/0004-637X/783/2/85}

\bibitem[{Trebitsch {et~al.}(2018)Trebitsch, Volonteri, Dubois, \& Madau}]{trebitsch2018}
Trebitsch, M., Volonteri, M., Dubois, Y., \& Madau, P. 2018, MNRAS, 478, 5607, \dodoi{10.1093/mnras/sty1406}

\bibitem[{Trussler {et~al.}(2020)Trussler, Maiolino, Maraston, Peng, Thomas, Goddard, \& Lian}]{trussler2020}
Trussler, J., Maiolino, R., Maraston, C., {et~al.} 2020, MNRAS, 491, 5406, \dodoi{10.1093/mnras/stz3286}

\bibitem[{Valentino {et~al.}(2020)Valentino, Tanaka, Davidzon, Toft, {G{\'o}mez-Guijarro}, Stockmann, Onodera, Brammer, Ceverino, Faisst, Gallazzi, Hayward, Ilbert, Kubo, Magdis, Selsing, Shimakawa, Sparre, Steinhardt, Yabe, \& Zabl}]{valentino2020}
Valentino, F., Tanaka, M., Davidzon, I., {et~al.} 2020, ApJ, 889, 93, \dodoi{10.3847/1538-4357/ab64dc}

\bibitem[{Valentino {et~al.}(2023)Valentino, Brammer, Gould, Kokorev, Fujimoto, Jespersen, Vijayan, Weaver, Ito, Tanaka, Ilbert, Magdis, Whitaker, Faisst, Gallazzi, Gillman, {Gimenez-Arteaga}, {Gomez-Guijarro}, Kubo, Heintz, Hirschmann, Oesch, Onodera, Rizzo, Lee, Strait, \& Toft}]{valentino2023}
Valentino, F., Brammer, G., Gould, K. M.~L., {et~al.} 2023, An {{Atlas}} of {{Color-selected Quiescent Galaxies}} at \$z{$>$}3\$ in {{Public}} \${{JWST}}\$ {{Fields}},  {arXiv}, \dodoi{10.48550/arXiv.2302.10936}

\bibitem[{Vijayan {et~al.}(2021)Vijayan, Lovell, Wilkins, Thomas, Barnes, Irodotou, Kuusisto, \& Roper}]{vijayan2021}
Vijayan, A.~P., Lovell, C.~C., Wilkins, S.~M., {et~al.} 2021, MNRAS, 501, 3289, \dodoi{10.1093/mnras/staa3715}

\bibitem[{Vijayaraghavan \& Ricker(2013)}]{vijayaraghavan2013}
Vijayaraghavan, R., \& Ricker, P.~M. 2013, MNRAS, 435, 2713, \dodoi{10.1093/mnras/stt1485}

\bibitem[{Vulcani {et~al.}(2021)Vulcani, Poggianti, Moretti, Franchetto, Bacchini, McGee, Jaff{\'e}, Mingozzi, Werle, Tomi{\v c}i{\'c}, Fritz, Bettoni, Wolter, \& Gullieuszik}]{vulcani2021}
Vulcani, B., Poggianti, B.~M., Moretti, A., {et~al.} 2021, ApJ, 914, 27, \dodoi{10.3847/1538-4357/abf655}

\bibitem[{Weaver {et~al.}(2022)Weaver, Kauffmann, Ilbert, McCracken, Moneti, Toft, Brammer, Shuntov, Davidzon, Hsieh, Laigle, Anastasiou, Jespersen, Vinther, Capak, Casey, McPartland, {Milvang-Jensen}, Mobasher, Sanders, Zalesky, Arnouts, Aussel, Dunlop, Faisst, Franx, Furtak, Fynbo, Gould, Greve, Gwyn, Kartaltepe, Kashino, Koekemoer, Kokorev, Le~F{\`e}vre, Lilly, Masters, Magdis, Mehta, Peng, Riechers, Salvato, Sawicki, Scarlata, Scoville, Shirley, Silverman, Sneppen, Smol{\u c}i{\'c}, Steinhardt, Stern, Tanaka, Taniguchi, Teplitz, Vaccari, Wang, \& Zamorani}]{weaver2022}
Weaver, J.~R., Kauffmann, O.~B., Ilbert, O., {et~al.} 2022, ApJ Suppl. Ser., 258, 11, \dodoi{10.3847/1538-4365/ac3078}

\bibitem[{Whitaker {et~al.}(2012)Whitaker, Kriek, {van Dokkum}, Bezanson, Brammer, Franx, \& Labb{\'e}}]{whitaker2012}
Whitaker, K.~E., Kriek, M., {van Dokkum}, P.~G., {et~al.} 2012, ApJ, 745, 179, \dodoi{10.1088/0004-637X/745/2/179}

\bibitem[{Whitaker {et~al.}(2010)Whitaker, {van Dokkum}, Brammer, Kriek, Franx, Labb{\'e}, Marchesini, Quadri, Bezanson, Illingworth, Lee, Muzzin, Rudnick, \& Wake}]{whitaker2010}
Whitaker, K.~E., {van Dokkum}, P.~G., Brammer, G., {et~al.} 2010, ApJ, 719, 1715, \dodoi{10.1088/0004-637X/719/2/1715}

\bibitem[{Whitaker {et~al.}(2011)Whitaker, Labb{\'e}, {van Dokkum}, Brammer, Kriek, Marchesini, Quadri, Franx, Muzzin, Williams, Bezanson, Illingworth, Lee, Lundgren, Nelson, Rudnick, Tal, \& Wake}]{whitaker2011}
Whitaker, K.~E., Labb{\'e}, I., {van Dokkum}, P.~G., {et~al.} 2011, ApJ, 735, 86, \dodoi{10.1088/0004-637X/735/2/86}

\bibitem[{Whitaker {et~al.}(2019)Whitaker, Ashas, Illingworth, Magee, Leja, Oesch, {van Dokkum}, Mowla, Bouwens, Franx, Holden, Labb{\'e}, Rafelski, Teplitz, \& Gonzalez}]{whitaker2019}
Whitaker, K.~E., Ashas, M., Illingworth, G., {et~al.} 2019, ApJ Suppl. Ser., 244, 16, \dodoi{10.3847/1538-4365/ab3853}

\bibitem[{Whitaker {et~al.}(2021{\natexlab{a}})Whitaker, Williams, Mowla, Spilker, Toft, Narayanan, Pope, Magdis, {van Dokkum}, Akhshik, Bezanson, Brammer, Leja, Man, Nelson, Richard, Pacifici, Sharon, \& Valentino}]{whitaker2021b}
Whitaker, K.~E., Williams, C.~C., Mowla, L., {et~al.} 2021{\natexlab{a}}, Nature, 597, 485, \dodoi{10.1038/s41586-021-03806-7}

\bibitem[{Whitaker {et~al.}(2021{\natexlab{b}})Whitaker, Narayanan, Williams, Li, Spilker, Dav{\'e}, Akhshik, Akins, Bezanson, Katz, Leja, Magdis, Mowla, Nelson, Pope, Privon, Toft, \& Valentino}]{whitaker2021}
Whitaker, K.~E., Narayanan, D., Williams, C.~C., {et~al.} 2021{\natexlab{b}}, ApJ, 922, L30, \dodoi{10.3847/2041-8213/ac399f}

\bibitem[{Wild {et~al.}(2016)Wild, Almaini, Dunlop, Simpson, Rowlands, Bowler, Maltby, \& McLure}]{wild2016}
Wild, V., Almaini, O., Dunlop, J., {et~al.} 2016, MNRAS, 463, 832, \dodoi{10.1093/mnras/stw1996}

\bibitem[{Wilkins {et~al.}(2011)Wilkins, Bunker, Stanway, Lorenzoni, \& Caruana}]{wilkins2011}
Wilkins, S.~M., Bunker, A.~J., Stanway, E., Lorenzoni, S., \& Caruana, J. 2011, MNRAS, 417, 717, \dodoi{10.1111/j.1365-2966.2011.19315.x}

\bibitem[{Williams {et~al.}(2021)Williams, Spilker, Whitaker, Dav{\'e}, Woodrum, Brammer, Bezanson, Narayanan, \& Weiner}]{williams2021}
Williams, C.~C., Spilker, J.~S., Whitaker, K.~E., {et~al.} 2021, ApJ, 908, 54, \dodoi{10.3847/1538-4357/abcbf6}

\bibitem[{Williams {et~al.}(2023)Williams, Alberts, Ji, Hainline, Lyu, Rieke, Endsley, Suess, Johnson, Florian, Shivaei, Rujopakarn, Baker, Bhatawdekar, Boyett, Bunker, Carniani, Charlot, {Curtis-Lake}, DeCoursey, {de Graaff}, Egami, Eisenstein, Gibson, Hausen, Helton, Maiolino, Maseda, Nelson, {Perez-Gonzalez}, Rieke, Robertson, Sun, Tacchella, Willmer, \& Willott}]{williams2023a}
Williams, C.~C., Alberts, S., Ji, Z., {et~al.} 2023, The Galaxies Missed by {{Hubble}} and {{ALMA}}: The Contribution of Extremely Red Galaxies to the Cosmic Census at 3, \dodoi{10.48550/arXiv.2311.07483}

\bibitem[{Williams {et~al.}(2009)Williams, Quadri, Franx, {van Dokkum}, \& Labb{\'e}}]{williams2009}
Williams, R.~J., Quadri, R.~F., Franx, M., {van Dokkum}, P., \& Labb{\'e}, I. 2009, ApJ, 691, 1879, \dodoi{10.1088/0004-637X/691/2/1879}

\bibitem[{Woodrum {et~al.}(2022)Woodrum, Williams, Rieke, Leja, Johnson, Bezanson, Kennicutt, Spilker, \& Tacchella}]{woodrum2022}
Woodrum, C., Williams, C.~C., Rieke, M., {et~al.} 2022, ApJ, 940, 39, \dodoi{10.3847/1538-4357/ac9af7}

\bibitem[{Xiao {et~al.}(2023)Xiao, Oesch, Elbaz, Bing, Nelson, Weibel, Naidu, Daddi, Bouwens, Matthee, Wuyts, Chisholm, Brammer, Dickinson, Magnelli, Leroy, {van Dokkum}, Schaerer, {Herard-Demanche}, Barrufet, Endsley, Fudamoto, {G{\'o}mez-Guijarro}, Gottumukkala, Illingworth, Labbe, Magee, Marchesini, Maseda, Qin, Reddy, Shapley, Shivaei, Shuntov, Stefanon, Whitaker, \& Wyithe}]{xiao2023}
Xiao, M., Oesch, P., Elbaz, D., {et~al.} 2023, Massive {{Optically Dark Galaxies Unveiled}} by {{JWST Challenge Galaxy Formation Models}}, \dodoi{10.48550/arXiv.2309.02492}

\bibitem[{Yang {et~al.}(2023)Yang, Papovich, Bagley, Ferguson, Finkelstein, Koekemoer, {P{\'e}rez-Gonz{\'a}lez}, Haro, Bisigello, Caputi, Cheng, Costantin, Dickinson, Fontana, Gardner, Grazian, Grogin, Harish, Holwerda, Iani, Kartaltepe, Kewley, Kirkpatrick, Kocevski, Kokorev, Lotz, Lucas, {Navarro-Carrera}, Pentericci, Pirzkal, Ravindranath, Rinaldi, Shen, Somerville, Trump, {de la Vega}, Wilkins, \& Yung}]{yang2023}
Yang, G., Papovich, C., Bagley, M., {et~al.} 2023, {{CEERS MIRI Imaging}}: {{Data Reduction}} and {{Quality Assessment}},  {arXiv}, \dodoi{10.48550/arXiv.2307.14509}

\end{thebibliography}
\bibliographystyle{aasjournal}



\end{document}